\documentclass[12pt,prd,onecolumn,showpacs,amsmath,amssymb,aps,floats,floatfix,nofootinbib]{revtex4-1}
\usepackage[colorlinks=true,urlcolor=blue,anchorcolor=blue,citecolor=blue,filecolor=blue,linkcolor=blue,menucolor=blue,pagecolor=blue,linktocpage=true]{hyperref} 

\usepackage[inline]{enumitem}
\usepackage[multidot]{grffile} % allow the name of figures to include dots
\usepackage{dcolumn}
\usepackage{bm}
\usepackage{amsmath}
\usepackage{amsfonts}
\usepackage{amssymb}
\usepackage{color}
\usepackage{latexsym}
\usepackage{slashed} % slash mark
\usepackage{pstricks}
\usepackage{indentfirst}
\usepackage{mathrsfs}
\usepackage{multirow}
\usepackage{epsfig,psfrag}
\usepackage{subfigure}
\usepackage{mathtools}
\usepackage{setspace} % spacing
\usepackage[utf8]{inputenc} % accept utf-8 input encoding
\usepackage[scientific-notation=true]{siunitx} % comprehensive units
\usepackage{verbatim}

\graphicspath{{fig/}}
\newcommand{\uvec}{\boldsymbol}
\newcommand{\ud}{\mathrm{d}}

\allowdisplaybreaks % allow eqnarray breaks
\makeatother

\begin{document}
%%%%%%%%%%%%%%%%%%%%%%%%%%%%%%%%%%%%%%%%%%%%%%%%%%%%%%%%%%%%%%%%%%%%%%%%%%%%%%%%%%
\title{Pion and nucleon relativistic electromagnetic four-current distributions}
\author{Yi Chen}
\affiliation{Interdisciplinary Center for Theoretical Study and Department of Modern Physics, University of Science and Technology of China, Hefei, Anhui 230026, China}
\affiliation{Shanghai Institute of Applied Physics, Chinese Academy of Sciences, Shanghai 201800, China}
\affiliation{University of Chinese Academy of Sciences, Beijing 100049, China}
\author{C\'edric Lorc\'e}~\email[Corresponding author: ]{cedric.lorce@polytechnique.edu}
\affiliation{CPHT, CNRS, \'Ecole polytechnique, Institut Polytechnique de Paris, 91120 Palaiseau, France}
%
%\author{Qun Wang}~\email[Corresponding author: ]{qunwang@ustc.edu.cn}
%\affiliation{Interdisciplinary Center for Theoretical Study and Department of Modern Physics, University of Science and Technology of China, Hefei, Anhui 230026, China}
%\affiliation{Peng Huanwu Center for Fundamental Theory, Hefei, Anhui 230026, China}

%%%%%%%%%%%%%%%%%%%%%%%%%%%%%%%%%%%%%%%%%%%%%%%%%%%%%%%%%%%%%%%%%%%%%%%%%%%%%%%%%%

\begin{abstract}

The quantum phase-space approach allows one to define relativistic spatial distributions inside a target with arbitrary spin and arbitrary average momentum. We apply this quasiprobabilistic formalism to the whole electromagnetic four-current operator in the case of spin-$0$ and spin-$\frac{1}{2}$ targets, study in detail the frame dependence of the corresponding spatial distributions, and compare our results with those from the light-front formalism. While former works focused on the charge distributions, we extend here the investigations to the current distributions. We clarify the role played by the Wigner rotation and argue that electromagnetic properties are most naturally understood in terms of Sachs form factors, contrary to what the light-front formalism previously suggested. Finally, we illustrate our results using the pion and nucleon electromagnetic form factors extracted from experimental data. 

\end{abstract}

\maketitle

\newpage
%%%%%%%%%%%%%%%%%%%%%%%%%%%%%%%%%%%%%%%%%%%%%%%%%%%%%%%%%%%%%%%%%%%%%%%%%%%%%%%%%%
\section{Introduction}
\label{Introduction}

Pions and nucleons are key systems to study for understanding quantum chromodynamics (QCD). Pions, the lightest bound states in QCD, play a special role since they are the (pseudo) Nambu-Goldstone bosons associated with the dynamical breakdown of chiral symmetry~\cite{Weinberg1995Vol1}. Nucleons are by far the most abundant (known) hadrons in nature, responsible for more than $99\%$ of the visible matter in the universe~\cite{Gao:2021sml}. Pions and nucleons have very different masses originating from their different, rich and complicated internal structures, which constitutes a fundamental puzzle for modern physics.

The electromagnetic structure of hadrons is encoded in Lorentz-invariant functions known as form factors (FFs). They have been measured with extreme precision in various scattering experiments over the past decades~\cite{NA7:1986vav,Anklin:1994ae,Qattan:2004ht,JeffersonLabFpi-2:2006ysh,Arrington:2007ux,JeffersonLab:2008jve,CLAS:2008idi,A1:2010nsl,Zhan:2011ji,Puckett:2011xg,BaBar:2012bdw,A1:2013fsc,Punjabi:2015bba,BESIII:2015equ,Puckett:2017flj,SANE:2018cub,Xiong:2019umf,Mihovilovic:2019jiz,PRad:2020oor,Atac:2021wqj,Zhou:2021gyh}. On the theory side, lattice QCD calculations of these FFs have witnessed tremendous progress in the last few years~\cite{Alexandrou:2017ypw,Hasan:2017wwt,Shintani:2018ozy,Alexandrou:2018sjm,Jang:2019jkn,Alexandrou:2020aja,Wang:2020nbf,Park:2021ypf,Ishikawa:2021eut,Bar:2021crj,Djukanovic:2021cgp,Djukanovic:2021qxp,Alexandrou:2021ztx}. Recent reviews on the extraction and the physics associated with electromagnetic FFs can be found in Refs.~\cite{Gao:2021sml,Arrington:2006zm,Perdrisat:2006hj,Pacetti:2014jai,Punjabi:2015bba}. 

According to textbooks, electromagnetic FFs can be interpreted as 3D Fourier transforms of charge and magnetization densities in the Breit frame (BF)~\cite{Ernst:1960zza,Sachs:1962zzc}. However, relativistic recoil corrections spoil their interpretation as probabilistic distributions~\cite{Yennie:1957rmp,Breit:1964ga,Kelly:2002if,Burkardt:2000za,Belitsky:2003nz,Jaffe:2020ebz}. Switching to the light-front formalism allows one to define alternative 2D charge densities free of these issues~\cite{Burkardt:2002hr,Miller:2007uy,Carlson:2007xd,Alexandrou:2008bn,Alexandrou:2009hs,Gorchtein:2009qq,Carlson:2009ovh,Miller:2010nz,Miller:2018ybm}, but there is a price to pay. Besides losing one spatial dimension these light-front distributions also display various distortions, which are sometimes hard to reconcile with an intuitive picture of the system at rest.

The concept of relativistic spatial distribution has recently been revisited in several works with the goal of clarifying the relation between 3D Breit frame and 2D light-front definitions; see e.g.~\cite{Panteleeva:2021iip,Freese:2021mzg,Kim:2021kum,Kim:2022bia,Epelbaum:2022fjc,Panteleeva:2022khw,Li:2022ldb,Carlson:2022eps,Freese:2022fat}. In this paper, we adopt the quantum phase-space approach where the physical interpretation is relaxed to a quasiprobabilistic\footnote{In the probabilistic picture, the state is perfectly localized in position space and the expectation value of an operator $O$ is written as $\langle O\rangle=\int\ud^3R\,|\Psi(\uvec R)|^2O(\uvec R)$ with $\Psi(\uvec R)$ the position space wave packet. In the quasiprobabilistic picture, the same expectation value is expressed as $\langle O\rangle=\int\frac{\ud^3P}{(2\pi)^3}\,\ud^3R\,\rho_\Psi(\uvec R,\uvec P)O(\uvec R,\uvec P)$ with $\rho_\Psi(\uvec R,\uvec P)$ the Wigner distribution, a real-valued function constructed from $\Psi(\uvec R)$ and describing the localization of the system in phase space~\cite{Wigner:1932eb,Hillery:1983ms}. Due to Heisenberg's uncertainty principle, Wigner distributions are negative in some regions and cannot therefore be interpreted as strict probability densities.} one, allowing one to define relativistic spatial distributions inside a target with arbitrary spin and arbitrary average momentum~\cite{Lorce:2017wkb,Lorce:2018zpf,Lorce:2018egm,Lorce:2020onh,Lorce:2022jyi,Lorce:2022cle}. This formalism is particularly appealing since it provides a natural and smooth connection between the Breit frame and the infinite-momentum frame pictures, and allows one to understand the distortions in the light-front distributions in terms of relativistic kinematical effects. While the discussions in the literature have essentially focused on the charge distributions, we extend here the study to the whole electromagnetic four-current and demonstrate the consistency of the phase-space picture.

This paper is organized as follows. In Sec.~\ref{sec:Elastic frame distributions}, we quickly review the concept of generic elastic frame distributions within the quantum phase-space approach. We start our analysis in Sec.~\ref{sec:Spin-0 case} with a spin-$0$ target, introducing relativistic electromagnetic four-current distributions and studying their frame dependence. Then we compare with the corresponding light-front distributions and illustrate our findings using the pion ($\pi^+$) electromagnetic form factor extracted from experimental data. We proceed in Sec.~\ref{sec:Spin-1/2 case} with a spin-$\frac{1}{2}$ target. We discuss in detail the non-trivial role played by the Wigner spin rotation and show that electromagnetic properties are most naturally understood in terms of the Sachs form factors. Here we also compare with the light-front formalism and illustrate our results using the nucleon electromagnetic form factors extracted from experimental data. Finally, we summarize our findings in Sec.~\ref{sec:Summary}, and provide further discussions and details in three Appendices.

%%%%%%%%%%%%%%%%%%%%%%%%%%%%%%%%%%%%%%%%%%%%%%%%%%%%%%%%%%%%%%%%%%%%%%%%%%%%%%%%%%
\section{Elastic frame distributions}
\label{sec:Elastic frame distributions}

Two-dimensional spatial distributions in a generic elastic frame (EF) have been introduced in~\cite{Lorce:2017wkb} to study the angular momentum inside the nucleon. They are fully relativistic objects that can be interpreted as the internal distributions associated with a target localized in phase-space (i.e.~with definite average momentum and position) in the Wigner sense~\cite{Lorce:2018zpf,Lorce:2018egm,Lorce:2021gxs}. Although they have in general only a quasiprobabilistic interpretation, they provide a natural connection between spatial distributions defined in the BF and in the infinite-momentum frame (IMF)~\cite{Kogut:1969xa}.

For convenience, we choose the $z$-axis along the average total momentum of the target $\uvec P=\frac{1}{2}(\uvec p'+\uvec p)=(\uvec 0_\perp,P_z)$. The spatial distributions of the electromagnetic four-current are then defined within the phase-space formalism as~\cite{Lorce:2020onh}
\begin{equation}\label{EF-distribution:def}
	J^\mu_\text{EF}(\uvec b_{\perp};P_z)\equiv\int\frac{\ud^2\Delta_\perp}{(2\pi)^2}\,e^{-i\uvec\Delta_\perp\cdot\uvec b_\perp}\,\frac{\langle p',s'|\hat j^\mu(0)|p,s\rangle}{2P^0}\bigg|_{\Delta_z=0},
\end{equation}
where $\hat j^\mu(x)$ is the electromagnetic four-current operator, $\Delta=p'-p$ is the four-momentum transfer, and $\uvec b_\perp$ is the transverse position relative to the average center of the target. The four-momentum eigenstates are normalized as $\langle p',s'|p,s\rangle=2p^0(2\pi)^3\delta^{(3)}(\uvec p'-\uvec p)\,\delta_{s's}$ with $s$ and $s'$ the usual canonical spin labels. The elastic condition $\Delta^0=\uvec P\cdot\uvec\Delta/P^0=0$ is automatically ensured by the restriction\footnote{In the BF there is no need to set $\Delta_z=0$ since by definition $\uvec P=\uvec 0$, and so one can define in that case a static three-dimensional spatial distribution.} $\Delta_z=0$ and implies that the resulting distribution does not depend on time~\cite{Lorce:2017wkb}. Note that the factor $2P^0$ in the denominator of Eq.~\eqref{EF-distribution:def} appears naturally in the phase-space formalism and ensures that the total electric charge
\begin{equation}
	q=\int\ud^2b_\perp\,J^0_\text{EF}(\uvec b_{\perp};P_z)\big|_{s'=s}= \frac{\langle p,s|\hat j^0(0)|p,s\rangle}{2p^0}  
\end{equation}
transforms as a Lorentz scalar~\cite{Friar:1975bk,Lorce:2020onh}. One can also formally write
\begin{equation}
    \frac{\langle p,s|\hat j^0(0)|p,s\rangle}{2p^0}=\frac{\langle p,s| \int\ud^3r\,\hat j^0(\uvec r)|p,s\rangle}{\langle p,s|p,s\rangle},
\end{equation}
indicating that the total charge does not depend on the particular choice made for the normalization of the four-momentum eigenstates. 

Using Poincar\'e and discrete spacetime symmetries, the off-forward matrix elements $\langle p',s'|\hat{j}^\mu(0)|p,s\rangle$ for a spin-$j$ target can be parameterized in terms of $(2j+1)$ Lorentz-invariant electromagnetic FFs~\cite{Durand:1962zza,Scadron:1968zz,Lorce:2009bs,Cotogno:2019vjb}. Different sets of FFs for a given spin value $j$ have been considered in the literature. These sets are all physically equivalent since they simply correspond to different choices for the basis of Lorentz tensors, and hence are linearly related to each other. A particularly convenient and physically transparent basis is provided by the multipole expansion in the BF, i.e.~for $\uvec P=\uvec 0$. In that frame, the multipole structure appears to be the same as in the non-relativistic theory: the charge distribution consists of a tower of electric multipoles of even order, while the electric current can be expressed in terms of a tower of magnetic multipoles of odd order~\cite{Scwhartz:1955,Kleefeld:2000nv,Lorce:2009bs}.

Poincar\'e symmetry can also be used to determine how matrix elements of the electromagnetic four-current operator in different Lorentz frames are related to each other. One can write in general~\cite{Jacob:1959at,Durand:1962zza}
\begin{equation}
	\begin{aligned}\label{EFLorentzTrans-Spinj}
		\langle p',s'|\hat j^\mu(0)|p,s\rangle &= \sum_{s_{B}',s_{B}} {D}^{*(j)}_{s_{B}'s'}(p_{B}',\Lambda){D}^{(j)}_{s_{B}s}(p_{B},\Lambda) \,\Lambda^{\mu}_{\phantom{\mu}\nu} \,\langle p_{B}',s_{B}'|\hat j^{\nu}(0)|p_{B},s_{B}\rangle,
	\end{aligned}
\end{equation}
where $\langle p_{B}',s_{B}'|\hat j^{\nu}(0)|p_{B},s_{B}\rangle$ is the BF matrix element, $\Lambda^{\mu}_{\phantom{\mu}\nu}$ is the Lorentz boost from the BF to a generic Lorentz frame, and ${D}^{(j)}$ is the Wigner rotation matrix for spin-$j$ targets, see Appendix~\ref{App-Wigner-Melosh Rotations}. Since temporal and spatial components of the electromagnetic four-current get mixed under a Lorentz boost, the odd magnetic multipoles in the BF will induce odd electric multipoles in a generic Lorentz frame. Similarly, even electric multipoles in the BF will induce even magnetic multipoles in a generic Lorentz frame. These odd electric and even magnetic multipoles do not break parity ($\mathsf{P}$) nor time-reversal ($\mathsf{T}$) symmetries, and should therefore not be confused with the $\mathsf{P}$- and $\mathsf{T}$-breaking ones which are not considered in this work. Wigner rotations complicate further the relation~\eqref{EFLorentzTrans-Spinj} by reorganizing the multipole weights. Namely, any particular multipole in the BF will usually generate a contribution to \emph{all}\,\footnote{A similar mechanism explains why relations between transverse-momentum dependent parton distributions and orbital angular momentum appear in various models of the nucleon~\cite{Lorce:2011zta,Lorce:2011kn}.} multipoles in a generic Lorentz frame~\cite{Lorce:2020onh}.
\newline

In the following we will focus on the spin-$0$ and spin-$\frac{1}{2}$ targets, and apply our formalism to map the electromagnetic four-current distributions inside a pion and a nucleon using the electromagnetic FFs extracted from experimental data. While the EF charge distributions $J^0_\text{EF}$ have already been discussed in Refs.~\cite{Lorce:2020onh,Kim:2021kum}, the EF currents $\uvec J_\text{EF}$ will be studied here for the first time.

%%%%%%%%%%%%%%%%%%%%%%%%%%%%%%%%%%%%%%%%%%%%%%%%%%%%%%%%%%%%%%%%%%%%%%%%%%%%%%%%%%
\section{Spin-$0$ target}
\label{sec:Spin-0 case}

Let us start with the simplest case, namely a spin-$0$ target. The matrix elements of the electromagnetic four-current operator are parametrized in terms of a single FF
\begin{equation}
	\langle p'|\hat j^\mu(0)|p\rangle=e\,2P^\mu F(Q^2)
\end{equation}
with $Q^2=-\Delta^2$ and $e$ the electric charge of a proton. 

%%%%%%%%%%%%%%%%%%%%%%%%%%%%%%%%%%%%%%%%%%%%%%%%%%%%%%%%%%%%%%%%%%%%%%%%%%%%%%%%%%
\subsection{Breit frame distributions}

The BF electromagnetic four-current distributions are defined as
\begin{equation}\label{BFdef0}
    J^\mu_B(\uvec r)\equiv\int\frac{\ud^3 \Delta}{(2\pi)^3}\,e^{-i\uvec\Delta\cdot\uvec r}\,\frac{\langle p'_B|\hat j^\mu(0)|p_B\rangle}{2P^0_B}
\end{equation}
with $\uvec p'_B=-\uvec p_B=\uvec\Delta/2$ and $p'^0_B=p^0_B=P^0_B=M\sqrt{1+\tau}$. We introduced for convenience the Lorentz invariant quantity $\tau=Q^2/4M^2=\uvec\Delta^2/4M^2$ which measures the magnitude of relativistic effects. 

The BF charge distribution is obtained by considering the $\mu=0$ component in Eq.~\eqref{BFdef0}. Like in the non-relativistic theory, it corresponds simply to the 3D Fourier transform of the electromagnetic FF~\cite{Hofstadter:1956qs}
\begin{equation}\label{spin0-J0B}
     J^0_B(\uvec r)=e\int\frac{\ud^3 \Delta}{(2\pi)^3}\,e^{-i\uvec\Delta\cdot\uvec r}\,F(\uvec\Delta^2).
\end{equation}
It is spherically symmetric since there is no preferred spatial direction when $\uvec P=\uvec 0$. 

The BF current distribution for a spin-$0$ target is directly proportional to $\uvec P$ and hence vanishes in the BF
\begin{equation}
    \uvec J_B(\uvec r)=\uvec 0,
\end{equation}
which is consistent with the interpretation of the BF as the average rest frame of the system within the quantum phase-space approach~\cite{Lorce:2018zpf,Lorce:2018egm,Lorce:2020onh}. 

The BF picture agrees with our naive expectation for a spin-$0$ system at rest. In order to see what happens when the system has non-zero average momentum, we need to switch to the concept of EF distributions.

%%%%%%%%%%%%%%%%%%%%%%%%%%%%%%%%%%%%%%%%%%%%%%%%%%%%%%%%%%%%%%%%%%%%%%%%%%%%%%%%%%
\subsection{Elastic frame distributions}

Applying the definition~\eqref{EF-distribution:def} to the case of a spin-$0$ target, we find that the EF electromagnetic four-current distributions can be expressed as
\begin{equation}
	J^\mu_\text{EF}(\uvec b_\perp;P_z)=e\int\frac{\ud^2 \Delta_\perp}{(2\pi)^2}\,e^{-i\uvec\Delta_\perp\cdot\uvec b_\perp}\,\frac{P^\mu}{P^0}\,F(\uvec\Delta_\perp^2).
\end{equation}
They are axially symmetric since there is no preferred transverse direction. At $P_z=0$, they coincide with the projection of the BF electromagnetic four-current distributions onto the transverse plane
\begin{equation}\label{BFreduc}
    J^\mu_\text{EF}(\uvec b_\perp;0)=\int\ud r_z \, J^\mu_B(\uvec r)
\end{equation}
with $\uvec r=(\uvec b_\perp,r_z)$.

\begin{figure}[h]
	\centering
	{\includegraphics[angle=0,scale=0.45]{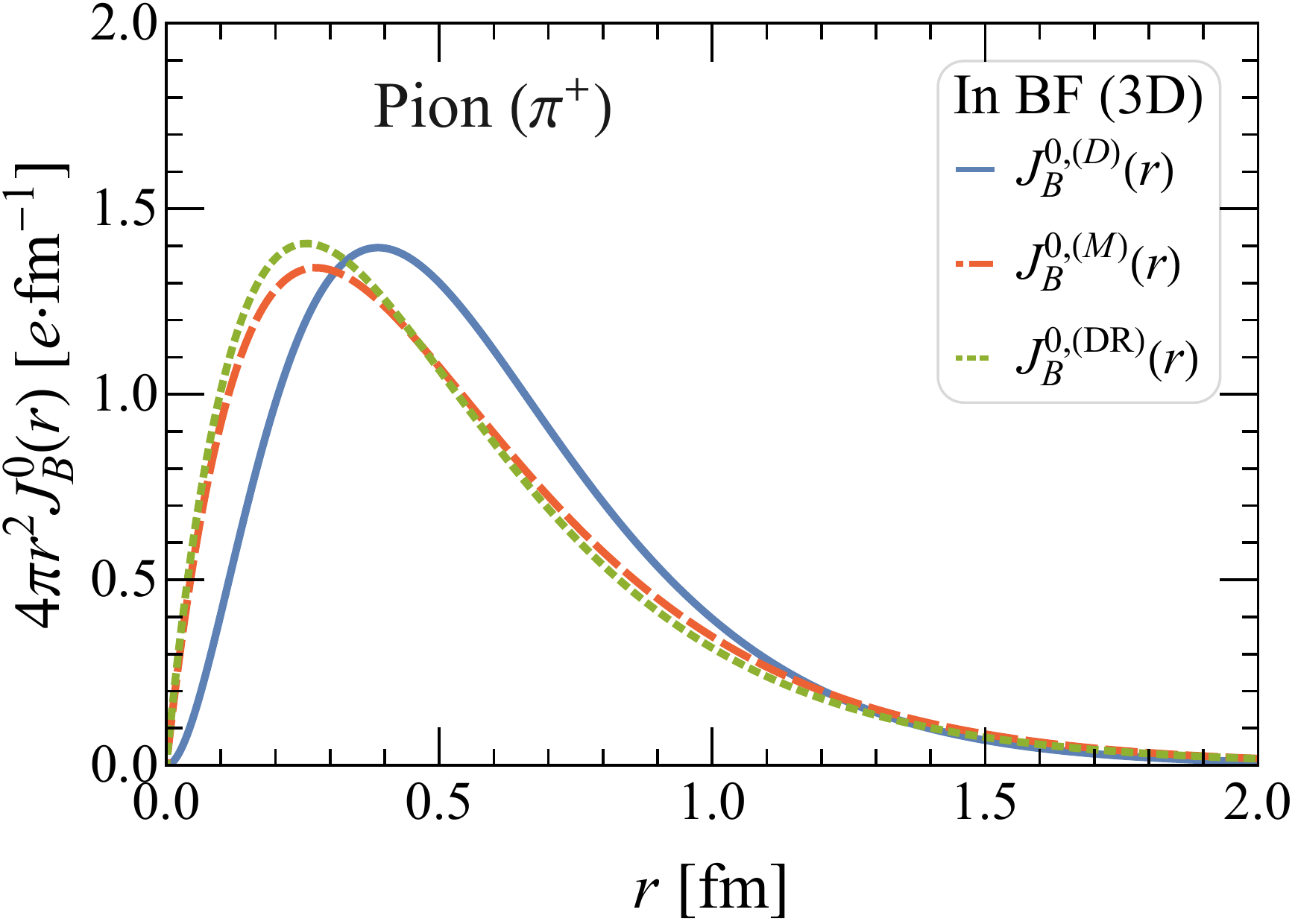}}
	{\includegraphics[angle=0,scale=0.45]{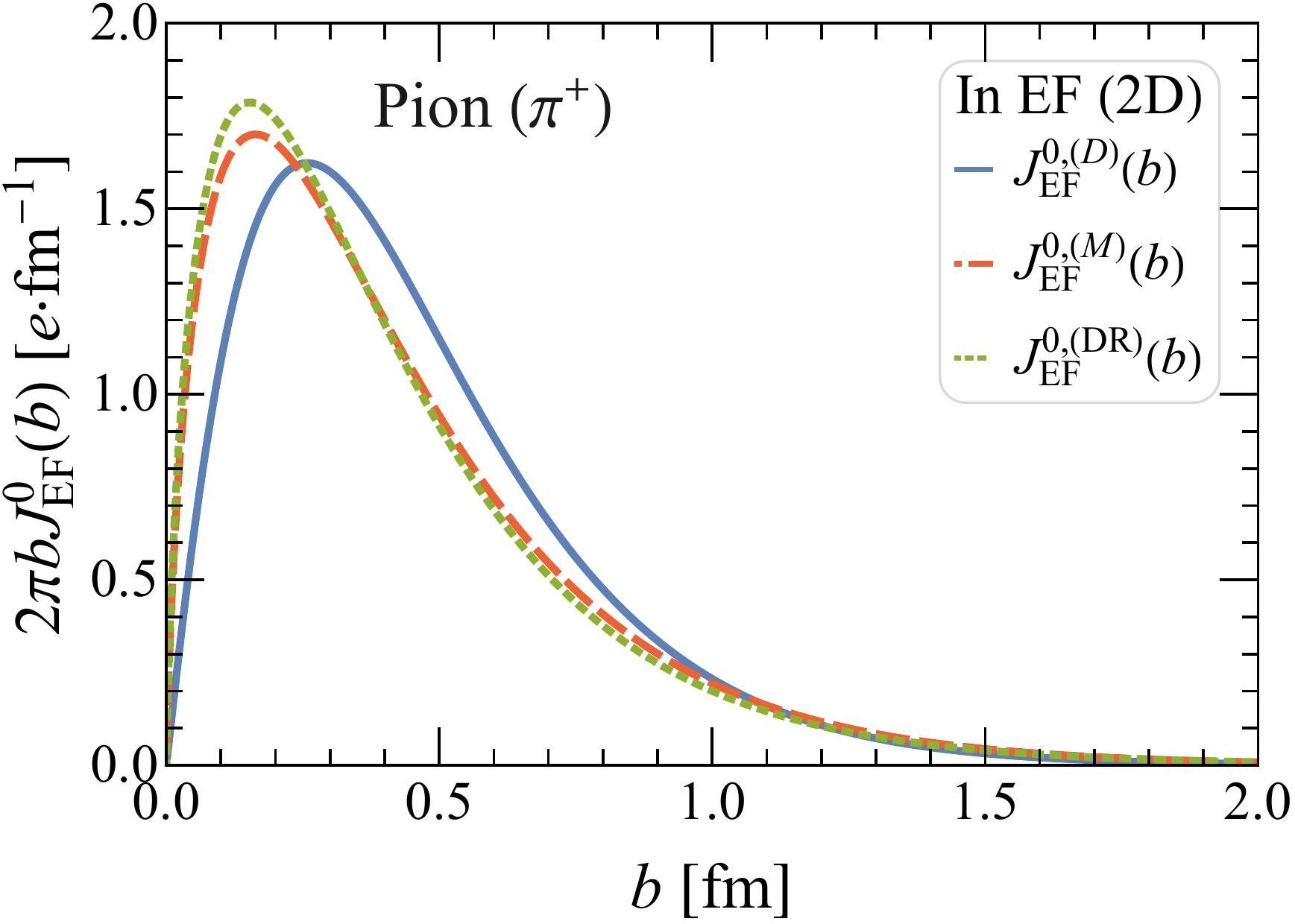}}
	\caption{(Color online) Comparison between the BF (left panel) and EF (right panel) radial charge distributions for a pion ($\pi^+$), based on the pion electromagnetic FFs from simple dipole (solid line) and monopole (dashed line) models in~(\ref{dipolemodel}) and (\ref{monopolemodel}), and modified dispersion relation (dotted line) in~(\ref{modified-DR}).}
	\label{Fig_Pion3DBF2DEFJ0}
\end{figure}

As noted in Ref.~\cite{Lorce:2020onh}, the EF charge distribution
\begin{equation}\label{spin0-J0EF}
	J^0_\text{EF}(\uvec b_\perp;P_z)=e\int\frac{\ud^2\Delta_\perp}{(2\pi)^2}\,e^{-i\uvec\Delta_\perp\cdot\uvec b_{\perp} }\,F(\uvec\Delta_\perp^2)
\end{equation}
does not depend on the target momentum $P_z$. This indicates in particular that the denominator in Eq.~\eqref{EF-distribution:def} properly accounts for Lorentz contraction effects. Indeed, the restriction $\Delta_z=0$ used to define the EF distributions corresponds in position space to an integration over the longitudinal coordinate $\int\ud r_z$. Contrary to its 3D counterpart, the 2D EF charge distribution does not get multiplied by a Lorentz factor $\gamma$ under a longitudinal boost because it is compensated by the Lorentz contraction factor $1/\gamma$ coming from the longitudinal measure $\ud r_z$. The $P_z$-independence implies in particular that the same EF charge distribution is found in the IMF, i.e.~when $P_z\to\infty$. In Fig.~\ref{Fig_Pion3DBF2DEFJ0}, we compare the BF and EF radial charge distributions for various parametrizations of the pion electromagnetic FF, see Appendix~\ref{sec:Pion form factors in both spacelike and timelike regions} for more details. 

\begin{figure}[!tb]
	\centering
	{\includegraphics[angle=0,scale=0.442]{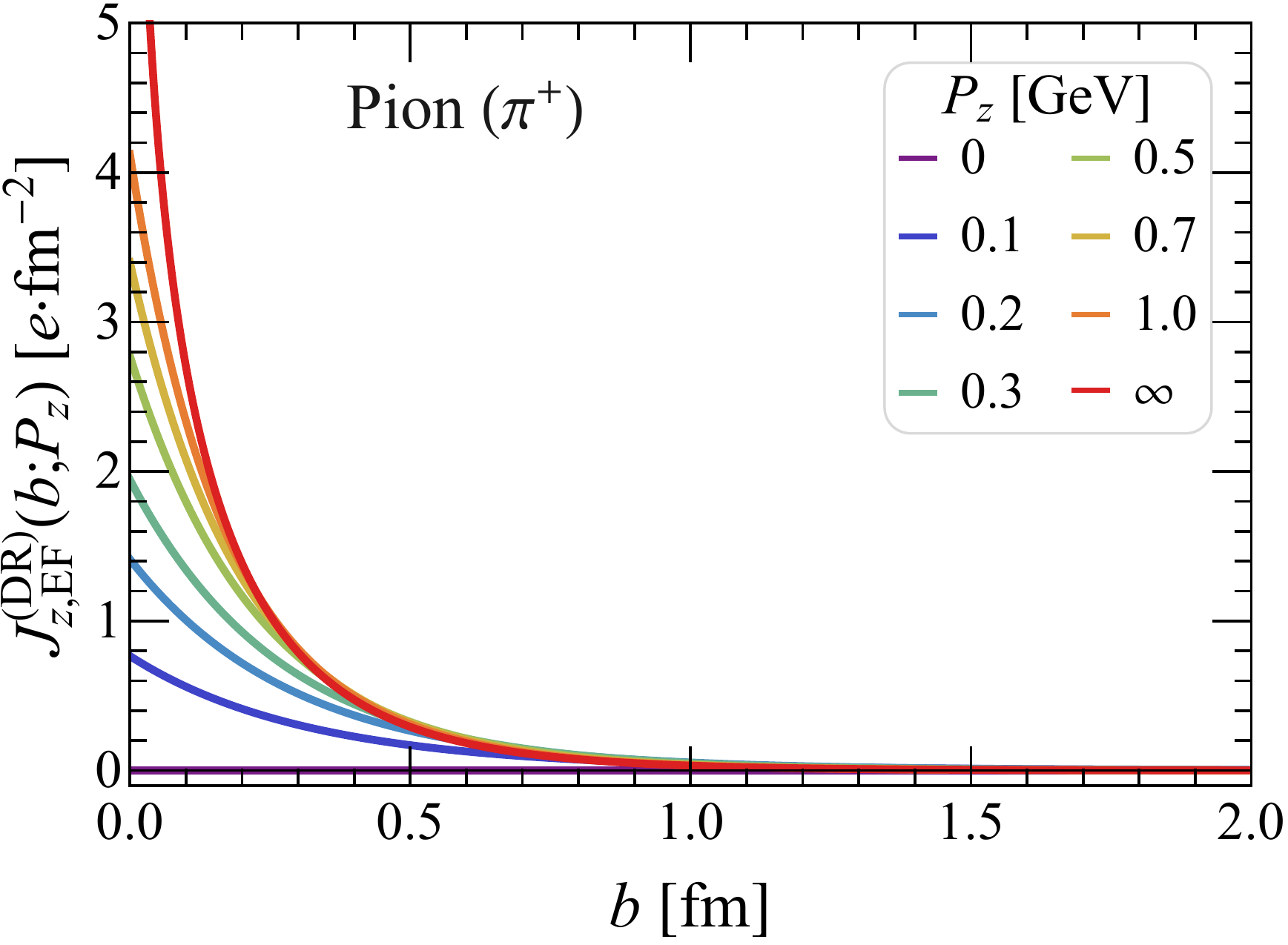}}
	{\includegraphics[angle=0,scale=0.456]{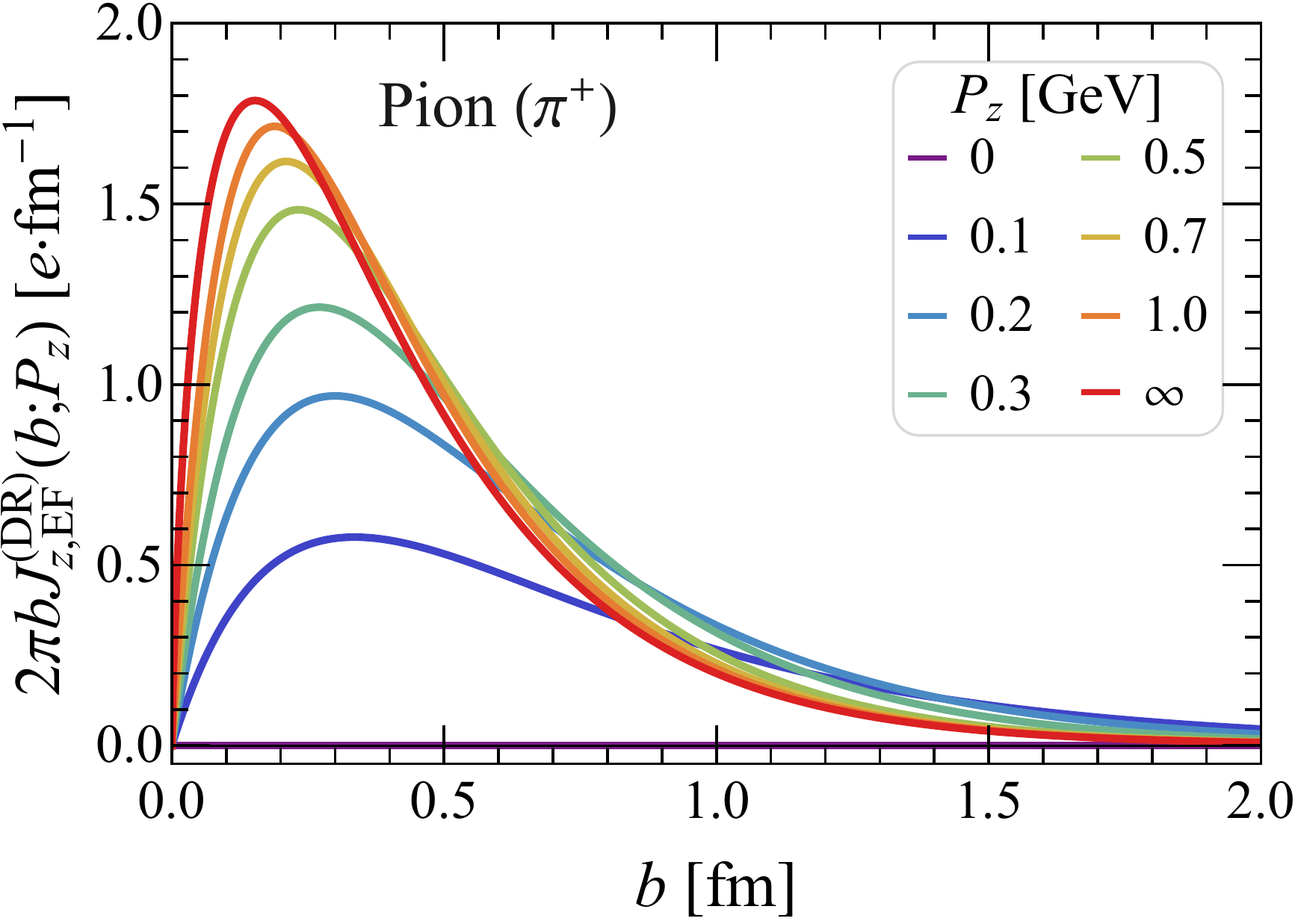}}
	\caption{(Color online) Momentum dependence of the 2D (left panel) and radial (right panel) longitudinal EF current distribution
	for a pion ($\pi^+$), based on the pion electromagnetic FF from modified dispersion relation~\eqref{modified-DR}.}
	\label{Fig_Pion2DEFJz}
\end{figure}

For the EF current distributions, we obtain
\begin{equation}\label{EFcurrent0}
	\uvec J_\text{EF}( \uvec b_\perp;P_z)=e\int\frac{\ud^2\Delta_\perp}{(2\pi)^2}\,e^{-i\uvec\Delta_\perp\cdot\uvec b_\perp}\,\frac{\uvec P}{P^0}\,F(\uvec\Delta_\perp^2).
\end{equation}
Since the average four-momentum $P^\mu$ is a timelike four-vector, we can interpret the quantity $\uvec P/P^0$ as a velocity. For a target of mass $M$, the EF inertia $P^0=p^0=p'^0$ is given by
\begin{equation}\label{P0EF}
	P^0=\sqrt{M^2(1+\tau)+\uvec P^2}=\sqrt{M^2(1+\tau)+P_z^2}.
\end{equation}
As a result of the $\tau$-dependence of $P^0$, the velocity $\uvec P/P^0$ cannot be pulled out of the integral in Eq.~\eqref{EFcurrent0}. Note however that in the IMF the longitudinal current distribution becomes equal to the charge distribution
\begin{equation}\label{spin0-JzJ0Equal}
    J_{z,\text{EF}}(\uvec b_\perp;\infty)=  J^0_\text{EF}(\uvec b_\perp;\infty)
\end{equation}
because the velocity $P_z/P^0$ tends to $1$ (i.e.~the speed of light) for all values of the momentum transfer. In Fig.~\ref{Fig_Pion2DEFJz}, we show how the longitudinal EF current distribution for the pion changes with $P_z$.

The indefiniteness of inertia (i.e.~the $\tau$-dependence of $P^0$) is a major impediment for the physical interpretation of relativistic spatial distributions. We observed at least three main responses to this problem in the literature. The first one consists in focusing only on those (``good'') components that do not involve $P^0$ explicitly, and ignoring the other (``bad'') components. The second response is to invoke ``relativistic corrections'' and introduce \emph{by hand} some factors, like $P^0/M$ or $P^0/E_P$ with $E_P=\sqrt{M^2+\uvec P^2}$, usually in an ambiguous and model-dependent way. The third option is to consider a particular limit, like e.g.~the non-relativistic (or static) limit where $P^0\approx M$ or the IMF where $P^0\approx P_z\gg M$. All these responses aimed at providing the most realistic representation of the system within a probabilistic density picture, even if it meant sacrificing part of the Lorentz covariance. By contrast, the phase-space approach adopted in this work provides a less restrictive quasiprobabilistic picture, allowing one to maintain a fully relativistic definition of spatial distributions for arbitrary values of the average target momentum. In particular, if we want to provide a physical interpretation of all the components of the electromagnetic four-current, treated in the same consistent way without considering a particular frame nor making assumptions about the dynamics of the system, then we are forced to accept that the EF distributions provide a picture of the target with definite average momentum $\uvec P$ but indefinite inertia (whence indefinite average velocity). As a result the analogy with a classical current should always be considered with a grain of salt (see Appendix~\ref{App-classical analogy} for some discussions), which was already the case because of the quasiprobabilistic nature of $J^\mu_\text{EF}$.

%%%%%%%%%%%%%%%%%%%%%%%%%%%%%%%%%%%%%%%%%%%%%%%%%%%%%%%%%%%%%%%%%%%%%%%%%%%%%%%%%%
\subsection{Light-front distributions}

In the light-front (LF) formalism, four-momentum eigenstates are normalized according to 
$_\text{LF}\langle p',\lambda'|p,\lambda\rangle_\text{LF}=2p^+(2\pi)^3\delta(p'^+-p^+)\delta^{(2)}(\uvec p'_\perp-\uvec p_\perp)\,\delta_{\lambda'\lambda}$,    
where $a^\pm=(a^0\pm a^3)/\sqrt{2}$ are the LF components, and $\lambda'$ and $\lambda$ are LF helicities. As a result, similarly to Eq.~\eqref{EF-distribution:def}, the 2D LF four-current distributions are defined as
\begin{equation}\label{LF-definition}
    J^\mu_\text{LF}(\uvec b_\perp;P^+)\equiv\int\frac{\ud^2\Delta_\perp}{(2\pi)^2}\,e^{-i\uvec\Delta_\perp\cdot\uvec b_\perp}\,\frac{_\text{LF}\langle p',\lambda'|\hat j^\mu(0)|p,\lambda\rangle_\text{LF}}{2P^+}\bigg|_{\Delta^+=0}.
\end{equation} 
In particular, for a spin-$0$ target we can write
\begin{equation}
J^\mu_\text{LF}(\uvec b_\perp;P^+)=e\int\frac{\ud^2\Delta_\perp}{(2\pi)^2}\,e^{-i\uvec\Delta_\perp\cdot\uvec b_\perp}\,\frac{P^\mu}{P^+}\,F(\uvec\Delta_\perp^2).
\end{equation}
In the literature, one has essentially focused on the $\mu=+$ component 
\begin{equation}\label{spin0-LFJp}
     J^+_\text{LF}(\uvec b_\perp;P^+)=e\int\frac{\ud^2\Delta_\perp}{(2\pi)^2}\,e^{-i\uvec\Delta_\perp\cdot\uvec b_\perp}\,F(\uvec\Delta_\perp^2),
\end{equation}
which is usually interpreted as the LF charge distribution, and which also allows a strict probabilistic interpretation thanks to the Galilean symmetry in the LF transverse plane~\cite{Susskind:1967rg,Burkardt:2002hr,Miller:2009sg,Miller:2010nz}. Although $J^+_\text{LF}(\uvec b_\perp;P^+)$ and $J^0_\text{EF}(\uvec b_\perp;P_z)$ correspond strictly speaking to different matrix elements and hence to different physical quantities, they lead to the same 2D spatial distribution. The reason is that when $P_z\to\infty$ we can write $P_z\approx P^0\approx P^+/\sqrt{2}$, leading in general to $J^+_\text{LF}(\uvec b_\perp;\infty)=J^0_\text{EF}(\uvec b_\perp;\infty)$. Then, since neither $J^+_\text{LF}$ nor $J^0_\text{EF}$ in the spin-$0$ case depends actually on the target momentum, we must have $J^+_\text{LF}(\uvec b_\perp;P^+)=J^0_\text{EF}(\uvec b_\perp;P_z)$. We will see later that the situation gets more complicated for spinning targets.

\begin{figure}[!h]
	\centering
	{\includegraphics[angle=0,scale=0.45]{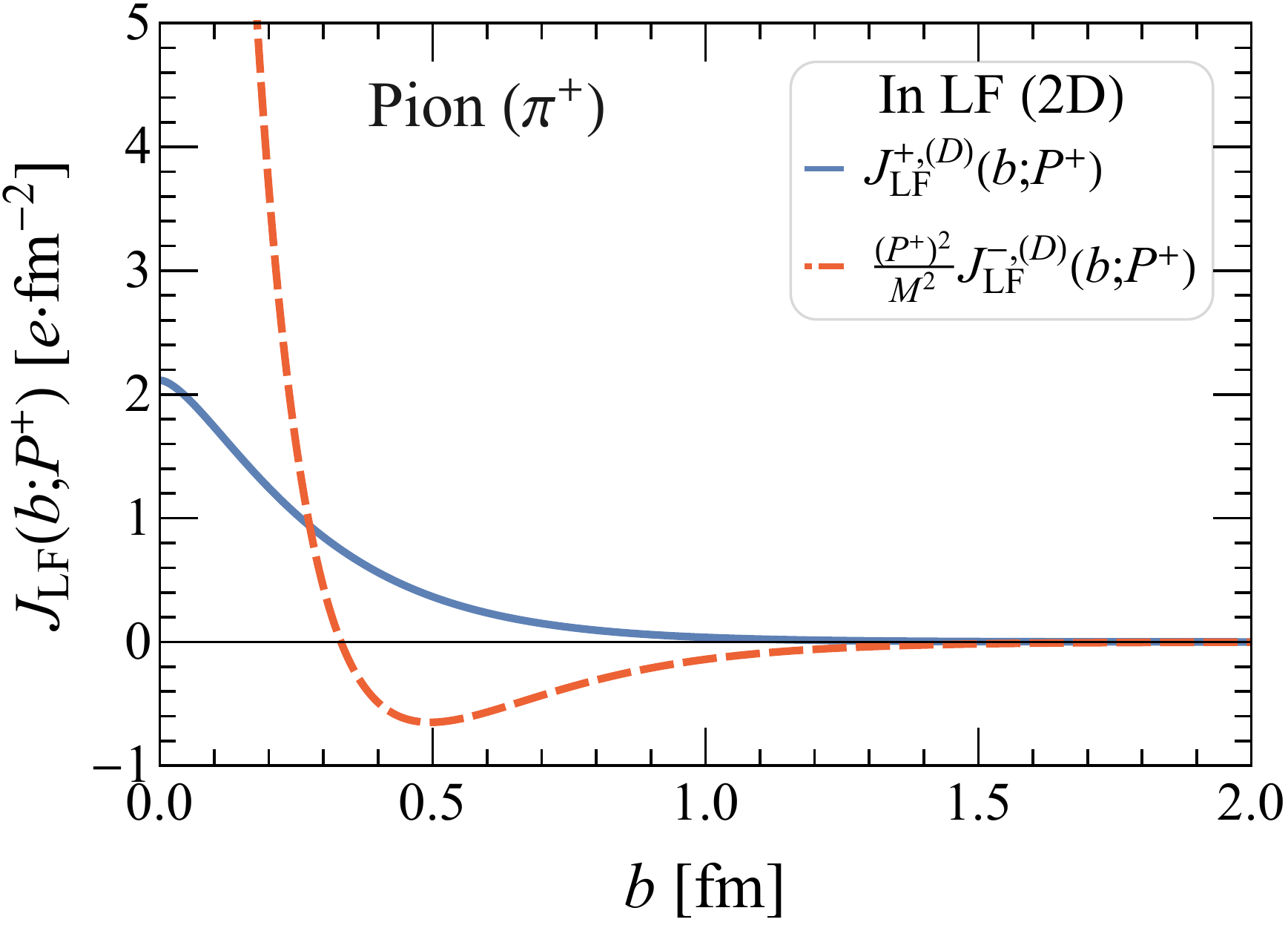}}
	{\includegraphics[angle=0,scale=0.45]{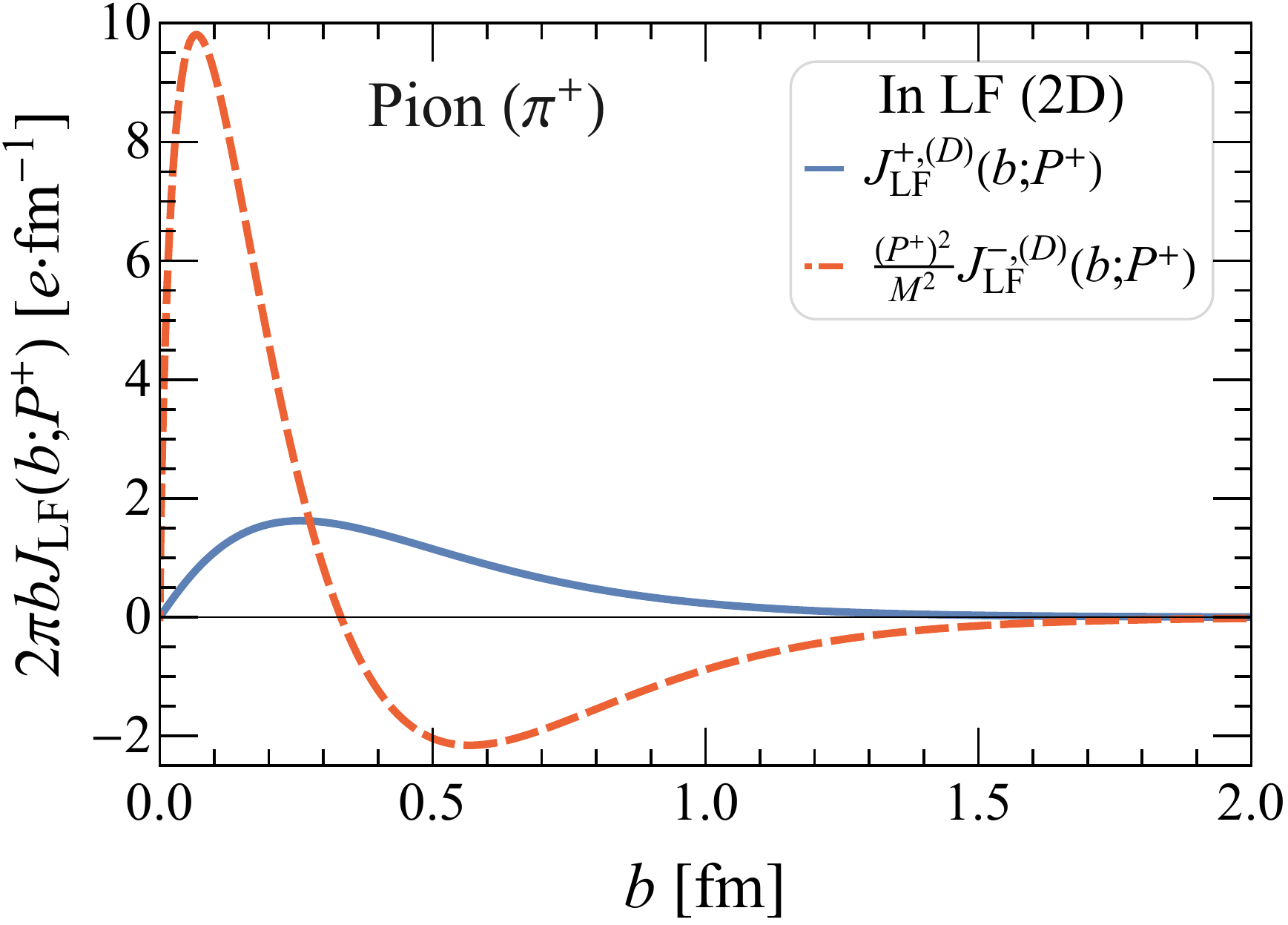}}
	\caption{(Color online) Comparison between the 2D (left panel) [or radial (right panel)] LF charge (solid line) and longitudinal current (dashed line) distributions for a pion ($\pi^+$), based on the dipole parametrization~\eqref{dipolemodel} of the pion electromagnetic FF. Factors of $P^+/M$ have been introduced so as to make the distributions $P^+$-independent.}
	\label{Fig_Pion2DLFJpJm}
\end{figure}

Similarly to the EF current distributions, the transverse LF current distributions always vanish, while the longitudinal LF current distribution depends on the target momentum
\begin{equation}\label{spin0-LFJm0}
     J^-_\text{LF}(\uvec b_\perp;P^+)=e\int\frac{\ud^2\Delta_\perp}{(2\pi)^2}\,e^{-i\uvec\Delta_\perp\cdot\uvec b_\perp}\,\frac{P^-}{P^+}\,F(\uvec\Delta_\perp^2).
\end{equation}
Since $P^2=2P^+P^-$ (remember that we chose our axes such that $\uvec P_\perp=\uvec 0_\perp$), we find that the $P^+$-dependence can be factored out
\begin{equation}\label{spin0-LFJm}
    J^-_\text{LF}(\uvec b_\perp;P^+)=e\,\frac{M^2}{(P^+)^2}\int\frac{\ud^2\Delta_\perp}{(2\pi)^2}\,e^{-i\uvec\Delta_\perp\cdot\uvec b_\perp}\,\frac{1+\tau}{2}\,F(\uvec\Delta_\perp^2).
\end{equation}
This is clearly a technical advantage of the LF formalism associated with the fact that LF boosts are kinematical operations. The drawback is that a system with definite average $P^+$ usually does not have definite average $P_z$~\cite{Diehl:2002he}, and hence is harder to picture physically unless one goes to the IMF (where the technical advantage over the usual instant-form formalism fades away). Note also that although $P^+$ (which plays the role of LF inertia) is treated as an independent kinematical variable in the LF formalism, the off-shellness of $P^\mu$ is transferred to $P^-$ and remains as a $\tau$-dependent kinematical factor under the integral in Eq.~\eqref{spin0-LFJm}. This kinematical factor unfortunately makes the Fourier transform in Eq.~\eqref{spin0-LFJm} ill-defined for the monopole and DR parametrizations of the pion electromagnetic FF. We therefore use in Fig.~\ref{Fig_Pion2DLFJpJm} the dipole parametrization to make comparison between the LF charge and longitudinal current distributions.

%%%%%%%%%%%%%%%%%%%%%%%%%%%%%%%%%%%%%%%%%%%%%%%%%%%%%%%%%%%%%%%%%%%%%%%%%%%%%%%%%%
\section{Spin-$\frac{1}{2}$ target}
\label{sec:Spin-1/2 case}

As usual, including spin will increase the complexity of a system~\cite{Lorce:2009bs,Alexandrou:2009hs,Carlson:2009ovh}. This is the reason why we started with a spin-$0$ target. We are now ready to move on to the next simplest case, namely a spin-$\frac{1}{2}$ target.

The matrix elements of the electromagnetic four-current operator for a spin-$\frac{1}{2}$ target
\begin{equation}\label{genparam}
	\langle p',s'|\hat j^\mu(0)|p,s\rangle=e\,\overline u(p',s')\Gamma^\mu(P,\Delta)u(p,s)
\end{equation}
can be parametrized in terms of the traditional Dirac and Pauli FFs
\begin{equation}
	\Gamma^\mu(P,\Delta)=\gamma^\mu\,F_1(Q^2)+\frac{i\sigma^{\mu\nu}\Delta_\nu}{2M}\,F_2(Q^2),
\end{equation}
or in terms of the electric and magnetic Sachs FFs~\cite{Lorce:2020onh}
\begin{equation}\label{sachsparam}
	\Gamma^\mu(P,\Delta)=\frac{MP^\mu}{P^2}\,G_E(Q^2)+\frac{i\epsilon^{\mu\alpha\beta\lambda}\Delta_\alpha P_\beta\gamma_\lambda\gamma_5}{2P^2}\,G_M(Q^2)
\end{equation}
with $\epsilon_{0123}=+1$. These two sets of FFs are related via~\cite{Ernst:1960zza,Sachs:1962zzc}
\begin{equation}
	\begin{aligned}
		G_E(Q^2)&=F_1(Q^2)-\tau F_2(Q^2),\\
		G_M(Q^2)&=F_1(Q^2)+F_2(Q^2).
	\end{aligned}
\end{equation}
We consider that the parametrization~\eqref{sachsparam} in terms of the Sachs FFs is physically more transparent\footnote{An even better parametrization would be in terms of $\bar G_{E,M}(Q^2)\equiv\frac{M}{\sqrt{P^2}}\,G_{E,M}(Q^2)=\frac{1}{\sqrt{1+\tau}}\,G_{E,M}(Q^2)$, but for historical reasons we stick to the traditional Sachs FFs.}, since its structure in momentum space is reminiscent of a classical current in a polarized medium~\cite{Yennie:1957rmp,Sheng:2022iyn,Li:2022ldb}. More specifically, the first term is directly proportional to the four-momentum and has therefore the structure of a convective current, while the second term involves the axial-vector Dirac current and hence can be interpreted as a polarization (or spin) current~\cite{Lorce:2020onh}.

%%%%%%%%%%%%%%%%%%%%%%%%%%%%%%%%%%%%%%%%%%%%%%%%%%%%%%%%%%%%%%%%%%%%%%%%%%%%%%%%%%
\subsection{Breit frame distributions}

It has been noticed long ago that the matrix elements of the electromagnetic four-current for a spin-$\frac{1}{2}$ target in the BF take the simple form~\cite{Yennie:1957rmp,Ernst:1960zza,Sachs:1962zzc}
\begin{equation}\label{BFampl}
	\begin{aligned}
		\langle p'_B,s'|\hat j^0(0)|p_B,s\rangle&=e\,2M\,\delta_{s's}\,G_E(Q^2),\\
		\langle p'_B,s'|\hat{\uvec j}(0)|p_B,s\rangle&=e\,(\uvec\sigma_{s's}\times i\uvec\Delta)\,G_M(Q^2),
	\end{aligned}
\end{equation}
with $\uvec\sigma$ the Pauli matrices. The BF charge distribution depends only on $G_E$ and hence is purely convective. Similarly, the BF current depends only on $G_M$ and has indeed the form of a spin current. The corresponding relativistic 3D spatial distributions are given by~\cite{Yennie:1957rmp,Friar:1975bk,Lorce:2020onh}
\begin{equation}\label{BFdensities}
	\begin{aligned}
		J^0_B(\uvec r)&=e\,\delta_{s's}\int\frac{\ud^3\Delta}{(2\pi)^3}\,e^{-i\uvec\Delta\cdot\uvec r}\,\frac{M}{P^0_B}\,G_E(\uvec\Delta^2),\\
		\uvec J_B(\uvec r)&=e\,\frac{\uvec\nabla\times\uvec\sigma_{s's}}{2M}\int\frac{\ud^3\Delta}{(2\pi)^3}\,e^{-i\uvec\Delta\cdot\uvec r}\,\frac{M}{P^0_B}\,G_M(\uvec\Delta^2),
	\end{aligned}
\end{equation}
with $P^0_B=M\sqrt{1+\tau}$. We stress here that these relativistic distributions differ from the conventional ones introduced by Sachs~\cite{Ernst:1960zza,Sachs:1962zzc}, where the Lorentz contraction factor $M/P^0_B=1/\sqrt{1+\tau}$ has been removed by hand for closer analogy with the non-relativistic expressions\footnote{The expressions in Eq.~\eqref{BFdensities} suggest in fact that the genuine electromagnetic FFs are given by $\bar G_{E,M}(Q^2)=\frac{M}{\sqrt{P^2}}\,G_{E,M}(Q^2)=\frac{1}{\sqrt{1+\tau}}\,G_{E,M}(Q^2)$.}. 

\begin{figure}[tb!]
	\centering
	{\includegraphics[angle=0,scale=0.425]{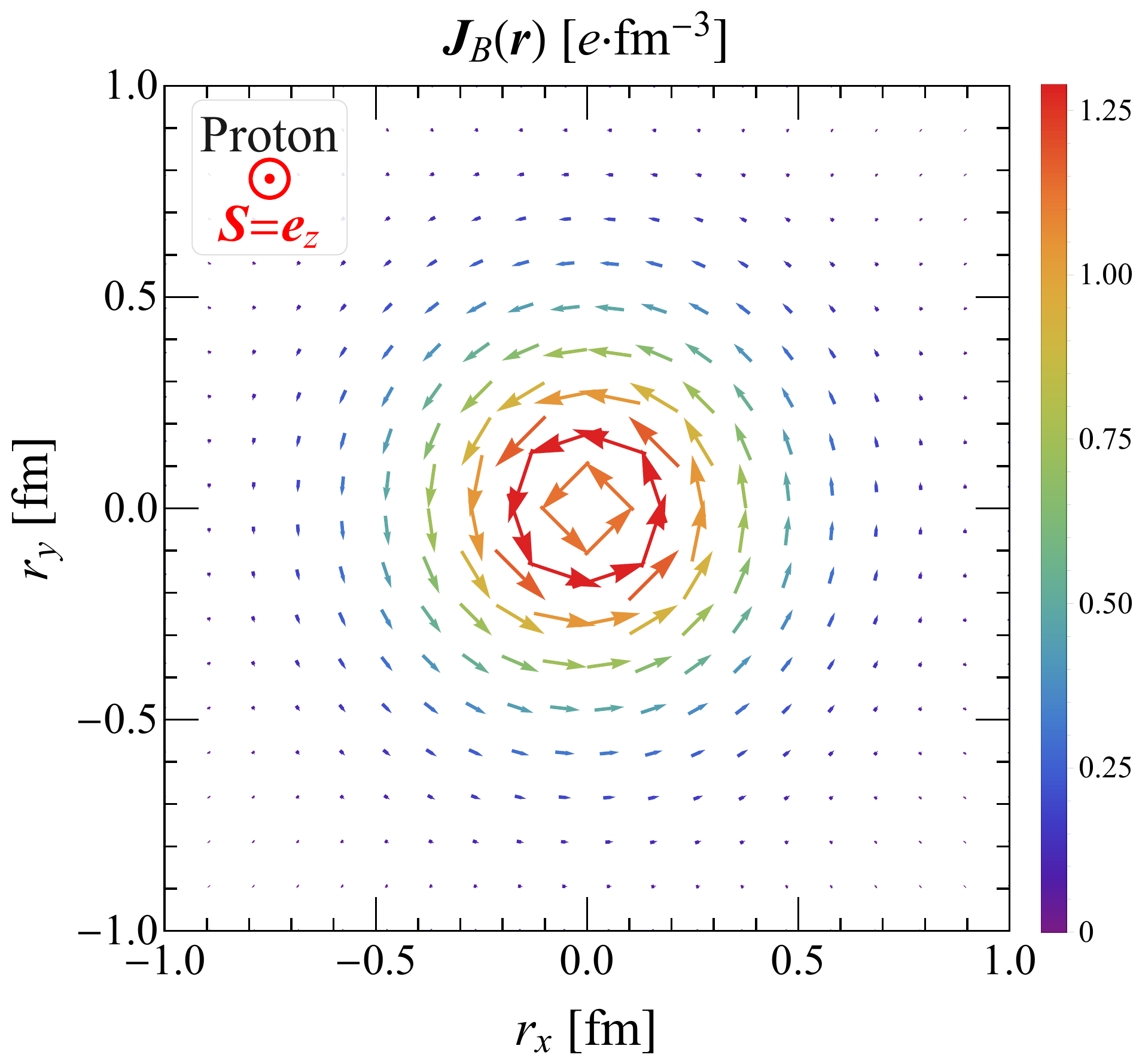}}
	{\includegraphics[angle=0,scale=0.425]{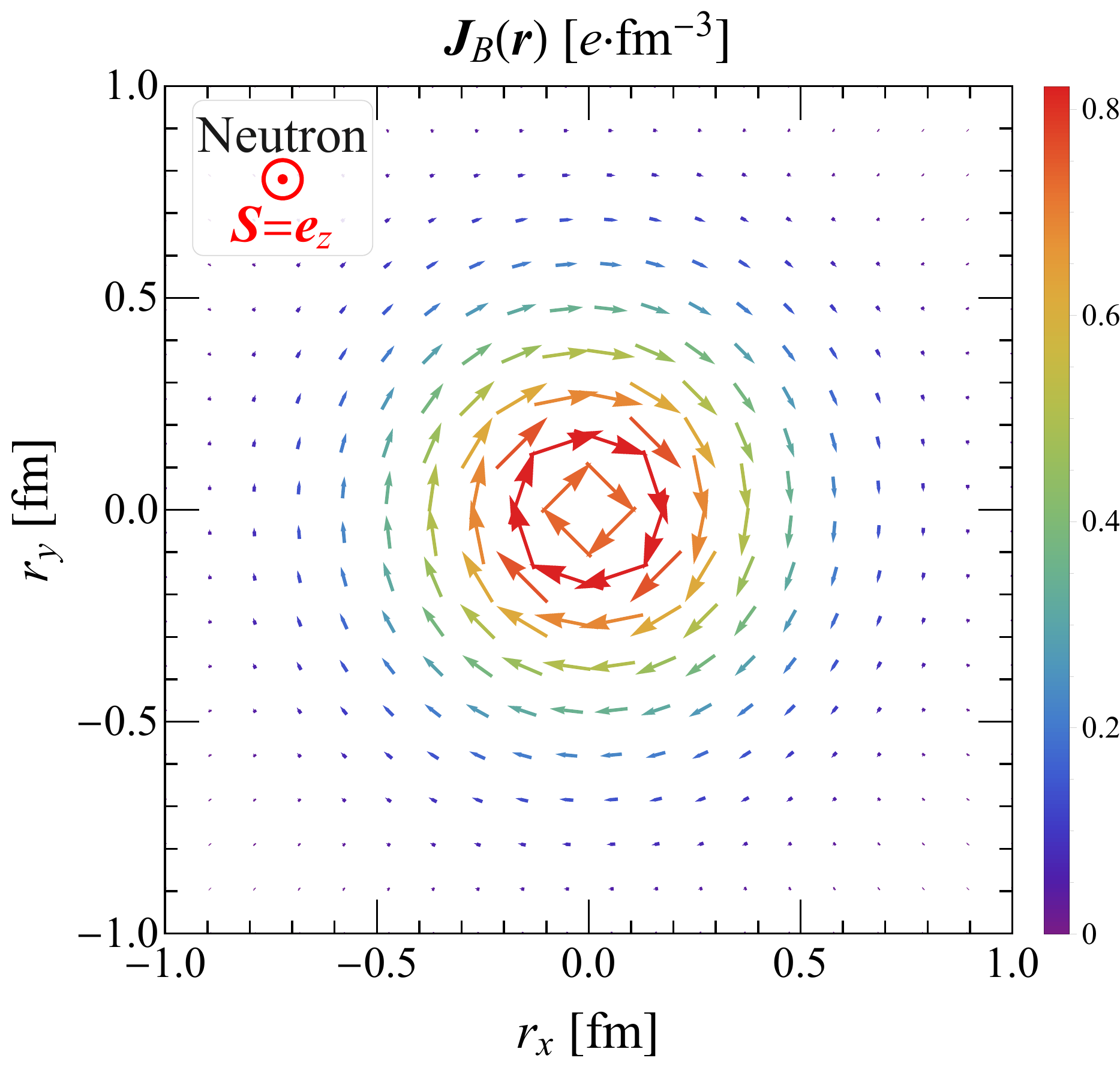}}
	\caption{(Color online) BF current distributions $\uvec J_B(\uvec r)$ in the transverse plane ($r_z=0$) for a proton (left panel) and a neutron (right panel) polarized along the $z$-axis, based on the parametrization for the nucleon electromagnetic FFs given in Ref.~\cite{Bradford:2006yz}. The colorbar and the arrow length indicate the magnitude of the current density.}
	\label{Fig_NucleonJvB3DBF}
\end{figure}

The BF charge distribution inside the nucleon has already been presented in Fig.~1 of Ref.~\cite{Lorce:2020onh}. We show in Fig.~\ref{Fig_NucleonJvB3DBF} the distribution of the BF current in the plane defined by $r_z=0$ for a nucleon polarized in the $z$-direction, based on the parametrization for the nucleon electromagnetic FFs given in Ref.~\cite{Bradford:2006yz}. The current swirls around the polarization axis in opposite directions for proton and neutron, in agreement with the sign of their magnetic moment.

%%%%%%%%%%%%%%%%%%%%%%%%%%%%%%%%%%%%%%%%%%%%%%%%%%%%%%%%%%%%%%%%%%%%%%%%%%%%%%%%%%
\subsection{Elastic frame distributions}

Contrary to the 3D BF distributions, the 2D EF distributions are defined for arbitrary values of $P_z$. Evaluating explicitly the Dirac bilinears in the parametrization~\eqref{genparam}, e.g.~with the aid of the results in Ref.~\cite{Lorce:2017isp}, the EF charge amplitudes in momentum space can be put in the following form:
\begin{equation}\label{EFj0ampl}
		\langle p',s'|\hat j^0(0)|p,s\rangle=e\,2P^0\left[\delta_{s's}\,\mathcal A_U+\frac{(\uvec\sigma_{s's}\times i\uvec\Delta)_z}{2M}\,\mathcal A_T\right],
\end{equation}
where the spin-independent amplitude,
\begin{equation}
        \mathcal A_U=\frac{P^0+M(1+\tau)}{(P^0+M)(1+\tau)}\,G_E(Q^2)+\frac{\tau P_z^2}{P^0(P^0+M)(1+\tau)}\,G_M(Q^2),
\end{equation}
was obtained in~\cite{Lorce:2020onh}, and the spin-dependent amplitude,
\begin{equation}
        \mathcal A_T=-\frac{P_z}{(P^0+M)(1+\tau)}\,G_E(Q^2)+\frac{P_z(P^0+M(1+\tau))}{P^0(P^0+M)(1+\tau)}\,G_M(Q^2),
\end{equation}
was derived in~\cite{Kim:2021kum}. Similarly, we find that the longitudinal EF current amplitudes take the following form
\begin{equation}
    \langle p',s'|\hat j_z(0)|p,s\rangle=e\,2P^0\left[\delta_{s's}\,\mathcal B_U+\frac{(\uvec\sigma_{s's}\times i\uvec\Delta)_z}{2M}\,\mathcal B_T\right],
\end{equation}
with
\begin{equation}\label{endEFresult}
\begin{aligned}
        \mathcal B_U&=\frac{P_z(P^0+M(1+\tau))}{P^0(P^0+M)(1+\tau)}\,G_E(Q^2)+\frac{\tau P_z}{(P^0+M)(1+\tau)}\,G_M(Q^2),\\
        \mathcal B_T&=-\frac{ P_z^2}{P^0(P^0+M)(1+\tau)}\,G_E(Q^2)+\frac{P^0+M(1+\tau)}{(P^0+M)(1+\tau)}\,G_M(Q^2).
\end{aligned}
\end{equation}
For the transverse EF current amplitudes, we get
\begin{equation}\label{transverseEFampl}
   	\langle p',s'|\hat{\uvec j}_\perp(0)|p,s\rangle=e\,(\sigma_z)_{s's}\,(\uvec e_z\times i\uvec \Delta)_\perp\,G_M(Q^2).
\end{equation}
Contrary to the transverse EF current amplitudes, the spin structures of the EF charge and longitudinal current amplitudes have two contributions and depend on the average momentum. The simplest form is obtained when $P_z=0$, which is a strong incentive for considering the Sachs FFs as the natural basis for studying the electromagnetic properties.

\begin{figure}[tb!]
	\centering
	{\includegraphics[angle=0,scale=0.357]{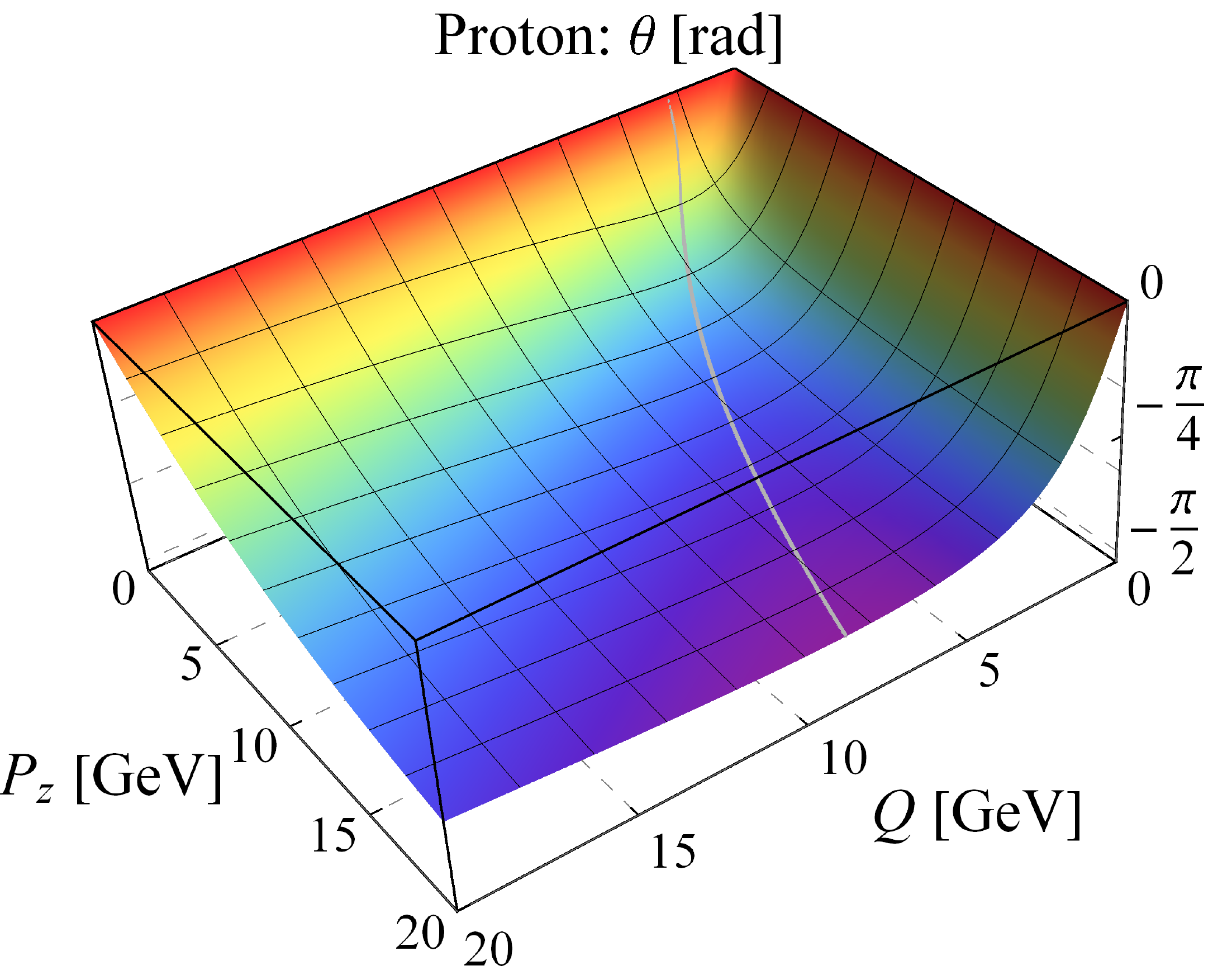}}
	{\includegraphics[angle=0,scale=0.468]{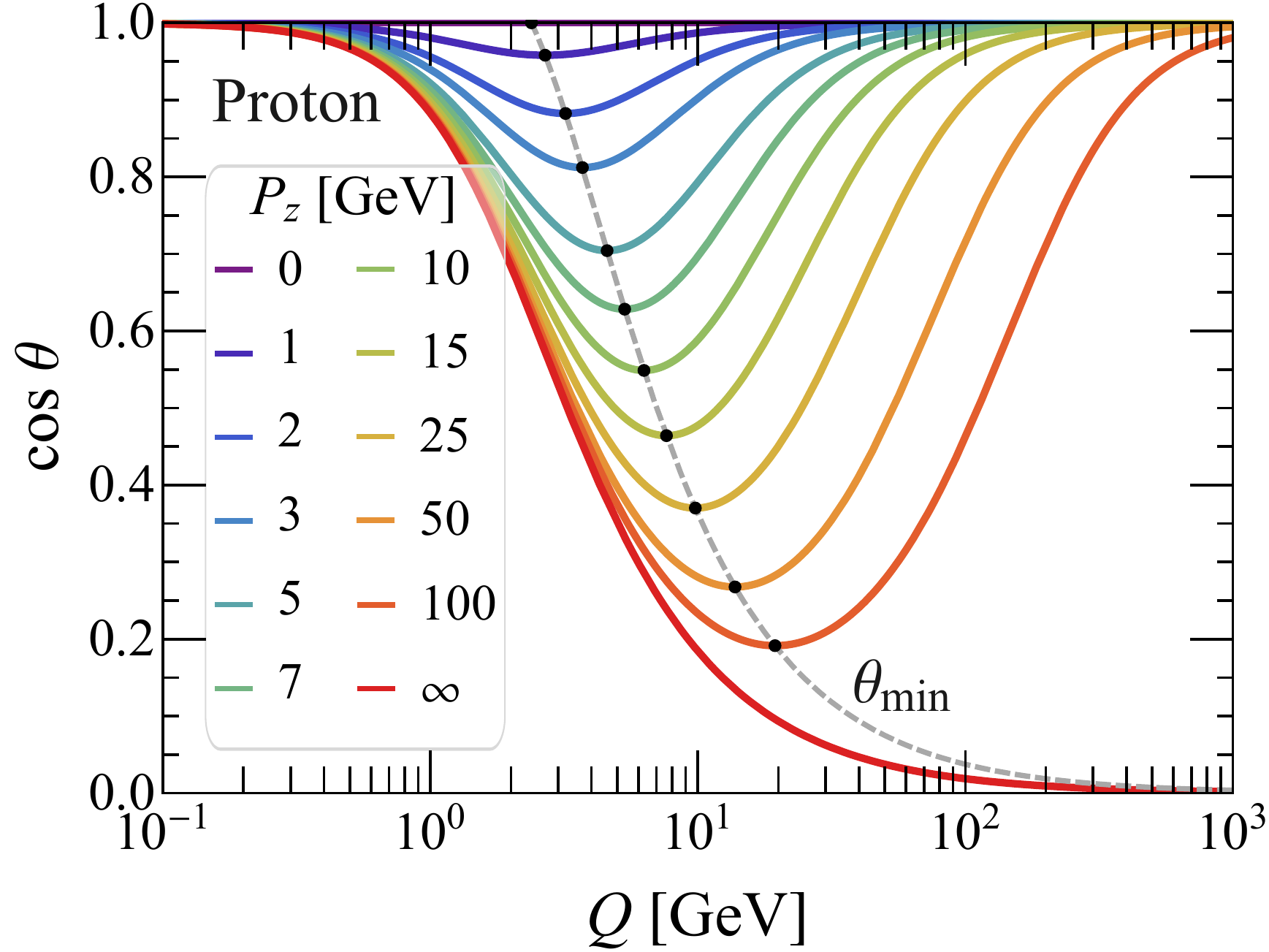}}
	\caption{(Color online) Wigner rotation angle $\theta$ (left panel) and $\cos\theta$ (right panel) for the proton as a function of the proton's average momentum $P_z$ and the magnitude of momentum transfer $Q$.}
	\label{Fig_WingerAngle}
\end{figure}

The results above are fully consistent with the generic Lorentz transformation~\eqref{EFLorentzTrans-Spinj} of the BF amplitudes~\eqref{BFampl}. Indeed, noting that the Wigner rotation matrix $D(p^{(\prime)}_B,\Lambda)$ describes a spin rotation by an angle $\theta$ in the $(\uvec P,\uvec\Delta)$-plane, we can write the EF amplitudes~\eqref{EFLorentzTrans-Spinj} more explicitly in the spin-$\frac{1}{2}$ case as
\begin{equation}\label{spinhalfexplicitLT}
    \begin{aligned}
        \langle p',s'|\hat j^0(0)|p,s\rangle&=e\,2M\,\gamma\,\bigg[\delta_{s's}\left(\cos\theta\,G_E(Q^2)-\beta\sin\theta\,\sqrt{\tau}\,G_M(Q^2)\right)\\
        &\qquad\qquad +\frac{(\uvec\sigma_{s's}\times i\uvec\Delta)_z}{2M\sqrt{\tau}}\left(\sin\theta\,G_E(Q^2)+\beta\cos\theta\,\sqrt{\tau}\,G_M(Q^2)\right)\bigg],\\
         \langle p',s'|\hat j_z(0)|p,s\rangle&=e\,2M\,\gamma\,\bigg[\delta_{s's}\left(\beta\cos\theta\,G_E(Q^2)-\sin\theta\,\sqrt{\tau}\,G_M(Q^2)\right)\\
        &\qquad\qquad +\frac{(\uvec\sigma_{s's}\times i\uvec\Delta)_z}{2M\sqrt{\tau}}\left(\beta\sin\theta\,G_E(Q^2)+\cos\theta\,\sqrt{\tau}\,G_M(Q^2)\right)\bigg],\\
        \langle p',s'|\hat{\uvec j}_\perp(0)|p,s\rangle&=e\,(\sigma_z)_{s's}\,(\uvec e_z\times i\uvec \Delta)_\perp\,G_M(Q^2).
    \end{aligned}
\end{equation}
First of all, we notice that the transverse current amplitudes remain invariant under a longitudinal boost, as confirmed by a comparison between Eqs.~(\ref{BFampl}) and \eqref{transverseEFampl}. Also, comparing Eq.~\eqref{spinhalfexplicitLT} with Eqs.~(\ref{EFj0ampl}-\ref{endEFresult}), we find that the Lorentz boost parameters are given by
\begin{equation}\label{boostparam}
    \gamma=\frac{P^0}{\sqrt{P^2}}=\frac{P^0}{P^0_B}=\frac{P^0}{M\sqrt{1+\tau}},\qquad \beta=\frac{P_z}{P^0},
\end{equation}
as one would have expected, and we conclude that the Wigner rotation angle $\theta$ satisfies
\begin{equation}\label{WignerAngleSinCos}
    \cos\theta=\frac{P^0+M(1+\tau)}{(P^0+M)\sqrt{1+\tau}},\qquad \sin\theta=-\frac{\sqrt{\tau}P_z}{(P^0+M)\sqrt{1+\tau}}.
\end{equation}
It is straightforward to check that $\cos^2\theta+\sin^2\theta=1$ thanks to Eq.~\eqref{P0EF}. Note that $\theta$ should not depend on the spin of the target. We indeed find that
\begin{equation}\label{WignerAngleTan}
    \tan\theta=-\frac{\sqrt{\tau}P_z}{P^0+M(1+\tau)}
\end{equation}
agrees with the result derived from the general angular condition for a spin-$1$ target~\cite{Lorce:2022jyi}. In Fig.~\ref{Fig_WingerAngle}, we show the dependence of the Wigner rotation angle $\theta$ on $P_z$ and $Q=|\uvec\Delta|$ for a proton with mass $M_p\approx 0.938$ GeV. We also present the $Q$-dependence of $\cos\theta$ at different values of $P_z$. For fixed value of $P_z>0$, the minimum value of this cosine is given by
\begin{equation}\label{WignerAngleMinima}
	\cos\theta_{\text{min}} = \frac{1+2\tau_\text{min}}{(1+\tau_\text{min})^{3/2}},\qquad \tau_\text{min}=\frac{1}{2}+\sqrt{\frac{P_z^2}{M^2}+\frac{5}{4}},
\end{equation}
represented by the gray lines in Fig.~\ref{Fig_WingerAngle}. Since $-\tfrac{\pi}{2} \leq \theta\leq 0$, $\theta_{\text{min}}$ actually corresponds to the largest Wigner spin rotation angle at a given $P_z>0$. For $P_z=0$, there is by definition no Wigner rotation.

We are now ready to discuss the EF distributions. We find that the EF charge distribution can be expressed as
\begin{equation}\label{spinhalfEFJ0}
    \begin{aligned}
        J^0_\text{EF}(\uvec b_\perp;P_z)&=e\int\frac{\ud^2\Delta_\perp}{(2\pi)^2}\,e^{-i\uvec\Delta_\perp\cdot\uvec b_\perp}\left[\delta_{s's}\,\cos\theta+\frac{(\uvec\sigma_{s's}\times i\uvec\Delta)_z}{2M\sqrt{\tau}}\,\sin\theta\right]\frac{G_E(\uvec\Delta_\perp^2)}{\sqrt{1+\tau}}\\
        &+e\int\frac{\ud^2\Delta_\perp}{(2\pi)^2}\,e^{-i\uvec\Delta_\perp\cdot\uvec b_\perp}\,\frac{P_z}{P^0}\left[-\delta_{s's}\,\sin\theta+\frac{(\uvec\sigma_{s's}\times i\uvec\Delta)_z}{2M\sqrt{\tau}}\,\cos\theta\right]\frac{\sqrt{\tau}\,G_M(\uvec\Delta_\perp^2)}{\sqrt{1+\tau}}.
    \end{aligned}
\end{equation}
The first line corresponds to the convective part, and the second line to the polarization part of the charge distribution. Similarly, the longitudinal EF current distribution reads
\begin{equation}\label{spinhalfEFJL}
    \begin{aligned}
        J_{z,\text{EF}}(\uvec b_\perp;P_z)&=e\int\frac{\ud^2\Delta_\perp}{(2\pi)^2}\,e^{-i\uvec\Delta_\perp\cdot\uvec b_\perp}\,\frac{P_z}{P^0}\left[\delta_{s's}\,\cos\theta+\frac{(\uvec\sigma_{s's}\times i\uvec\Delta)_z}{2M\sqrt{\tau}}\,\sin\theta\right]\frac{G_E(\uvec\Delta_\perp^2)}{\sqrt{1+\tau}}\\
        &+e\int\frac{\ud^2\Delta_\perp}{(2\pi)^2}\,e^{-i\uvec\Delta_\perp\cdot\uvec b_\perp}\left[-\delta_{s's}\,\sin\theta+\frac{(\uvec\sigma_{s's}\times i\uvec\Delta)_z}{2M\sqrt{\tau}}\,\cos\theta\right]\frac{\sqrt{\tau}\,G_M(\uvec\Delta_\perp^2)}{\sqrt{1+\tau}}.
    \end{aligned}
\end{equation}
For the transverse EF current distributions, we obtain the simpler expression
\begin{equation}\label{spinhalfEFJT}
    \uvec J_{\perp,\text{EF}}(\uvec b_\perp;P_z)=e\,(\sigma_z)_{s's}\int\frac{\ud^2\Delta_\perp}{(2\pi)^2}\,e^{-i\uvec\Delta_\perp\cdot\uvec b_\perp}\,\frac{(\uvec e_z\times i\uvec\Delta)_\perp}{2P^0}\,G_M(\uvec\Delta_\perp^2).
\end{equation}
Just like in the spin-$0$ case, the spin-$\frac{1}{2}$ EF four-current distribution reduces at $P_z=0$ to the projection of the 3D BF four-current distribution onto the transverse plane; see, e.g., Eq.~\eqref{BFreduc}.

\begin{figure}[htb!]
	\centering
	{\includegraphics[angle=0,scale=0.450]{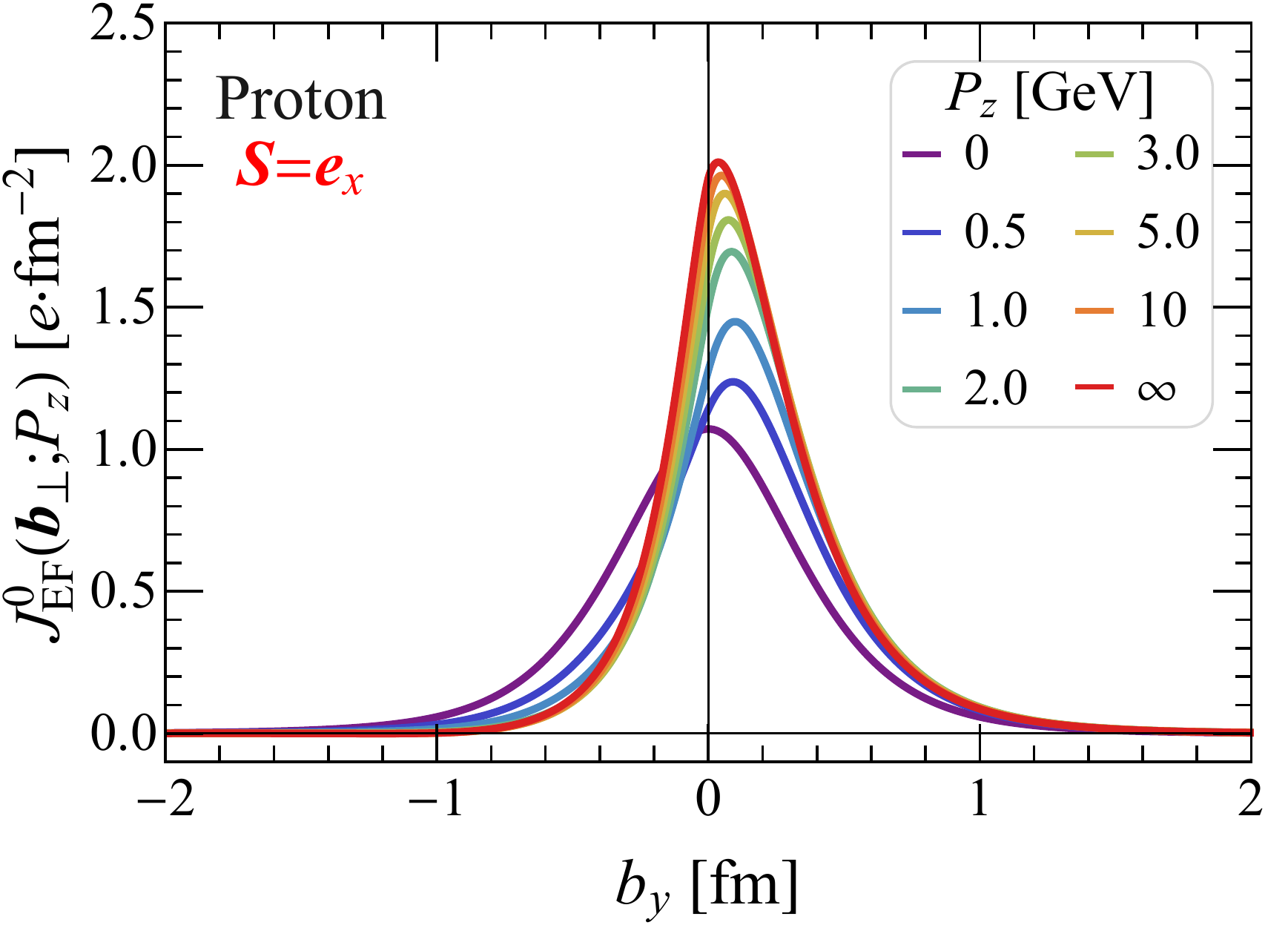}}
	{\includegraphics[angle=0,scale=0.462]{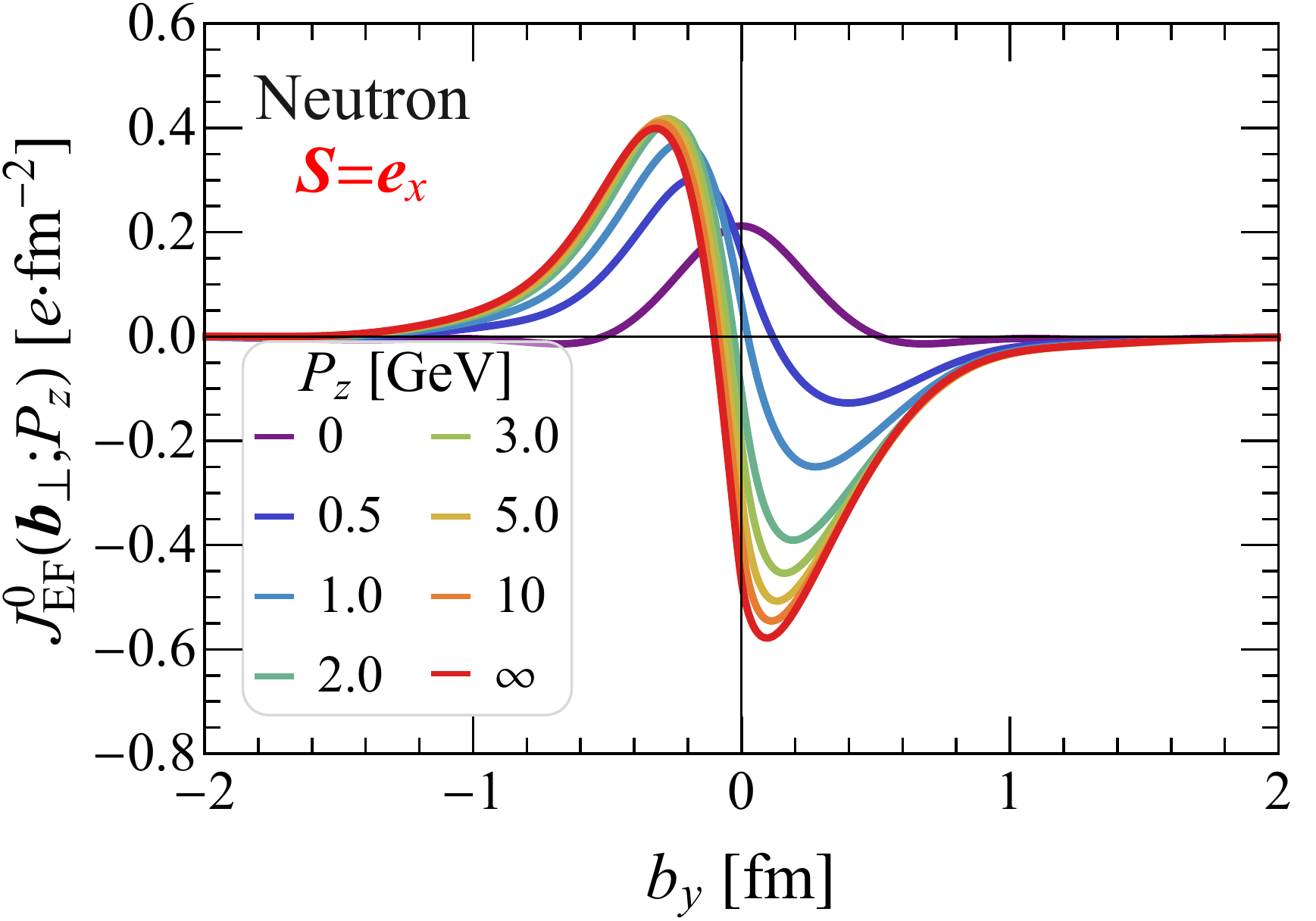}}
	{\includegraphics[angle=0,scale=0.462]{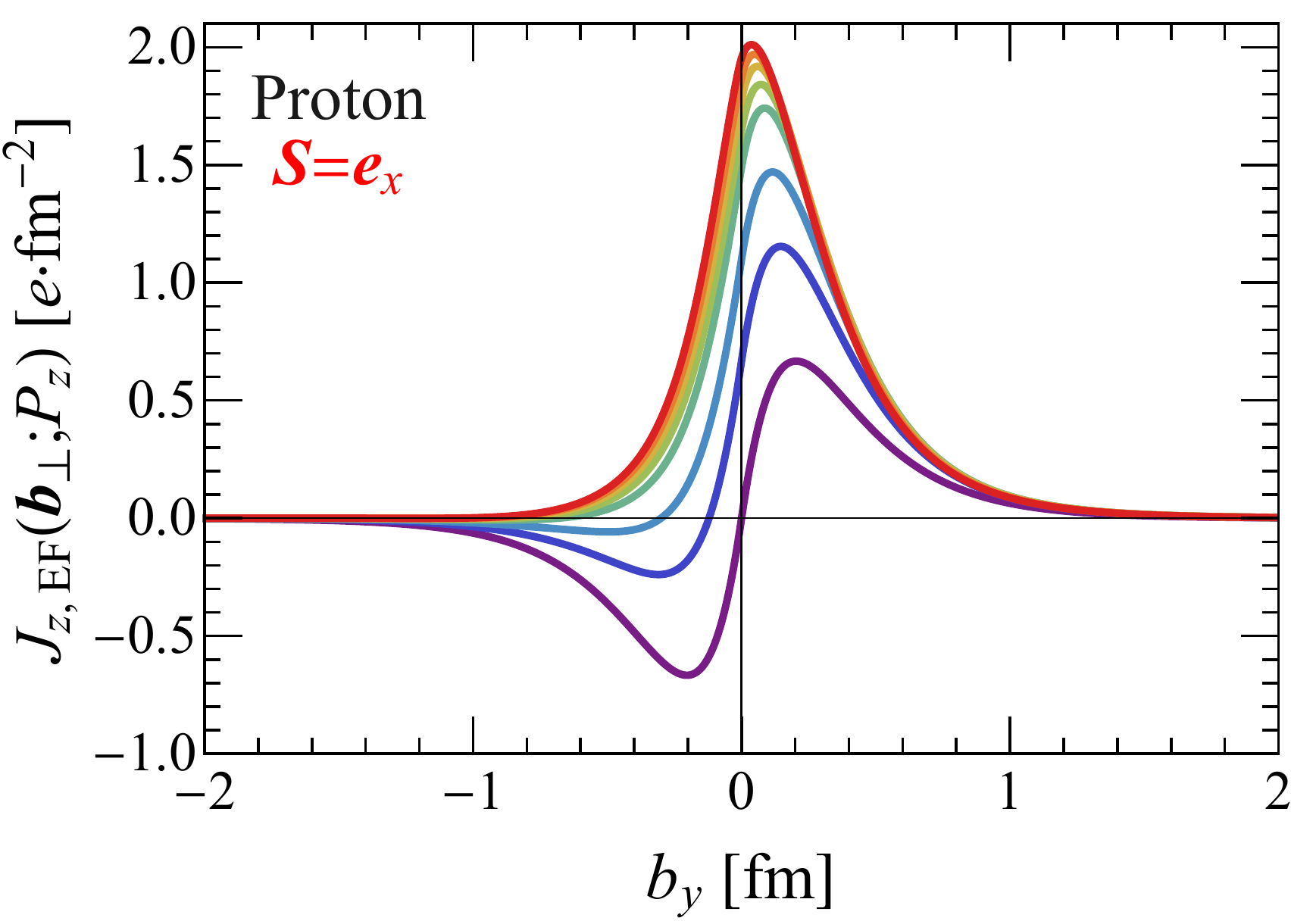}}
	{\includegraphics[angle=0,scale=0.462]{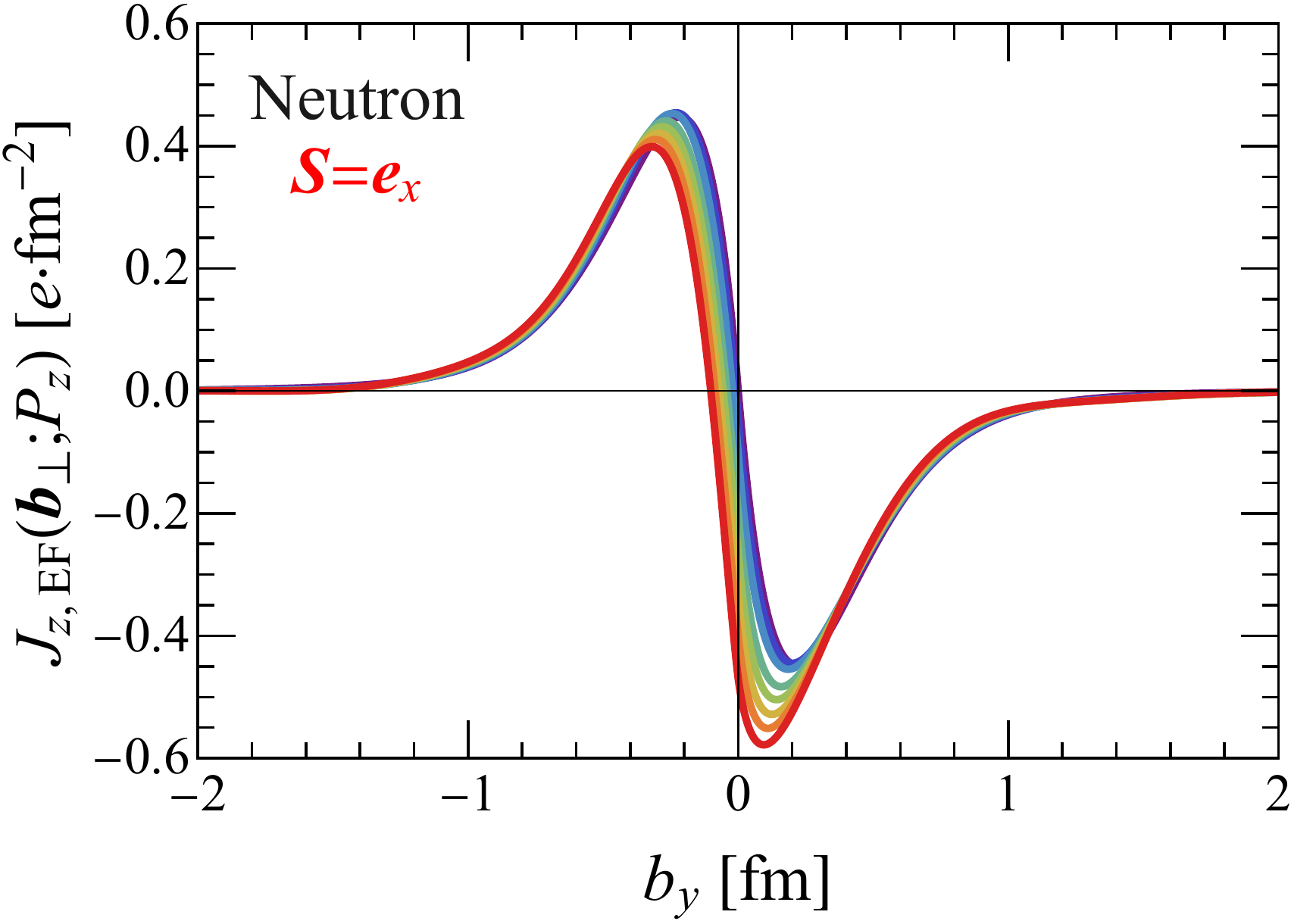}}
	{\includegraphics[angle=0,scale=0.462]{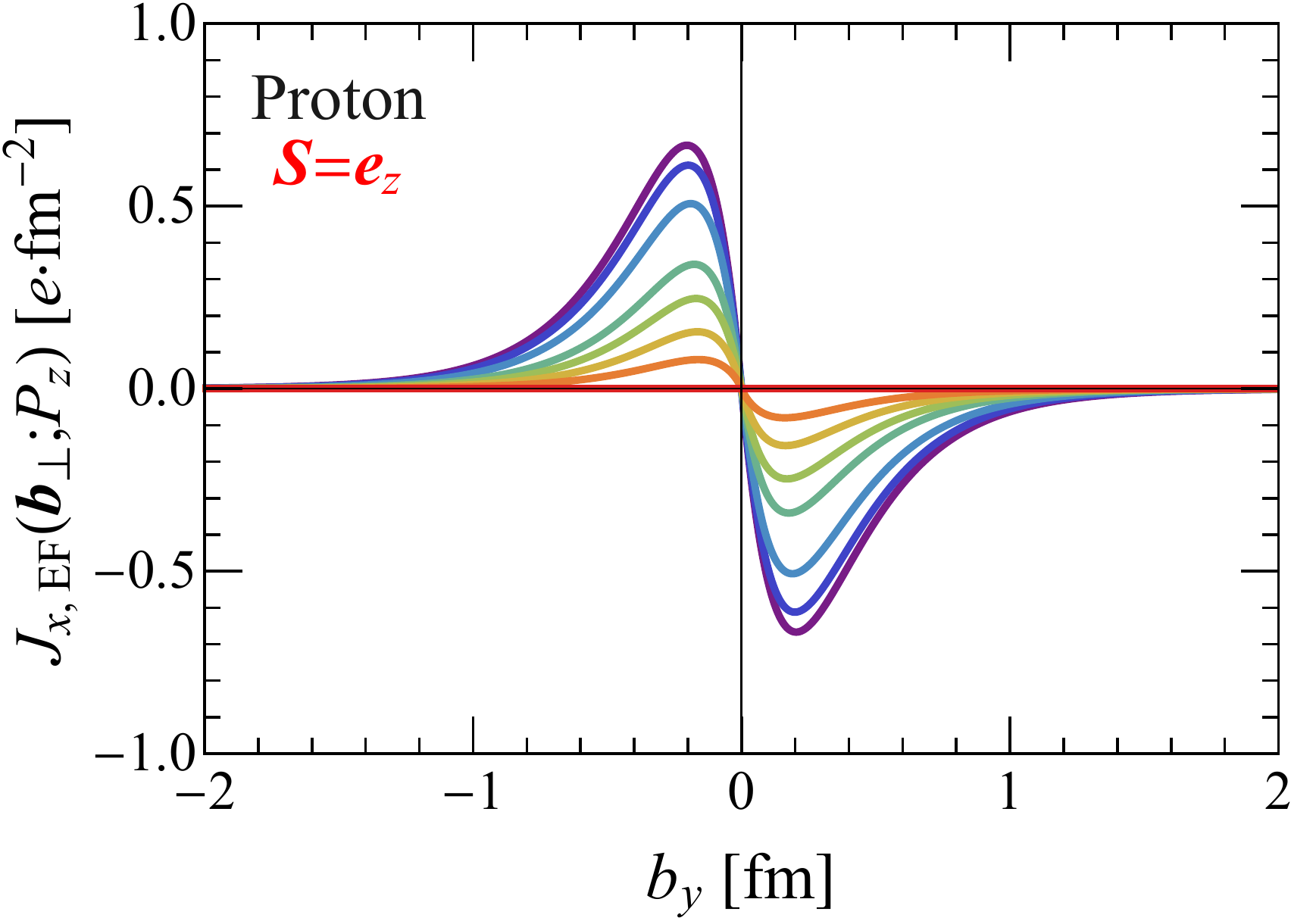}}
	{\includegraphics[angle=0,scale=0.462]{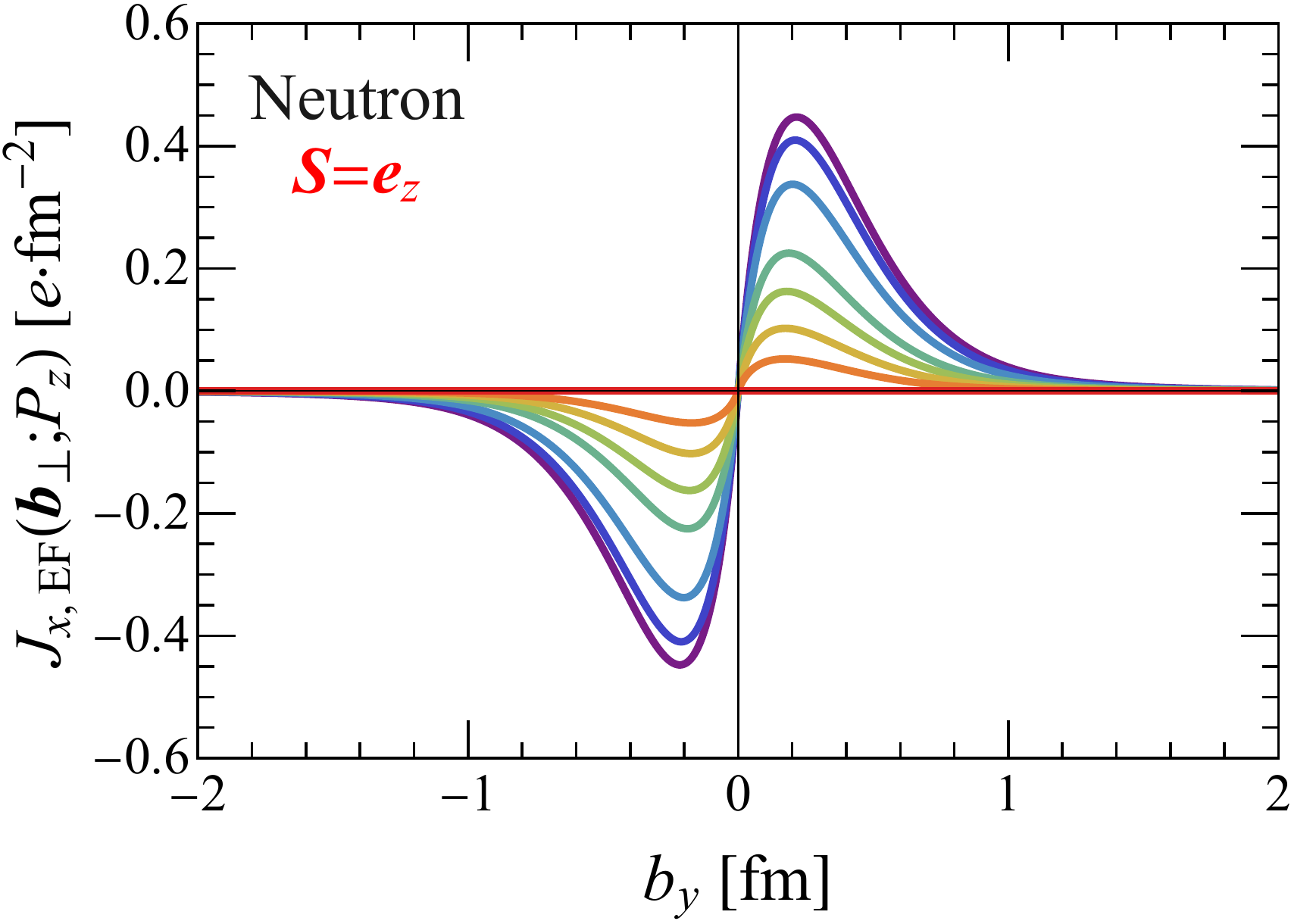}}
	\caption{(Color online) Momentum dependence of the 2D EF four-current distributions at $b_x=0$ for a proton (left panels) and a neutron (right panels), based on the parametrization for the nucleon electromagnetic FFs given in Ref.~\cite{Bradford:2006yz}. The EF charge (first row) and longitudinal current (second row) distribution are shown for a transversely polarized nucleon, while the transverse EF current distribution (third row) is shown for a longitudinally polarized nucleon.}
	\label{Fig_Nucleon2DEFJ0JzJx}
\end{figure}

For an unpolarized target, only the terms proportional to $\delta_{s's}$ do contribute. Like in the spin-$0$ case, the transverse EF current distributions vanish and both the EF charge and longitudinal current distributions are axially symmetric. The difference is that the latter two distributions are now driven by $P_z$-dependent linear combinations of $G_E$ and $G_M$. While the $G_E$ contribution associated with the convective part of the current is expected, the $G_M$ contribution may be surprising. It is in fact a consequence of the Wigner rotation, which disappears if one sets $\theta=0$ by hand in these expressions. When the target is polarized, dipolar distortions correlated with the target polarization arise. These dipolar distortions are naturally attributed to the polarization part of the current driven by $G_M$, but their magnitude also depends on $G_E$ as a result of the Wigner rotation. We observe that the magnitude of $\uvec J_{\perp,\text{EF}}$ decreases for increasing values of $P_z$ owing to the factor $P^0$ in the denominator of Eq.~\eqref{spinhalfEFJT}, while $J^0_\text{EF}$ and $J_{z,\text{EF}}$ tend toward the same distribution as $P_z\to \infty$ like in the spin-$0$ case in Eq.~\eqref{spin0-JzJ0Equal}. In Fig.~\ref{Fig_Nucleon2DEFJ0JzJx}, we show some components of the EF four-current distributions for different polarizations and momentum values of the nucleon.

%%%%%%%%%%%%%%%%%%%%%%%%%%%%%%%%%%%%%%%%%%%%%%%%%%%%%%%%%%%%%%%%%%%%%%%%%%%%%%%%%%
\subsection{Infinite-momentum limit and light-front distributions}

The 2D EF distributions being defined for arbitrary values of $P_z$, they provide a natural interpolation between the 3D BF distributions projected onto the transverse plane and the 2D IMF distributions. Since longitudinal boosts simply rescale the LF components of a Lorentz four-vector, we can decouple the four-vector boost from the Wigner rotation in Eq.~\eqref{spinhalfexplicitLT} by considering the following combinations of amplitudes
\begin{equation}\label{LFspinhalfexplicitPM}
    \begin{aligned}
        \langle p',s'|\hat j^+(0)|p,s\rangle&=e\,\frac{2MP^+}{\sqrt{P^2}}\,\bigg[\delta_{s's}\left(\cos\theta\,G_E(Q^2)-\sin\theta\,\sqrt{\tau}\,G_M(Q^2)\right)\\
        &\qquad\qquad +\frac{(\uvec\sigma_{s's}\times i\uvec\Delta)_z}{2M\sqrt{\tau}}\left(\sin\theta\,G_E(Q^2)+\cos\theta\,\sqrt{\tau}\,G_M(Q^2)\right)\bigg],\\
         \langle p',s'|\hat j^-(0)|p,s\rangle&=e\,\frac{2MP^-}{\sqrt{P^2}}\,\bigg[\delta_{s's}\left(\cos\theta\,G_E(Q^2)+\sin\theta\,\sqrt{\tau}\,G_M(Q^2)\right)\\
        &\qquad\qquad +\frac{(\uvec\sigma_{s's}\times i\uvec\Delta)_z}{2M\sqrt{\tau}}\left(\sin\theta\,G_E(Q^2)-\cos\theta\,\sqrt{\tau}\,G_M(Q^2)\right)\bigg].
    \end{aligned}
\end{equation}
Note that these are not proper LF amplitudes since they are defined in terms of the usual (or instant-form) polarization states instead of the LF helicity states. 

In the IMF (i.e.~$P_z \to \infty$), the amplitude $\langle p',s'|\hat j^+(0)|p,s\rangle$ is enhanced while the amplitude $\langle p',s'|\hat j^-(0)|p,s\rangle$ is suppressed, owing to the global factor of $P^+$ and $P^-$, respectively. We can also clearly see how the Wigner rotation mixes $G_E$ with $\sqrt{\tau}\,G_M$. Using the formulas \eqref{WignerAngleSinCos} and \eqref{WignerAngleTan} for the Wigner rotation angle $\theta$, we find
\begin{equation}\label{WignerAngleIMFlim}
	\lim_{P_z \to \infty} \cos\theta=\frac{1}{\sqrt{1+\tau}},\qquad 	\lim_{P_z \to \infty}\sin\theta=-\frac{\sqrt{\tau}}{\sqrt{1+\tau}},\qquad \lim_{P_z \to \infty}\tan\theta=-\frac{1}{\sqrt{\tau}},
\end{equation}
see, e.g., the lowest solid line for $\cos\theta$ with $P_z \to \infty$ in the right panel of Fig.~\ref{Fig_WingerAngle}.
The spin-independent contribution to $\langle p',s'|\hat j^+(0)|p,s\rangle$ is then driven in the IMF by the Dirac FF,
\begin{equation}
    \lim_{P_z\to\infty}\frac{\cos\theta\,G_E(Q^2)-\sin\theta\,\sqrt{\tau}\,G_M(Q^2)}{\sqrt{1+\tau}}=\frac{G_E(Q^2)+\tau\, G_M(Q^2)}{1+\tau}=F_1(Q^2),
\end{equation}
while the spin-dependent contribution is driven by the Pauli FF,
\begin{equation}
    \lim_{P_z\to\infty}\frac{\sin\theta\,G_E(Q^2)+\cos\theta\,\sqrt{\tau}\,G_M(Q^2)}{\sqrt{\tau}\,\sqrt{1+\tau}}=\frac{G_M(Q^2)- G_E(Q^2)}{1+\tau}=F_2(Q^2), 
\end{equation}
as observed in Refs.~\cite{Rinehimer:2009yv,Lorce:2020onh,Kim:2021kum}. Interestingly, the same structure also appears in the LF formalism~\cite{Chung:1988my,Burkardt:2002hr,Miller:2007uy,Carlson:2007xd},
\begin{equation}\label{LF-amplitude-Jp}
        _\text{LF}\langle p',\lambda'|\hat j^+(0)|p,\lambda\rangle_\text{LF}\big|_{\Delta^+=0}=e\,2P^+\,\bigg[\delta_{\lambda'\lambda}\,F_1(Q^2)+\frac{(\uvec\sigma_{\lambda'\lambda}\times i\uvec\Delta)_z}{2M}\,F_2(Q^2)\bigg],
\end{equation}
without having recourse to the IMF. The reason is that the Melosh rotation~\cite{Melosh:1974cu,Lorce:2011zta} relating the LF polarization states $|p,\lambda\rangle_\text{LF}$ to the usual canonical spin states $|p,s\rangle$ precisely coincides in the BF for $\Delta_z=0$ with the IMF Wigner rotation, see Appendix~\ref{App-Wigner-Melosh Rotations}.

Owing to Eq.~\eqref{LF-amplitude-Jp}, the Dirac and Pauli FFs are often considered in the LF formalism as the ``physical'' electric and magnetic FFs. We observe however that the matrix elements of the longitudinal LF current density operator $\hat j^-$ are usually not discussed. We find from Eq.~\eqref{LFspinhalfexplicitPM} that the spin-independent contribution to $\langle p',s'|\hat j^-(0)|p,s\rangle$ is driven by
\begin{equation}
    \lim_{P_z\to\infty}\frac{\cos\theta\,G_E(Q^2)+\sin\theta\,\sqrt{\tau}\,G_M(Q^2)}{\sqrt{1+\tau}}=\frac{(1-\tau)F_1(Q^2)-2\tau F_2(Q^2)}{1+\tau}\equiv G_1(Q^2),
\end{equation}
and the spin-dependent contribution is driven by
\begin{equation}
    \lim_{P_z\to\infty}\frac{\sin\theta\,G_E(Q^2)-\cos\theta\,\sqrt{\tau}\,G_M(Q^2)}{\sqrt{\tau}\,\sqrt{1+\tau}}=-\frac{2F_1(Q^2)+(1-\tau)F_2(Q^2)}{1+\tau}\equiv G_2(Q^2).
\end{equation}
From the standard LF perspective, these combinations of $F_1$ and $F_2$ do not seem to have any clear physical meaning. This is usually not considered as a problem since only the ``good'' LF component $\hat j^+$ allows a probabilistic interpretation~\cite{Soper:1976jc,Burkardt:2002hr,Miller:2010nz}, while the ``bad'' LF component $\hat j^-$ is regarded as a complicated object without clear physical interpretation, and is therefore often just ignored. From a covariant perspective, we find however this situation unsatisfactory. 

\begin{figure}[!tb]
	\centering
	{\includegraphics[angle=0,scale=0.4500]{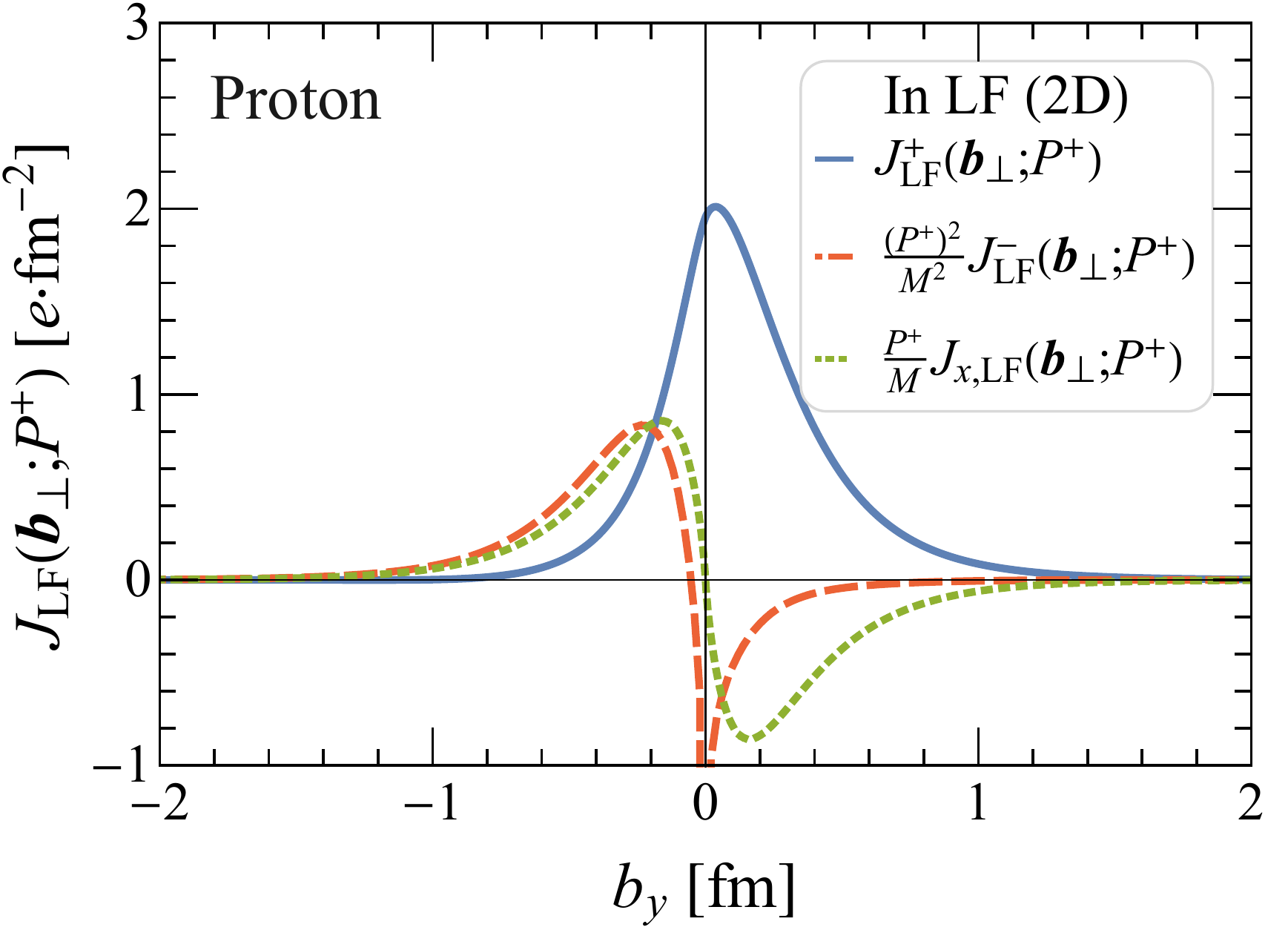}}
	{\includegraphics[angle=0,scale=0.4632]{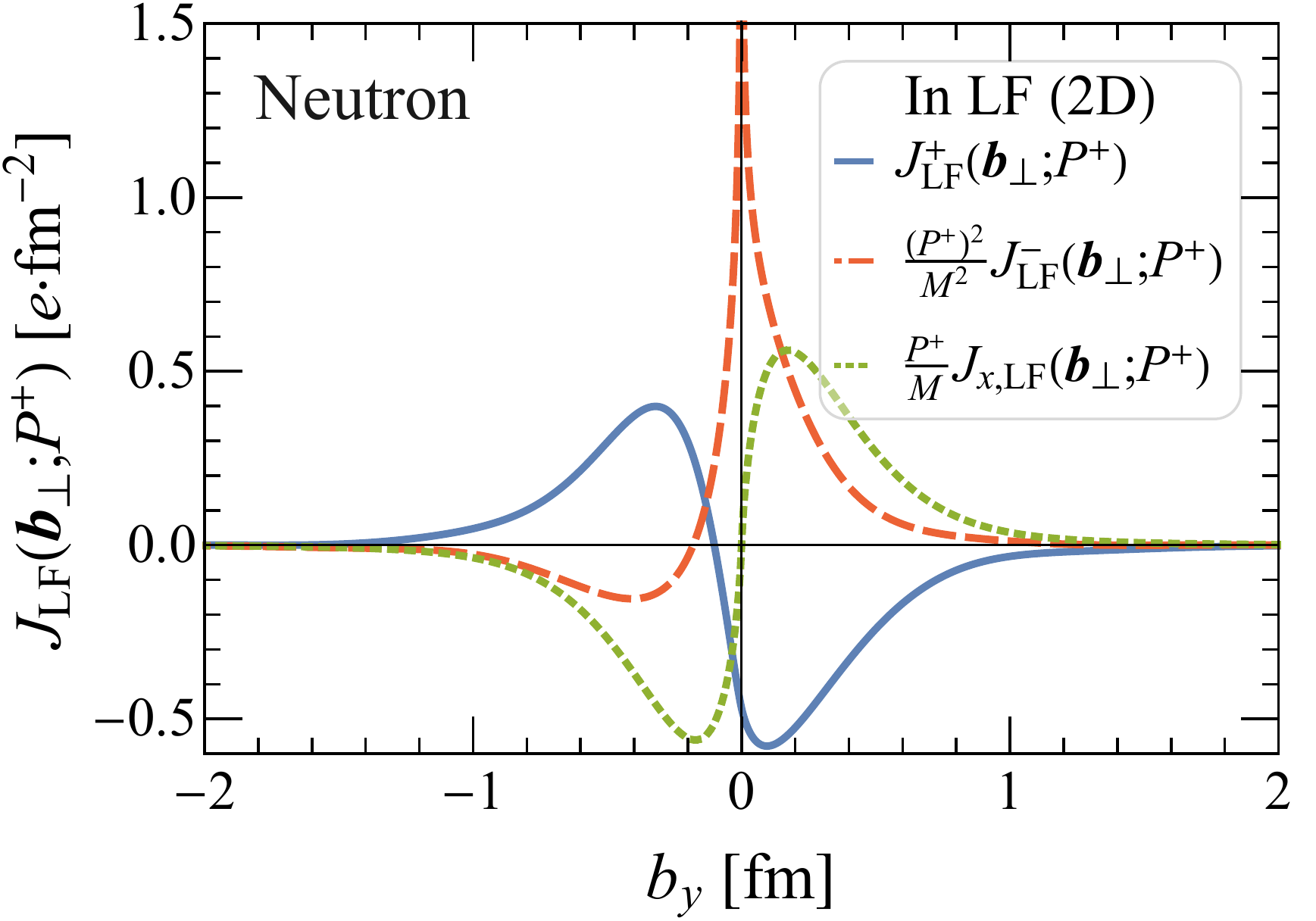}}
	\caption{(Color online) Comparisons between different components of LF four-current distributions at $b_x=0$ for a proton (left panel) and a neutron (right panel), based on the parametrization of the nucleon electromagnetic FFs given in Ref.~\cite{Bradford:2006yz}. Factors of $P^+/M$ have been introduced so as to make the distributions $P^+$-independent. The LF charge (solid lines) and longitudinal current (dashed lines) distributions are shown for a transversely polarized nucleon, and the transverse LF current distribution (dotted lines) is shown for a longitudinally polarized nucleon, similarly to Fig.~\ref{Fig_Nucleon2DEFJ0JzJx}.}
	\label{Fig_Nucleon2DLFJpJmJx}
\end{figure}

Applying the general definition~\eqref{LF-definition} for the LF four-current distributions to the case of a spin-$\frac{1}{2}$ target, we obtain 
\begin{equation}
    \begin{aligned}\label{spinhalf-LF-JpJmJT}
        J^+_\text{LF}(\uvec b_\perp;P^+) &= e\int\frac{\ud^2\Delta_\perp}{(2\pi)^2}\,e^{-i\uvec\Delta_\perp\cdot\uvec b_\perp} \left[\delta_{\lambda'\lambda}\,F_1(\uvec\Delta_\perp^2) + \frac{(\uvec\sigma_{\lambda'\lambda}\times i\uvec\Delta)_z}{2M}\,F_2(\uvec\Delta_\perp^2)\right],\\
         J^-_\text{LF}(\uvec b_\perp;P^+) &= e \int\frac{\ud^2\Delta_\perp}{(2\pi)^2}\,e^{-i\uvec\Delta_\perp\cdot\uvec b_\perp} \,\frac{P^-}{P^+}\left[\delta_{\lambda'\lambda}\,G_1(\uvec\Delta_\perp^2) + \frac{(\uvec\sigma_{\lambda'\lambda}\times i\uvec\Delta)_z}{2M}\,G_2(\uvec\Delta_\perp^2)\right],\\
        \uvec J_{\perp,\text{LF}}(\uvec b_\perp;P^+) &= e\,(\sigma_z)_{\lambda'\lambda} \int\frac{\ud^2\Delta_\perp}{(2\pi)^2}\,e^{-i\uvec\Delta_\perp\cdot\uvec b_\perp} \,\frac{(\uvec e_z\times i\uvec\Delta)_\perp}{2P^+}\,G_M(\uvec\Delta^2_\perp).
    \end{aligned}
\end{equation}
Like in the spin-$0$ case~\eqref{spin0-LFJm}, we can use the relation $P^2=2P^+P^-$ to factor out the $P^+$-dependence in $J^-_\text{LF}(\uvec b_\perp;P^+)$. We also observe that the transverse LF current distributions differ from the transverse EF current distributions~\eqref{spinhalfEFJT}, though both of them vanish in the IMF. Finally, the LF charge distribution is independent of $P^+$ and coincides with the IMF charge and longitudinal current distributions
\begin{equation}
    J^+_\text{LF}(\uvec b_\perp;P^+)=J^0_\text{EF}(\uvec b_\perp;\infty)=J_{z,\text{EF}}(\uvec b_\perp;\infty).
\end{equation}
In Fig.~\ref{Fig_Nucleon2DLFJpJmJx}, we show some components of LF four-current distributions for different polarizations of the nucleon.

As a final remark, we point out that the spin structure of the LF distributions~\eqref{spinhalf-LF-JpJmJT} does not depend on the average momentum, a feature achieved thanks to the Melosh rotation which converts canonical polarization into LF helicity, see Appendix~\ref{App-Wigner-Melosh Rotations}. While this may a priori be considered as an advantage, it comes with the price that the electromagnetic four-current is now described in terms of five linearly dependent FFs, viz.~$F_{1,2}(Q^2)$, $G_{1,2}(Q^2)$ and $G_M(Q^2)$. This is to be contrasted with the EF distributions~\eqref{spinhalfEFJ0}-\eqref{spinhalfEFJT}, where the spin structure requires only $G_{E,M}(Q^2)$ and a $P_z$-dependent Wigner rotation. In particular, the spin structure of the EF distributions becomes simple in the BF, while it remains complicated in the same frame for the  LF distributions. Hence, contrary to the traditional LF picture which focuses only on the LF component $\hat j^+$, a more general perspective based on the full electromagnetic four-current indicates that it is the Sachs FFs that should be considered as the physical FFs.

%%%%%%%%%%%%%%%%%%%%%%%%%%%%%%%%%%%%%%%%%%%%%%%%%%%%%%%%%%%%%%%%%%%%%%%%%%%%%%%%%%
\section{Summary}
\label{sec:Summary}

In this paper, we extended the study of the relativistic charge distributions within the quantum phase-space formalism to the whole electromagnetic four-current. We treated in detail the spin-$0$ and spin-$\frac{1}{2}$ cases, discussed their frame dependence and compared with the corresponding light-front distributions. We confirm that all the relativistic distortions arising for a target with non-vanishing momentum can be understood as a combination of the familiar Lorentz four-vector transformation of the current and the Wigner spin rotation.

In the spin-$0$ case, the situation is simple since there is no Wigner rotation. We found that the charge and transverse current distributions are the same in both phase-space and light-front formalisms. They differ however for the longitudinal current distribution, which is the only component that depends on the target average momentum. We noted in particular that the elastic frame distributions (i.e.~those defined within the quantum phase-space approach) should be interpreted as giving a picture of the target with definite average momentum, and not with definite average velocity. The reason is that the mass-shell constraint implies that the various Fourier components contributing to an elastic frame distribution have different inertias, and hence different velocities for a given momentum.

The picture gets more complicated for a spin-$\frac{1}{2}$ target because of the polarization contribution to the current and the Wigner rotation. Of all the possible elastic frames, the Breit frame (interpreted from the phase-space perspective as the average rest frame) leads to the simplest multipole structure, and hence strongly suggests that the Sachs form factors should be interpreted as the physical electric and magnetic form factors. This contrasts with the light-front formalism where the Dirac and Pauli form factors are often presented as the physical ones. While the latter appear in a natural way when studying the light-front density, they do not provide a clear interpretation for the structure of the other components of the current. In this work, we demonstrated explicitly that keeping track of the spin rotation clarifies the general multipole structure of the full electromagnetic four-current in any frame. In the phase-space formalism, the spin rotation arises from boosting a spinning system from the Breit frame to another frame, whereas in the light-front formalism it arises from switching from canonical polarization to light-front helicity states.

Relativistic charge and three-current distributions are in general frame-dependent. To illustrate our results, we used convenient parametrizations of the pion ($\pi^+$) and nucleon electromagnetic form factors fitted to the experimental data. Like in the relativistic charge distributions, we observe significant distortions in the other components of the four-current distributions. We emphasize that our results and physical interpretations are of course applicable to any physical spin-$0$ or spin-$\tfrac{1}{2}$ target (including their antiparticles), e.g., $\overline{p}$, $\overline{n}$, $K^{\pm}$, $\Lambda$, $\Delta(1750)$, $\Sigma^0$, $\Xi^{-}$, etc., as long as their electromagnetic form factors are available. Moreover, our analysis can easily be generalized to higher-spin targets.

%%%%%%%%%%%%%%%%%%%%%%%%%%%%%%%%%%%%%%%%%%%%%%%%%%%%%%%%%%%%%%%%%%%%%%%%%%%%%%%%%%
\begin{acknowledgements}

We are grateful to Christoph Kopper for his careful reading and helpful comments on the manuscript. Y.~C. is grateful to Prof.~Qun Wang, Prof.~Shi Pu, and the Department of Modern Physics for their very kind hospitality and help during his visit to the University of Science and Technology of China. Y.~C. thanks Prof.~Qun Wang, Prof.~Guang-Peng Zhang, Prof.~Jian Zhou, Prof.~Bo-Wen Xiao, Prof.~Dao-Neng Gao and Prof.~Yang Li for insightful discussions, as well as Dr.~Xin-Li Sheng and Dr.~Ren-Jie Wang for helpful communications. This work is supported in part by the National Natural Science Foundation of China (NSFC) under Grant Nos.~12135011, 11890713 (a sub-Grant of 11890710), and by the Strategic Priority Research Program of the Chinese Academy of Sciences (CAS) under Grant No.~XDB34030102.

\end{acknowledgements}

%%%%%%%%%%%%%%%%%%%%%%%%%%%%%%%%%%%%%%%%%%%%%%%%%%%%%%%%%%%%%%%%%%%%%%%%%%%%%%%%%%
\appendix
\section{Pion electromagnetic form factors}
\label{sec:Pion form factors in both spacelike and timelike regions}

As a simple ansatz, one can use the following dipole model for the spacelike pion ($\pi^+$) FF,
\begin{equation}\label{dipolemodel}
	F^{(D)}_{\pi}(Q^2) = \frac{1}{\left( 1+Q^2R^2/12\right)^2},
\end{equation}
where the pion root-mean-square charge radius $R=\sqrt{\langle \uvec r_{\pi}^2\rangle}=(0.672\pm 0.008)~\text{fm}$ has been extracted from a fit to the world data~\cite{ParticleDataGroup:2012pjm,Carmignotto:2014rqa}. From~\eqref{dipolemodel}, one can easily obtain the corresponding 3D BF and 2D EF charge distributions (with $r=|\uvec r|$ and $b=|\uvec b_{\perp} |$)
\begin{equation}\label{3D2D-Dipole-model}
	\begin{aligned}
		J_B^{0,(D)}(r) &= \frac{3\sqrt{3}}{\pi R^3}\, \exp\!\left(- \frac{2\sqrt{3}\,r}{R}\right) ,\qquad
		J^{0,(D)}_{\text{EF}}(b) = \frac{6\sqrt{3}\,b}{\pi R^3}\, K_{1}\!\left(\frac{2\sqrt{3}\,b}{R} \right) ,\\
	\end{aligned}
\end{equation}
where $K_{\nu}(z)$ is the $\nu$-th order modified Bessel function of the second kind. Likewise, one can use the monopole model,
\begin{equation}\label{monopolemodel}
	F^{(M)}_{\pi}(Q^2) = \frac{1}{1+Q^2R^2/6},
\end{equation}
where $R$ is the same as in~\eqref{dipolemodel} and the corresponding 3D BF and 2D EF charge distributions now respectively read
\begin{equation}\label{3D2D-Monopole-model}
	\begin{aligned}
		J_B^{0,(M)}(r) &= \frac{3}{2\pi r R^2}\, \exp\!\left(- \frac{\sqrt{6}\,r}{R}\right) ,\qquad
		J^{0,(M)}_{\text{EF}}(b) = \frac{3}{\pi R^2}\, K_{0}\!\left(\frac{\sqrt{6}\,b}{R} \right).
	\end{aligned}
\end{equation}
Note in particular the manifest singular behaviors of $J_B^{0,(M)}(r)$ as $r \to0$ and $J^{0,(M)}_{\text{EF}}(b)$ as $b \to 0$, if a monopole FF like~\eqref{monopolemodel} is involved~\cite{Carmignotto:2014rqa,Miller:2009qu}.

The pion FF in the timelike region ($s=q^2>4m_{\pi}^2$ with $m_{\pi}$ the pion mass) over a large $q^2$ range has been precisely measured by the BABAR Collaboration~\cite{BaBar:2012bdw}, using the initial-state radiation method. Thanks to the dispersion theory, one can in principle obtain the corresponding pion FF in the spacelike region, especially for large $Q^2$ regions. The modular square of the timelike pion FF $|F_{\pi}(q^2)|^2$ is given by~\cite{BaBar:2012bdw}
\begin{equation}\label{BaBar-Pion-FFs}
	\begin{aligned}
		|F_{\pi}(s)|^2 =\frac{\sigma_{\pi\pi}(s)}{\sigma_{\text{point}}(s)},\qquad
		\sigma_{\pi\pi}(s) = \frac{\sigma^0_{\pi\pi(\gamma)}(s)}{1+\delta^{\pi\pi}_{\text{FSR}}(s) } \left(\frac{\alpha_{\text{EM}}(s)}{\alpha_{\text{EM}}(0)} \right)^2, 
	\end{aligned}
\end{equation}
where $\sqrt{s}$ is the net center-of-mass energy of the produced $\pi^+\pi^-$ pair with each particle moving at speed $\beta_{\pi}(s)=\sqrt{1-4m_{\pi}^2/s}$, $\alpha_{\text{EM}}(0)=e^2/(4\pi)\approx 1/137$ is the vacuum fine structure constant in the low-energy limit. Note that the lowest-order spin-0 pointlike charged particle pair-production cross section is $\sigma_{\text{point}}(s)=\pi \alpha^2_{\text{EM}}(0)\beta_{\pi}^3(s)/(3s)$, and $\sigma_{\pi\pi}(s)$ is the total dressed cross section, which differs from the experimentally measured bare cross section $\sigma^0_{\pi\pi(\gamma)}(s)$ by two corrections: one is the final-state radiation correction $1/(1+\delta^{\pi\pi}_{\text{FSR}}(s))$, and the other is the vacuum polarization correction $( \alpha_{\text{EM}}(s)/\alpha_{\text{EM}}(0)) ^2$.

\begin{figure}[!tb]
	\centering
	\subfigure[~Timelike pion form factor.\label{fig:FFPipBaBarFitExp}]
	{\includegraphics[angle=0,scale=0.4515]{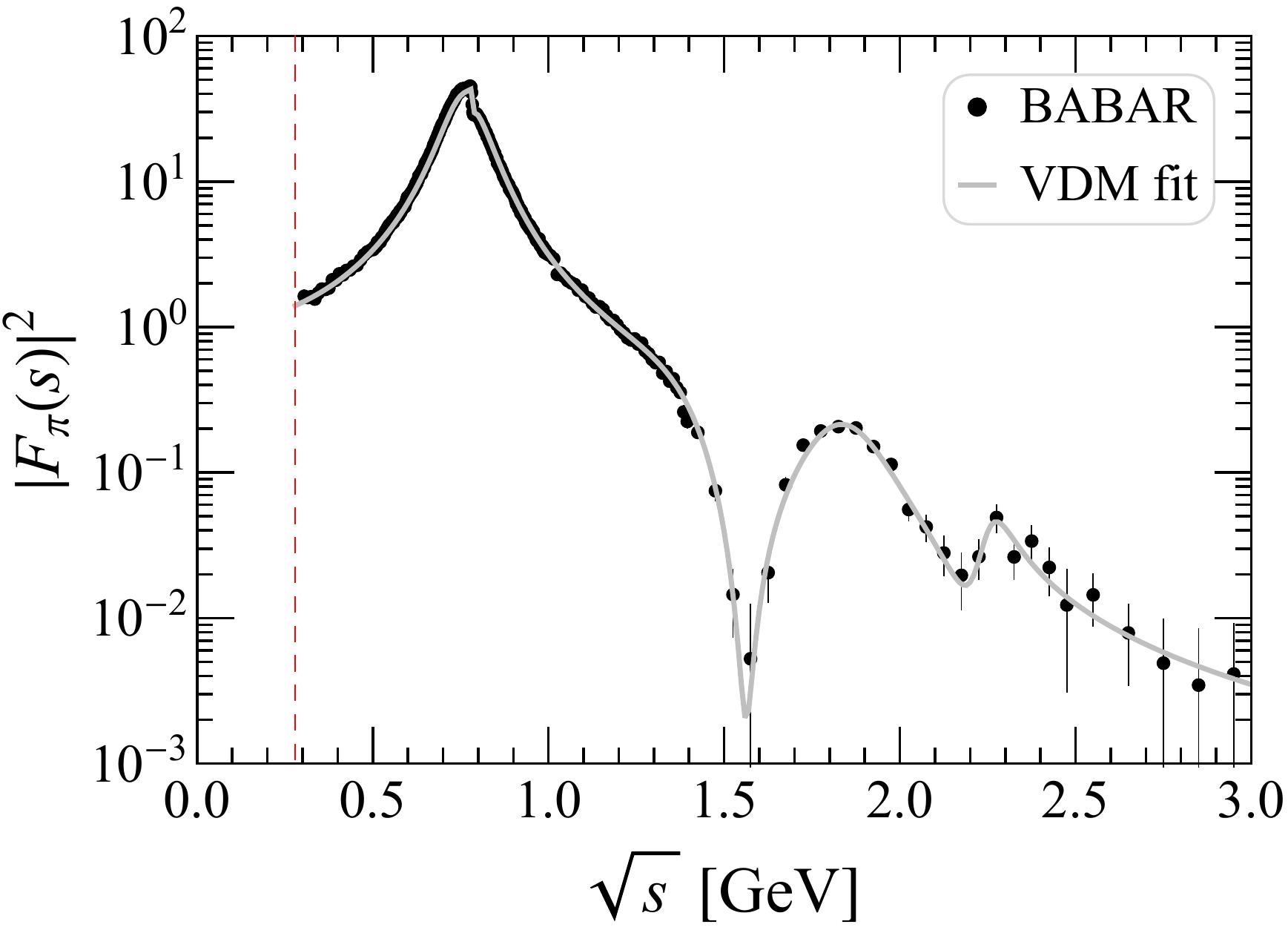}}
	\subfigure[~Spacelike pion form factor.\label{fig:FFPipSLTLAllFig}]
	{\includegraphics[angle=0,scale=0.4500]{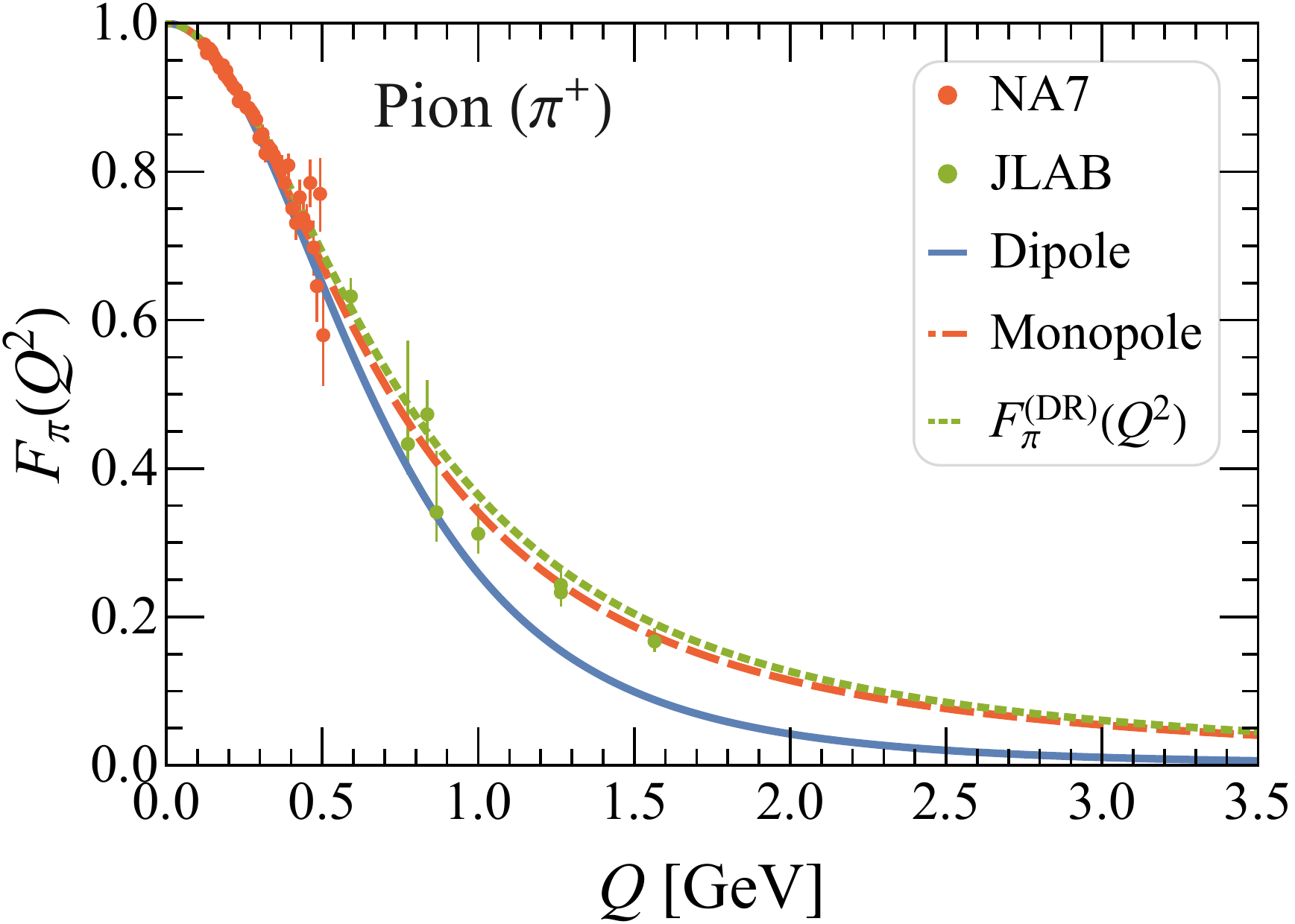}}
	\caption{(Color online) (a) A VDM fit of the timelike pion ($\pi^+$) FF measured by BABAR~\cite{BaBar:2012bdw} and (b) comparison between pion FFs in the spacelike region from three different approaches: simple dipole \eqref{dipolemodel} and monopole \eqref{monopolemodel} models, and modified dispersion relation \eqref{modified-DR}, along with the NA7~\cite{NA7:1986vav} and Jefferson Lab (JLAB) measurements~\cite{JeffersonLab:2008jve}.}
	\label{Fig_PionSLTLFF}
\end{figure}

In dispersion theory, the standard dispersion relation (DR) for the pion electromagnetic FF reads
\begin{equation}
	\begin{aligned}
		F^{\text{(DR)}}_{\pi}(Q^2) &= \frac{1}{\pi} \int_{s_0}^{\infty} \mathrm{d}s\, \frac{\text{Im}\, F_{\pi}(s)}{s-q^2-i\epsilon},
	\end{aligned}
\end{equation}
which connects the spacelike pion FF $F^{\text{(DR)}}_{\pi}(Q^2)$ at $Q^2=-q^2>0$ with the imaginary part of the timelike FF $F_{\pi}(s)$ integrated over $s$ above the two-pion threshold energy $\sqrt{s_0}=2m_{\pi}$ [indicated by a vertical dashed red line in Fig.~\ref{fig:FFPipBaBarFitExp}]. It has also been suggested to use the following modified DR for the spacelike pion FF $F^{\text{(DR)}}_{\pi}(Q^2)$~\cite{Geshkenbein:1998gu,Cheng:2020vwr,Chai:2022ipu},
\begin{equation}
	\begin{aligned}\label{modified-DR}
		F^{\text{(DR)}}_{\pi}(Q^2) &= \exp\!\left[-\frac{Q^2\sqrt{s_0 + Q^2}}{2\pi}\int_{s_0}^{\infty}\mathrm{d}s\,  \frac{2\ln|F_{\pi}(s)|}{s(s+Q^2)\sqrt{s-s_0}} \right],
	\end{aligned}
\end{equation}
which automatically ensures $F^{\text{(DR)}}_{\pi}(0)=1$ for the total charge of a $\pi^+$ meson in units of $e$.

It has been known for quite long time from a perturbative QCD (pQCD) analysis in the $Q^2 \to \infty$ limit that the {\it leading} asymptotic behavior of the spacelike pion FF $F_{\pi}(Q^2)$ reads~\cite{Jackson:1977mta,Farrar:1979aw,Efremov:1978fi,Efremov:1978rn,Efremov:1979sn,Efremov:1979qk,Lepage:1979zb,Lepage:1979za,Lepage:1980fj,Parisi:1979jp}
\begin{equation}
	\begin{aligned}\label{FPi-Asymptotic-SL}
		F_{\pi}(Q^2) &\underset{Q^2\to\infty}{\approx} \frac{4\pi C_{F}}{Q^2} \,\alpha_{s}(Q^2)\left|\,a_{0} \left(\ln \frac{Q^2}{\Lambda_{\text{QCD}}^2 } \right)^{-\gamma_{0}}\, \right|^{2} \\
		&\,\,\,\,\,\approx \frac{8\pi  }{Q^2} \,\alpha_{s}(Q^2)\,f_{\pi}^2  \propto \frac{1}{Q^2}\,\frac{1 }{\ln ( Q^2/\Lambda_{\text{QCD}}^2  )},
	\end{aligned}
\end{equation}
where $C_F=(N_{c}^2-1)/(2N_c)=4/3$ is the Casimir operator for the fundamental representation of the $\text{SU(3)}_c$ gauge field theory with $N_c=3$ quark colors, $f_{\pi} \approx (130.2\pm 1.7)\,\text{MeV}$ is the pion decay constant for the $\pi^{+}\to \mu^{+}\nu_\mu$ reaction channel~\cite{Workman:2022ynf}, $\Lambda_{\text{QCD}}\approx 200\,\text{MeV}$ is the QCD energy scale, $\gamma_{0}=0$ is the leading anomalous dimension, and $a_{0}$ is related to the pion decay constant $f_{\pi}$ via $a_{0}=3f_{\pi}/\sqrt{2N_c}=\sqrt{\frac{3}{2}}\,f_{\pi}$. At one-loop order, the QCD normalized strong coupling constant $\alpha_{s}(Q^2)=g_s^2(Q^2)/(4\pi)$ is explicitly given by
\begin{equation}
	\begin{aligned}\label{AlphaSOneLoop-SL}
		\alpha_{s}(Q^2) &= \frac{4\pi}{\beta_{0} \ln( Q^2/\Lambda_{\text{QCD}}^2) },
	\end{aligned}
\end{equation}
where $\beta_0=11-2N_f/3$ with $N_f$ the number of quark flavors. It was noted in Ref.~\cite{Efremov:1978rn} that the leading short-distance interactions are reflected through $8\pi\alpha_{s}(Q^2)/Q^2$ in \eqref{FPi-Asymptotic-SL}, whereas all long-distance effects are absorbed in $f_{\pi}^2$. Moreover, we should remember that Eq.~(\ref{FPi-Asymptotic-SL}) actually predicts the same positive sign as that from the usual vector-meson dominance models (VDMs). Explicitly, we see from Eq.~(\ref{FPi-Asymptotic-SL}) that the asymptotic leading behavior of $F_{\pi}(Q^2 )$ indeed appears like $F_{\pi}(Q^2 ) \sim Q^{-2}$, well consistent with the prediction of asymptotic {\it scaling law}~\cite{Brodsky:1973kr,Matveev:1973ra,Brodsky:1979nc}, which is modulated by the logarithmic behavior inherited from the QCD running coupling constant $\alpha_{s}(Q^2)$ in the pQCD analysis.

To ensure that the asymptotic behavior of $F^{\text{(DR)}}_{\pi}(Q^2)$ agrees with the pQCD prediction \eqref{FPi-Asymptotic-SL}, we follow the method in Refs.~\cite{Cheng:2020vwr,Dominguez:2001zu,Bruch:2004py} by using a large-$N_c$ infinite set of equidistant $\rho$ resonances for the timelike pion FF $F_{\pi}(s)$ in the region $s \geq s_{\text{max}}\equiv (2.95~\text{GeV})^2$, and using the $\rho$-resonance VDM parametrization (including $\rho-\omega$ interference) of BABAR data~\cite{BaBar:2012bdw,Cheng:2020vwr} in the region $s_0 \leq s \leq s_{\text{max}}$, with the charge normalization condition $F_{\pi}(0)=1$ automatically ensured. We eventually obtained the full spacelike pion FF $F^{\text{(DR)}}_{\pi}(Q^2)$ via \eqref{modified-DR}; see the green dotted line in Fig.~\ref{fig:FFPipSLTLAllFig}. 

%%%%%%%%%%%%%%%%%%%%%%%%%%%%%%%%%%%%%%%%%%%%%%%%%%%%%%%%%%%%%%%%%%%%%%%%%%%%%%%%%%
\section{Effective velocity and charge distributions}
\label{App-classical analogy}

By analogy with a classical current $\uvec J(x)=J^0(x)\uvec v(x)$, one can define an effective EF charge velocity distribution as
\begin{equation}\label{ClassVelocity}
	\uvec v^\text{eff}_\text{EF}(\uvec b_{\perp};P_z) \equiv \frac{\uvec J_\text{EF}(\uvec  b_{\perp};P_z)}{J^0_\text{EF}(\uvec b_\perp;P_z)}.
\end{equation}
For a spin-$0$ target, this effective velocity distribution is purely along the $z$-axis and is usually non-uniform in the transverse plane, suggesting that there is some dispersion in the charge distribution along the $z$-direction when $P_z\neq 0$. This dispersion does not show up however in the EF charge distribution since the longitudinal coordinate is integrated over. 

Alternatively, one can use the average center-of-mass velocity $\uvec v_\text{CM}=\uvec P/E_P$ with $E_P=\sqrt{M^2+\uvec P^2}$ and define a non-dispersive effective charge distribution in the EF for a spin-0 target
\begin{equation}\label{EFRhoeffective}
	\rho^\text{eff}_\text{EF}(\uvec b_\perp;P_z)\equiv e\int\frac{\ud^2\Delta_\perp}{(2\pi)^2}\,e^{-i\uvec\Delta_\perp\cdot\uvec b_\perp}\,\frac{E_P}{P^0}\,F(\uvec\Delta_\perp^2),
\end{equation}
such that $\uvec J_\text{EF}(\uvec b_\perp;P_z)= \rho^\text{eff}_\text{EF}(\uvec b_\perp;P_z)\uvec v_\text{CM}$. Since integrating over $\uvec b_\perp$ amounts to setting $\uvec\Delta_\perp=\uvec 0_\perp$ in momentum space, the total electric charge is given by
\begin{equation}
	q=\int\ud^2b_\perp\,J^0_\text{EF}( \uvec b_\perp;P_z)=\int\ud^2b_\perp\,\rho^\text{eff}_\text{EF}(\uvec b_\perp;P_z)=e\,F(0).
\end{equation}
Contrary to $J^0_\text{EF}$, the effective charge distribution does depend on $P_z$. Interestingly, it becomes equal to $J^0_\text{EF}$ when $P_z\to\infty$:
\begin{equation}
	\rho^\text{eff}_\text{EF}(\uvec b_\perp;\infty)=J^0_\text{EF}(\uvec b_\perp;\infty),
\end{equation}
which is to be expected since in this case all Fourier components of the effective charge distribution have the same velocity close to the speed of light.

\begin{figure}[t]
	\centering
	{\includegraphics[angle=0,scale=0.443]{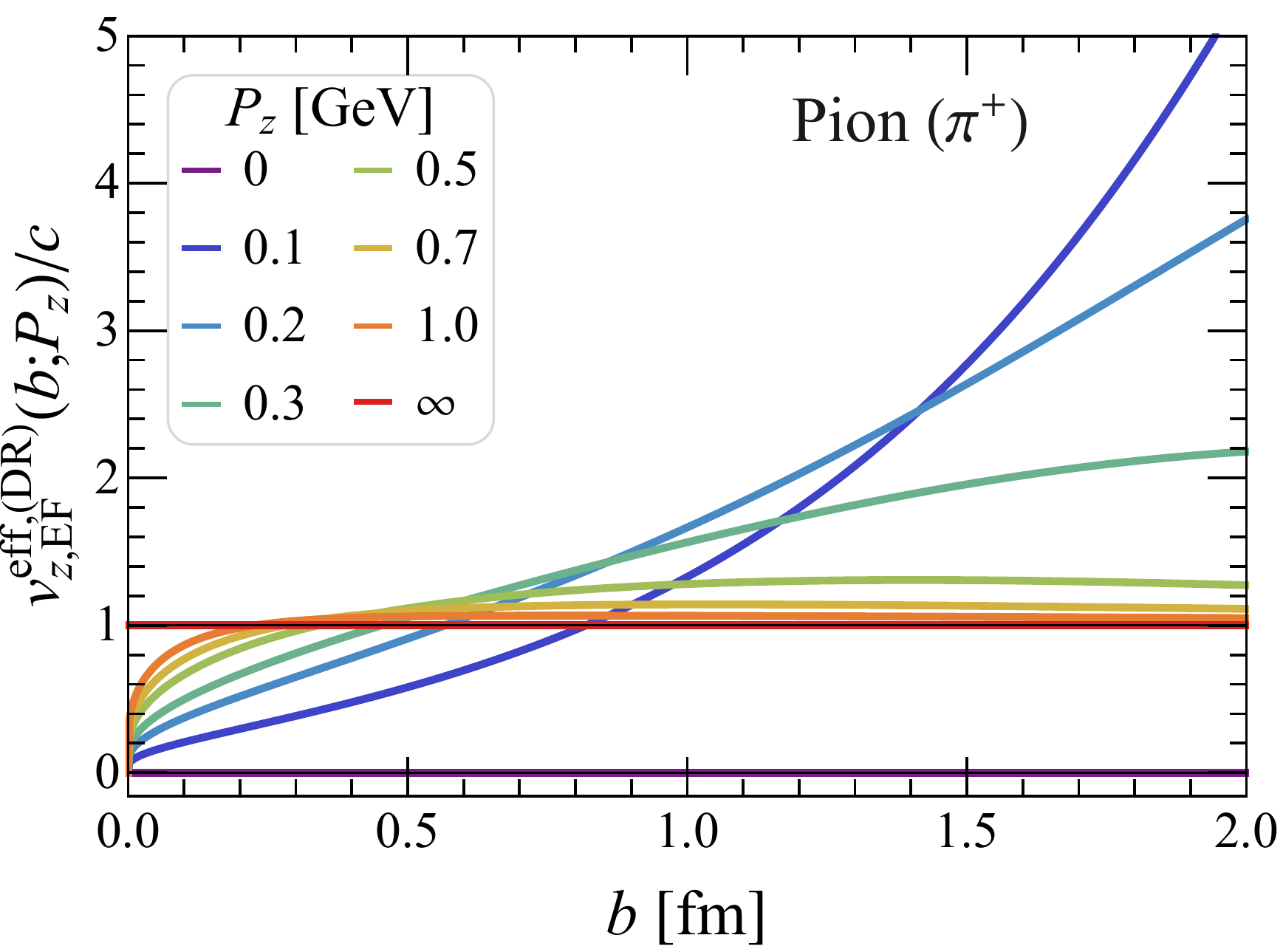}}
	{\includegraphics[angle=0,scale=0.455]{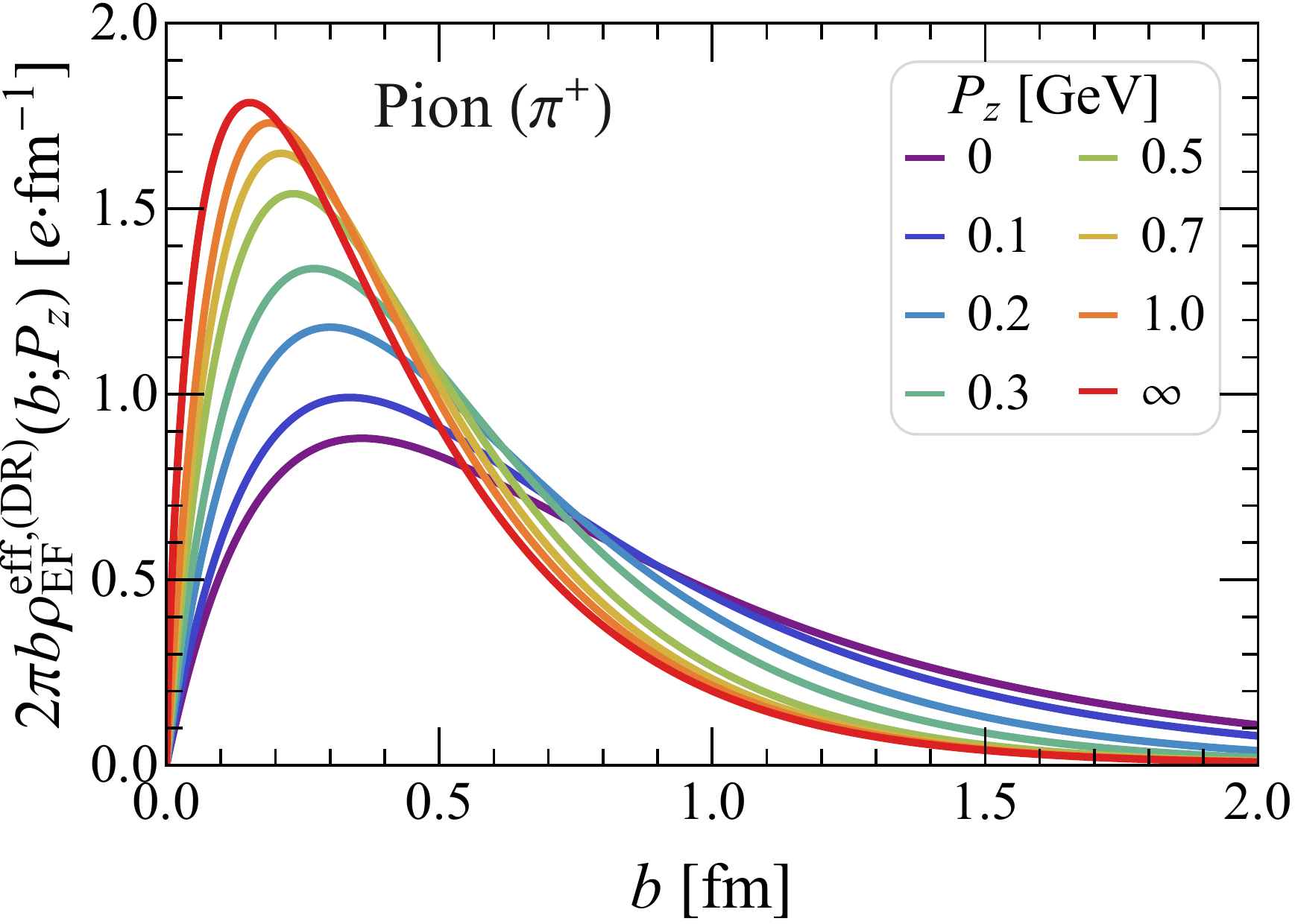}}
	\caption{(Color online) Momentum dependence of the longitudinal effective velocity distribution (left panel) and radial effective charge distribution (right panel) for a pion ($\pi^+$), based on the pion electromagnetic FF from modified dispersion relation~\eqref{modified-DR}.}
	\label{Fig_Pion2DEFVzRhoEff}
\end{figure}

 In Fig.~\ref{Fig_Pion2DEFVzRhoEff}, we show the radial effective velocity and charge distributions for various values of $P_z$, using the pion electromagnetic FF obtained within the dispersion relation framework. When $0<P_z<\infty$, we observe that there is a critical value for $b=|\uvec b_\perp|$ above which the effective velocity becomes superluminal. As $P_z$ increases, this critical value decreases towards $0$. In the IMF, the effective velocity becomes however equal to $1$ for all values of $b$. Superluminal effective velocities do not seem to be an artifact resulting from a poor choice of parametrization for the pion electromagnetic FF. We think it has in fact to do with the phase-space formalism itself. Indeed, position eigenstates are constructed following the Newton-Wigner approach~\cite{Newton:1949cq}. While at the initial time they correspond to perfectly localized states in position space, they will spread outside the light cone at later times. This is usually considered as a problematic feature, but as stressed in Ref.~\cite{Pavsic:2017orp} there is no actual information carried by this superluminal spreading, and hence no fundamental clash with the relativistic causality (a similar argument as for entangled states in the EPR paradox). In any case, we remind that the effective velocity was simply constructed by analogy with a classical four-current. While the analogy may be valid for the expectation value $\langle\Psi|\hat j^\mu(x)|\Psi\rangle$, it should be considered with a grain of salt in the case of $J^\mu_\text{EF}(\uvec b_\perp;P_z)$ since wave-packets have been factored out~\cite{Lorce:2018zpf,Lorce:2018egm,Lorce:2021gxs}. The same precaution applies to the effective charge distribution, since a genuine spin-$0$ charge distribution projected onto the transverse plane should in principle not depend on the target momentum, according to Special Relativity.

%%%%%%%%%%%%%%%%%%%%%%%%%%%%%%%%%%%%%%%%%%%%%%%%%%%%%%%%%%%%%%%%%%%%%%%%%%%%%%%%%%
\section{Wigner and Melosh rotations}
\label{App-Wigner-Melosh Rotations}

Let us consider a Lorentz transformation such that $p'^\mu=\Lambda^\mu_{\phantom{ \mu}\nu} p^\nu$. The Wigner rotation describes how the canonical polarization gets rotated\footnote{In the literature, the Wigner rotation matrix is often denoted as $\mathcal D_{s's}=(D^\dag)_{s's}= D^*_{ss'}$.}
\begin{align}
U(\Lambda)|p,s\rangle&=\sum_{s'}|p',s'\rangle\, D^*_{ss'}(p,\Lambda)\nonumber\\
    \Rightarrow\quad |p',s'\rangle&=\sum_s U(\Lambda)|p,s\rangle\, D_{ss'}(p,\Lambda),
\end{align}
where $U(\Lambda)$ is the unitary operator implementing the Lorentz transformation in Hilbert space. For a spin-$\frac{1}{2}$ system and a pure boost along the $z$-direction, we can write
\begin{equation}\label{Dthetahalf}
     D_{ss'}(p,\Lambda)=\cos\frac{\theta}{2}\,\delta_{ss'}+i\,\sin\frac{\theta}{2}\,\frac{(\uvec p\times\uvec\sigma_{ss'})_z}{|\uvec p_\perp|}.
\end{equation}
Using now the expressions we found for $\cos\theta$ and $\sin\theta$ in Eq.~\eqref{WignerAngleSinCos}, we conclude from the half-angle formulas that
\begin{equation}
\begin{aligned}
    \cos\frac{\theta}{2}&=\sqrt{\frac{(p^0+M\sqrt{1+\tau})(\sqrt{1+\tau}+1)}{2(p^0+M)\sqrt{1+\tau}}},\\
    \sin\frac{\theta}{2}&=-\sqrt{ \frac{ (p^0-M\sqrt{1+\tau})(\sqrt{1+\tau}-1)}{2(p^0+M)\sqrt{1+\tau}}}.
    \end{aligned}
\end{equation}
In particular, these expressions reduce in the IMF to
\begin{equation}\label{IMFtheta}
  \lim_{p_z\to\infty}  \cos\frac{\theta}{2}=\sqrt{\frac{\sqrt{1+\tau}+1}{2\sqrt{1+\tau}}},\qquad
    \lim_{p_z\to\infty}\sin\frac{\theta}{2}=-\sqrt{\frac{\sqrt{1+\tau}-1}{2\sqrt{1+\tau}}}.
\end{equation}

The Melosh rotation relates the LF helicity states to the canonical spin states as follows
\begin{equation}
    |p,\lambda\rangle_\text{LF}=\sum_s |p,s\rangle\,\mathcal{M}_{s\lambda},
\end{equation}
with
\begin{equation}\label{Melosh-MatrixElements}
    \mathcal{M}_{s\lambda}=\frac{(\sqrt{2}p^++M)\,\delta_{s\lambda }-i(\uvec p_\perp\times\uvec\sigma_{s\lambda})_z}{\sqrt{2\sqrt{2}p^+(p^0+M)}}.
\end{equation}
Introducing a Melosh rotation angle $\theta_M$ similarly to Eq.~\eqref{Dthetahalf}, we get
\begin{equation}\label{Meloshangle}
    \cos\frac{\theta_M}{2}=\frac{\sqrt{2}p^++M}{\sqrt{2\sqrt{2}p^+(p^0+M)}},\qquad
    \sin\frac{\theta_M}{2}=-\frac{|\uvec p_\perp|}{\sqrt{2\sqrt{2}p^+(p^0+M)}}.
\end{equation}
Since $(\sqrt{2}p^++M)^2+\uvec p^2_\perp=2\sqrt{2}p^+(p^0+M)$ it is easy to see that $\cos^2\frac{\theta_M}{2}+\sin^2\frac{\theta_M}{2}=1$. If we now consider a BF with purely transverse momentum transfer, the initial four-momentum reads $p^\mu_B=(M\sqrt{1+\tau},-\frac{\uvec\Delta_\perp}{2},0)$ and Eq.~\eqref{Meloshangle} reduces to
\begin{equation}
    \cos\frac{\theta_M}{2}\Big|_B=\frac{\sqrt{1+\tau}+1}{\sqrt{2\sqrt{1+\tau}(\sqrt{1+\tau}+1)}},\qquad
    \sin\frac{\theta_M}{2}\Big|_B=-\frac{\sqrt{\tau}}{\sqrt{2\sqrt{1+\tau}(\sqrt{1+\tau}+1)}}.
\end{equation}
Comparing this with Eq.~\eqref{IMFtheta}, we conclude that the Melosh rotation in the BF with $\Delta_z=0$ is equal to the IMF Wigner rotation. In other words, switching to the LF formalism amounts in some sense to using the IMF canonical polarization basis without having to consider the limit $p_z\to\infty$.

\newpage
%%%%%%%%%%%%%%%%%%%%%%%%%%%%%%%%%%%%%%%%%%%%%%%%%%%%%%%%%%%%%%%%%%%%%%%%%%%%%%%%%%
%
\bibliography{Relativistic_Charge_Four_Current_Refs.bib}

\begin{thebibliography}{120}%
\makeatletter
\providecommand \@ifxundefined [1]{%
 \@ifx{#1\undefined}
}%
\providecommand \@ifnum [1]{%
 \ifnum #1\expandafter \@firstoftwo
 \else \expandafter \@secondoftwo
 \fi
}%
\providecommand \@ifx [1]{%
 \ifx #1\expandafter \@firstoftwo
 \else \expandafter \@secondoftwo
 \fi
}%
\providecommand \natexlab [1]{#1}%
\providecommand \enquote  [1]{``#1''}%
\providecommand \bibnamefont  [1]{#1}%
\providecommand \bibfnamefont [1]{#1}%
\providecommand \citenamefont [1]{#1}%
\providecommand \href@noop [0]{\@secondoftwo}%
\providecommand \href [0]{\begingroup \@sanitize@url \@href}%
\providecommand \@href[1]{\@@startlink{#1}\@@href}%
\providecommand \@@href[1]{\endgroup#1\@@endlink}%
\providecommand \@sanitize@url [0]{\catcode `\\12\catcode `\$12\catcode
  `\&12\catcode `\#12\catcode `\^12\catcode `\_12\catcode `\%12\relax}%
\providecommand \@@startlink[1]{}%
\providecommand \@@endlink[0]{}%
\providecommand \url  [0]{\begingroup\@sanitize@url \@url }%
\providecommand \@url [1]{\endgroup\@href {#1}{\urlprefix }}%
\providecommand \urlprefix  [0]{URL }%
\providecommand \Eprint [0]{\href }%
\providecommand \doibase [0]{http://dx.doi.org/}%
\providecommand \selectlanguage [0]{\@gobble}%
\providecommand \bibinfo  [0]{\@secondoftwo}%
\providecommand \bibfield  [0]{\@secondoftwo}%
\providecommand \translation [1]{[#1]}%
\providecommand \BibitemOpen [0]{}%
\providecommand \bibitemStop [0]{}%
\providecommand \bibitemNoStop [0]{.\EOS\space}%
\providecommand \EOS [0]{\spacefactor3000\relax}%
\providecommand \BibitemShut  [1]{\csname bibitem#1\endcsname}%
\let\auto@bib@innerbib\@empty
%</preamble>
\bibitem [{\citenamefont {Weinberg}(1995)}]{Weinberg1995Vol1}%
  \BibitemOpen
  \bibfield  {author} {\bibinfo {author} {\bibfnamefont {S.}~\bibnamefont
  {Weinberg}},\ }\href {https://inspirehep.net/literature/406190} {\emph
  {\bibinfo {title} {The Quantum Theory of Fields. Vol. 1: Foundations}}}\
  (\bibinfo  {publisher} {Cambridge University Press, Cambridge, England},\
  \bibinfo {year} {1995})\BibitemShut {NoStop}%
\bibitem [{\citenamefont {Gao}\ and\ \citenamefont
  {Vanderhaeghen}(2022)}]{Gao:2021sml}%
  \BibitemOpen
  \bibfield  {author} {\bibinfo {author} {\bibfnamefont {H.}~\bibnamefont
  {Gao}}\ and\ \bibinfo {author} {\bibfnamefont {M.}~\bibnamefont
  {Vanderhaeghen}},\ }\href {\doibase 10.1103/RevModPhys.94.015002} {\bibfield
  {journal} {\bibinfo  {journal} {Rev. Mod. Phys.}\ }\textbf {\bibinfo {volume}
  {94}},\ \bibinfo {pages} {015002} (\bibinfo {year} {2022})},\ \Eprint
  {http://arxiv.org/abs/2105.00571} {arXiv:2105.00571 [hep-ph]} \BibitemShut
  {NoStop}%
\bibitem [{\citenamefont {Amendolia}\ \emph {et~al.}(1986)\citenamefont
  {Amendolia} \emph {et~al.}}]{NA7:1986vav}%
  \BibitemOpen
  \bibfield  {author} {\bibinfo {author} {\bibfnamefont {S.~R.}\ \bibnamefont
  {Amendolia}} \emph {et~al.} (\bibinfo {collaboration} {NA7}),\ }\href
  {\doibase 10.1016/0550-3213(86)90437-2} {\bibfield  {journal} {\bibinfo
  {journal} {Nucl. Phys. B}\ }\textbf {\bibinfo {volume} {277}},\ \bibinfo
  {pages} {168} (\bibinfo {year} {1986})}\BibitemShut {NoStop}%
\bibitem [{\citenamefont {Anklin}\ \emph {et~al.}(1994)\citenamefont {Anklin}
  \emph {et~al.}}]{Anklin:1994ae}%
  \BibitemOpen
  \bibfield  {author} {\bibinfo {author} {\bibfnamefont {H.}~\bibnamefont
  {Anklin}} \emph {et~al.},\ }\href {\doibase 10.1016/0370-2693(94)90538-X}
  {\bibfield  {journal} {\bibinfo  {journal} {Phys. Lett. B}\ }\textbf
  {\bibinfo {volume} {336}},\ \bibinfo {pages} {313} (\bibinfo {year}
  {1994})}\BibitemShut {NoStop}%
\bibitem [{\citenamefont {Qattan}\ \emph {et~al.}(2005)\citenamefont {Qattan}
  \emph {et~al.}}]{Qattan:2004ht}%
  \BibitemOpen
  \bibfield  {author} {\bibinfo {author} {\bibfnamefont {I.~A.}\ \bibnamefont
  {Qattan}} \emph {et~al.},\ }\href {\doibase 10.1103/PhysRevLett.94.142301}
  {\bibfield  {journal} {\bibinfo  {journal} {Phys. Rev. Lett.}\ }\textbf
  {\bibinfo {volume} {94}},\ \bibinfo {pages} {142301} (\bibinfo {year}
  {2005})},\ \Eprint {http://arxiv.org/abs/nucl-ex/0410010}
  {arXiv:nucl-ex/0410010} \BibitemShut {NoStop}%
\bibitem [{\citenamefont {Horn}\ \emph {et~al.}(2006)\citenamefont {Horn} \emph
  {et~al.}}]{JeffersonLabFpi-2:2006ysh}%
  \BibitemOpen
  \bibfield  {author} {\bibinfo {author} {\bibfnamefont {T.}~\bibnamefont
  {Horn}} \emph {et~al.} (\bibinfo {collaboration} {Jefferson Lab F(pi)-2}),\
  }\href {\doibase 10.1103/PhysRevLett.97.192001} {\bibfield  {journal}
  {\bibinfo  {journal} {Phys. Rev. Lett.}\ }\textbf {\bibinfo {volume} {97}},\
  \bibinfo {pages} {192001} (\bibinfo {year} {2006})},\ \Eprint
  {http://arxiv.org/abs/nucl-ex/0607005} {arXiv:nucl-ex/0607005} \BibitemShut
  {NoStop}%
\bibitem [{\citenamefont {Arrington}\ \emph
  {et~al.}(2007{\natexlab{a}})\citenamefont {Arrington}, \citenamefont
  {Melnitchouk},\ and\ \citenamefont {Tjon}}]{Arrington:2007ux}%
  \BibitemOpen
  \bibfield  {author} {\bibinfo {author} {\bibfnamefont {J.}~\bibnamefont
  {Arrington}}, \bibinfo {author} {\bibfnamefont {W.}~\bibnamefont
  {Melnitchouk}}, \ and\ \bibinfo {author} {\bibfnamefont {J.~A.}\ \bibnamefont
  {Tjon}},\ }\href {\doibase 10.1103/PhysRevC.76.035205} {\bibfield  {journal}
  {\bibinfo  {journal} {Phys. Rev. C}\ }\textbf {\bibinfo {volume} {76}},\
  \bibinfo {pages} {035205} (\bibinfo {year} {2007}{\natexlab{a}})},\ \Eprint
  {http://arxiv.org/abs/0707.1861} {arXiv:0707.1861 [nucl-ex]} \BibitemShut
  {NoStop}%
\bibitem [{\citenamefont {Huber}\ \emph {et~al.}(2008)\citenamefont {Huber}
  \emph {et~al.}}]{JeffersonLab:2008jve}%
  \BibitemOpen
  \bibfield  {author} {\bibinfo {author} {\bibfnamefont {G.~M.}\ \bibnamefont
  {Huber}} \emph {et~al.} (\bibinfo {collaboration} {Jefferson Lab}),\ }\href
  {\doibase 10.1103/PhysRevC.78.045203} {\bibfield  {journal} {\bibinfo
  {journal} {Phys. Rev. C}\ }\textbf {\bibinfo {volume} {78}},\ \bibinfo
  {pages} {045203} (\bibinfo {year} {2008})},\ \Eprint
  {http://arxiv.org/abs/0809.3052} {arXiv:0809.3052 [nucl-ex]} \BibitemShut
  {NoStop}%
\bibitem [{\citenamefont {Lachniet}\ \emph {et~al.}(2009)\citenamefont
  {Lachniet} \emph {et~al.}}]{CLAS:2008idi}%
  \BibitemOpen
  \bibfield  {author} {\bibinfo {author} {\bibfnamefont {J.}~\bibnamefont
  {Lachniet}} \emph {et~al.} (\bibinfo {collaboration} {CLAS}),\ }\href
  {\doibase 10.1103/PhysRevLett.102.192001} {\bibfield  {journal} {\bibinfo
  {journal} {Phys. Rev. Lett.}\ }\textbf {\bibinfo {volume} {102}},\ \bibinfo
  {pages} {192001} (\bibinfo {year} {2009})},\ \Eprint
  {http://arxiv.org/abs/0811.1716} {arXiv:0811.1716 [nucl-ex]} \BibitemShut
  {NoStop}%
\bibitem [{\citenamefont {Bernauer}\ \emph {et~al.}(2010)\citenamefont
  {Bernauer} \emph {et~al.}}]{A1:2010nsl}%
  \BibitemOpen
  \bibfield  {author} {\bibinfo {author} {\bibfnamefont {J.~C.}\ \bibnamefont
  {Bernauer}} \emph {et~al.} (\bibinfo {collaboration} {A1}),\ }\href {\doibase
  10.1103/PhysRevLett.105.242001} {\bibfield  {journal} {\bibinfo  {journal}
  {Phys. Rev. Lett.}\ }\textbf {\bibinfo {volume} {105}},\ \bibinfo {pages}
  {242001} (\bibinfo {year} {2010})},\ \Eprint {http://arxiv.org/abs/1007.5076}
  {arXiv:1007.5076 [nucl-ex]} \BibitemShut {NoStop}%
\bibitem [{\citenamefont {Zhan}\ \emph {et~al.}(2011)\citenamefont {Zhan} \emph
  {et~al.}}]{Zhan:2011ji}%
  \BibitemOpen
  \bibfield  {author} {\bibinfo {author} {\bibfnamefont {X.}~\bibnamefont
  {Zhan}} \emph {et~al.},\ }\href {\doibase 10.1016/j.physletb.2011.10.002}
  {\bibfield  {journal} {\bibinfo  {journal} {Phys. Lett. B}\ }\textbf
  {\bibinfo {volume} {705}},\ \bibinfo {pages} {59} (\bibinfo {year} {2011})},\
  \Eprint {http://arxiv.org/abs/1102.0318} {arXiv:1102.0318 [nucl-ex]}
  \BibitemShut {NoStop}%
\bibitem [{\citenamefont {Puckett}\ \emph {et~al.}(2012)\citenamefont {Puckett}
  \emph {et~al.}}]{Puckett:2011xg}%
  \BibitemOpen
  \bibfield  {author} {\bibinfo {author} {\bibfnamefont {A.~J.~R.}\
  \bibnamefont {Puckett}} \emph {et~al.},\ }\href {\doibase
  10.1103/PhysRevC.85.045203} {\bibfield  {journal} {\bibinfo  {journal} {Phys.
  Rev. C}\ }\textbf {\bibinfo {volume} {85}},\ \bibinfo {pages} {045203}
  (\bibinfo {year} {2012})},\ \Eprint {http://arxiv.org/abs/1102.5737}
  {arXiv:1102.5737 [nucl-ex]} \BibitemShut {NoStop}%
\bibitem [{\citenamefont {Lees}\ \emph {et~al.}(2012)\citenamefont {Lees} \emph
  {et~al.}}]{BaBar:2012bdw}%
  \BibitemOpen
  \bibfield  {author} {\bibinfo {author} {\bibfnamefont {J.~P.}\ \bibnamefont
  {Lees}} \emph {et~al.} (\bibinfo {collaboration} {BaBar}),\ }\href {\doibase
  10.1103/PhysRevD.86.032013} {\bibfield  {journal} {\bibinfo  {journal} {Phys.
  Rev. D}\ }\textbf {\bibinfo {volume} {86}},\ \bibinfo {pages} {032013}
  (\bibinfo {year} {2012})},\ \Eprint {http://arxiv.org/abs/1205.2228}
  {arXiv:1205.2228 [hep-ex]} \BibitemShut {NoStop}%
\bibitem [{\citenamefont {Bernauer}\ \emph {et~al.}(2014)\citenamefont
  {Bernauer} \emph {et~al.}}]{A1:2013fsc}%
  \BibitemOpen
  \bibfield  {author} {\bibinfo {author} {\bibfnamefont {J.~C.}\ \bibnamefont
  {Bernauer}} \emph {et~al.} (\bibinfo {collaboration} {A1}),\ }\href {\doibase
  10.1103/PhysRevC.90.015206} {\bibfield  {journal} {\bibinfo  {journal} {Phys.
  Rev. C}\ }\textbf {\bibinfo {volume} {90}},\ \bibinfo {pages} {015206}
  (\bibinfo {year} {2014})},\ \Eprint {http://arxiv.org/abs/1307.6227}
  {arXiv:1307.6227 [nucl-ex]} \BibitemShut {NoStop}%
\bibitem [{\citenamefont {Punjabi}\ \emph {et~al.}(2015)\citenamefont
  {Punjabi}, \citenamefont {Perdrisat}, \citenamefont {Jones}, \citenamefont
  {Brash},\ and\ \citenamefont {Carlson}}]{Punjabi:2015bba}%
  \BibitemOpen
  \bibfield  {author} {\bibinfo {author} {\bibfnamefont {V.}~\bibnamefont
  {Punjabi}}, \bibinfo {author} {\bibfnamefont {C.~F.}\ \bibnamefont
  {Perdrisat}}, \bibinfo {author} {\bibfnamefont {M.~K.}\ \bibnamefont
  {Jones}}, \bibinfo {author} {\bibfnamefont {E.~J.}\ \bibnamefont {Brash}}, \
  and\ \bibinfo {author} {\bibfnamefont {C.~E.}\ \bibnamefont {Carlson}},\
  }\href {\doibase 10.1140/epja/i2015-15079-x} {\bibfield  {journal} {\bibinfo
  {journal} {Eur. Phys. J. A}\ }\textbf {\bibinfo {volume} {51}},\ \bibinfo
  {pages} {79} (\bibinfo {year} {2015})},\ \Eprint
  {http://arxiv.org/abs/1503.01452} {arXiv:1503.01452 [nucl-ex]} \BibitemShut
  {NoStop}%
\bibitem [{\citenamefont {Ablikim}\ \emph {et~al.}(2016)\citenamefont {Ablikim}
  \emph {et~al.}}]{BESIII:2015equ}%
  \BibitemOpen
  \bibfield  {author} {\bibinfo {author} {\bibfnamefont {M.}~\bibnamefont
  {Ablikim}} \emph {et~al.} (\bibinfo {collaboration} {BESIII}),\ }\href
  {\doibase 10.1016/j.physletb.2015.11.043} {\bibfield  {journal} {\bibinfo
  {journal} {Phys. Lett. B}\ }\textbf {\bibinfo {volume} {753}},\ \bibinfo
  {pages} {629} (\bibinfo {year} {2016})},\ \bibinfo {note} {[Erratum:
  Phys.Lett.B 812, 135982 (2021)]},\ \Eprint {http://arxiv.org/abs/1507.08188}
  {arXiv:1507.08188 [hep-ex]} \BibitemShut {NoStop}%
\bibitem [{\citenamefont {Puckett}\ \emph {et~al.}(2017)\citenamefont {Puckett}
  \emph {et~al.}}]{Puckett:2017flj}%
  \BibitemOpen
  \bibfield  {author} {\bibinfo {author} {\bibfnamefont {A.~J.~R.}\
  \bibnamefont {Puckett}} \emph {et~al.},\ }\href {\doibase
  10.1103/PhysRevC.96.055203} {\bibfield  {journal} {\bibinfo  {journal} {Phys.
  Rev. C}\ }\textbf {\bibinfo {volume} {96}},\ \bibinfo {pages} {055203}
  (\bibinfo {year} {2017})},\ \bibinfo {note} {[Erratum: Phys.Rev.C 98, 019907
  (2018)]},\ \Eprint {http://arxiv.org/abs/1707.08587} {arXiv:1707.08587
  [nucl-ex]} \BibitemShut {NoStop}%
\bibitem [{\citenamefont {Liyanage}\ \emph {et~al.}(2020)\citenamefont
  {Liyanage} \emph {et~al.}}]{SANE:2018cub}%
  \BibitemOpen
  \bibfield  {author} {\bibinfo {author} {\bibfnamefont {A.}~\bibnamefont
  {Liyanage}} \emph {et~al.} (\bibinfo {collaboration} {SANE}),\ }\href
  {\doibase 10.1103/PhysRevC.101.035206} {\bibfield  {journal} {\bibinfo
  {journal} {Phys. Rev. C}\ }\textbf {\bibinfo {volume} {101}},\ \bibinfo
  {pages} {035206} (\bibinfo {year} {2020})},\ \Eprint
  {http://arxiv.org/abs/1806.11156} {arXiv:1806.11156 [nucl-ex]} \BibitemShut
  {NoStop}%
\bibitem [{\citenamefont {Xiong}\ \emph {et~al.}(2019)\citenamefont {Xiong}
  \emph {et~al.}}]{Xiong:2019umf}%
  \BibitemOpen
  \bibfield  {author} {\bibinfo {author} {\bibfnamefont {W.}~\bibnamefont
  {Xiong}} \emph {et~al.},\ }\href {\doibase 10.1038/s41586-019-1721-2}
  {\bibfield  {journal} {\bibinfo  {journal} {Nature}\ }\textbf {\bibinfo
  {volume} {575}},\ \bibinfo {pages} {147} (\bibinfo {year}
  {2019})}\BibitemShut {NoStop}%
\bibitem [{\citenamefont {Mihovilovi\v{c}}\ \emph {et~al.}(2021)\citenamefont
  {Mihovilovi\v{c}} \emph {et~al.}}]{Mihovilovic:2019jiz}%
  \BibitemOpen
  \bibfield  {author} {\bibinfo {author} {\bibfnamefont {M.}~\bibnamefont
  {Mihovilovi\v{c}}} \emph {et~al.},\ }\href {\doibase
  10.1140/epja/s10050-021-00414-x} {\bibfield  {journal} {\bibinfo  {journal}
  {Eur. Phys. J. A}\ }\textbf {\bibinfo {volume} {57}},\ \bibinfo {pages} {107}
  (\bibinfo {year} {2021})},\ \Eprint {http://arxiv.org/abs/1905.11182}
  {arXiv:1905.11182 [nucl-ex]} \BibitemShut {NoStop}%
\bibitem [{\citenamefont {Gasparian}\ \emph {et~al.}(2020)\citenamefont
  {Gasparian} \emph {et~al.}}]{PRad:2020oor}%
  \BibitemOpen
  \bibfield  {author} {\bibinfo {author} {\bibfnamefont {A.}~\bibnamefont
  {Gasparian}} \emph {et~al.} (\bibinfo {collaboration} {PRad}),\ }\href@noop
  {} {\  (\bibinfo {year} {2020})},\ \Eprint {http://arxiv.org/abs/2009.10510}
  {arXiv:2009.10510 [nucl-ex]} \BibitemShut {NoStop}%
\bibitem [{\citenamefont {Atac}\ \emph {et~al.}(2021)\citenamefont {Atac},
  \citenamefont {Constantinou}, \citenamefont {Meziani}, \citenamefont
  {Paolone},\ and\ \citenamefont {Sparveris}}]{Atac:2021wqj}%
  \BibitemOpen
  \bibfield  {author} {\bibinfo {author} {\bibfnamefont {H.}~\bibnamefont
  {Atac}}, \bibinfo {author} {\bibfnamefont {M.}~\bibnamefont {Constantinou}},
  \bibinfo {author} {\bibfnamefont {Z.~E.}\ \bibnamefont {Meziani}}, \bibinfo
  {author} {\bibfnamefont {M.}~\bibnamefont {Paolone}}, \ and\ \bibinfo
  {author} {\bibfnamefont {N.}~\bibnamefont {Sparveris}},\ }\href {\doibase
  10.1038/s41467-021-22028-z} {\bibfield  {journal} {\bibinfo  {journal}
  {Nature Commun.}\ }\textbf {\bibinfo {volume} {12}},\ \bibinfo {pages} {1759}
  (\bibinfo {year} {2021})},\ \Eprint {http://arxiv.org/abs/2103.10840}
  {arXiv:2103.10840 [nucl-ex]} \BibitemShut {NoStop}%
\bibitem [{\citenamefont {Zhou}\ \emph {et~al.}(2021)\citenamefont {Zhou},
  \citenamefont {Khachatryan}, \citenamefont {Gao}, \citenamefont {Gorbaty},\
  and\ \citenamefont {Higinbotham}}]{Zhou:2021gyh}%
  \BibitemOpen
  \bibfield  {author} {\bibinfo {author} {\bibfnamefont {J.}~\bibnamefont
  {Zhou}}, \bibinfo {author} {\bibfnamefont {V.}~\bibnamefont {Khachatryan}},
  \bibinfo {author} {\bibfnamefont {H.}~\bibnamefont {Gao}}, \bibinfo {author}
  {\bibfnamefont {S.}~\bibnamefont {Gorbaty}}, \ and\ \bibinfo {author}
  {\bibfnamefont {D.~W.}\ \bibnamefont {Higinbotham}},\ }\href@noop {} {\
  (\bibinfo {year} {2021})},\ \Eprint {http://arxiv.org/abs/2110.02557}
  {arXiv:2110.02557 [hep-ph]} \BibitemShut {NoStop}%
\bibitem [{\citenamefont {Alexandrou}\ \emph {et~al.}(2017)\citenamefont
  {Alexandrou}, \citenamefont {Constantinou}, \citenamefont {Hadjiyiannakou},
  \citenamefont {Jansen}, \citenamefont {Kallidonis}, \citenamefont {Koutsou},\
  and\ \citenamefont {Vaquero Aviles-Casco}}]{Alexandrou:2017ypw}%
  \BibitemOpen
  \bibfield  {author} {\bibinfo {author} {\bibfnamefont {C.}~\bibnamefont
  {Alexandrou}}, \bibinfo {author} {\bibfnamefont {M.}~\bibnamefont
  {Constantinou}}, \bibinfo {author} {\bibfnamefont {K.}~\bibnamefont
  {Hadjiyiannakou}}, \bibinfo {author} {\bibfnamefont {K.}~\bibnamefont
  {Jansen}}, \bibinfo {author} {\bibfnamefont {C.}~\bibnamefont {Kallidonis}},
  \bibinfo {author} {\bibfnamefont {G.}~\bibnamefont {Koutsou}}, \ and\
  \bibinfo {author} {\bibfnamefont {A.}~\bibnamefont {Vaquero Aviles-Casco}},\
  }\href {\doibase 10.1103/PhysRevD.96.034503} {\bibfield  {journal} {\bibinfo
  {journal} {Phys. Rev. D}\ }\textbf {\bibinfo {volume} {96}},\ \bibinfo
  {pages} {034503} (\bibinfo {year} {2017})},\ \Eprint
  {http://arxiv.org/abs/1706.00469} {arXiv:1706.00469 [hep-lat]} \BibitemShut
  {NoStop}%
\bibitem [{\citenamefont {Hasan}\ \emph {et~al.}(2018)\citenamefont {Hasan},
  \citenamefont {Green}, \citenamefont {Meinel}, \citenamefont {Engelhardt},
  \citenamefont {Krieg}, \citenamefont {Negele}, \citenamefont {Pochinsky},\
  and\ \citenamefont {Syritsyn}}]{Hasan:2017wwt}%
  \BibitemOpen
  \bibfield  {author} {\bibinfo {author} {\bibfnamefont {N.}~\bibnamefont
  {Hasan}}, \bibinfo {author} {\bibfnamefont {J.}~\bibnamefont {Green}},
  \bibinfo {author} {\bibfnamefont {S.}~\bibnamefont {Meinel}}, \bibinfo
  {author} {\bibfnamefont {M.}~\bibnamefont {Engelhardt}}, \bibinfo {author}
  {\bibfnamefont {S.}~\bibnamefont {Krieg}}, \bibinfo {author} {\bibfnamefont
  {J.}~\bibnamefont {Negele}}, \bibinfo {author} {\bibfnamefont
  {A.}~\bibnamefont {Pochinsky}}, \ and\ \bibinfo {author} {\bibfnamefont
  {S.}~\bibnamefont {Syritsyn}},\ }\href {\doibase 10.1103/PhysRevD.97.034504}
  {\bibfield  {journal} {\bibinfo  {journal} {Phys. Rev. D}\ }\textbf {\bibinfo
  {volume} {97}},\ \bibinfo {pages} {034504} (\bibinfo {year} {2018})},\
  \Eprint {http://arxiv.org/abs/1711.11385} {arXiv:1711.11385 [hep-lat]}
  \BibitemShut {NoStop}%
\bibitem [{\citenamefont {Shintani}\ \emph {et~al.}(2019)\citenamefont
  {Shintani}, \citenamefont {Ishikawa}, \citenamefont {Kuramashi},
  \citenamefont {Sasaki},\ and\ \citenamefont {Yamazaki}}]{Shintani:2018ozy}%
  \BibitemOpen
  \bibfield  {author} {\bibinfo {author} {\bibfnamefont {E.}~\bibnamefont
  {Shintani}}, \bibinfo {author} {\bibfnamefont {K.-I.}\ \bibnamefont
  {Ishikawa}}, \bibinfo {author} {\bibfnamefont {Y.}~\bibnamefont {Kuramashi}},
  \bibinfo {author} {\bibfnamefont {S.}~\bibnamefont {Sasaki}}, \ and\ \bibinfo
  {author} {\bibfnamefont {T.}~\bibnamefont {Yamazaki}},\ }\href {\doibase
  10.1103/PhysRevD.99.014510} {\bibfield  {journal} {\bibinfo  {journal} {Phys.
  Rev. D}\ }\textbf {\bibinfo {volume} {99}},\ \bibinfo {pages} {014510}
  (\bibinfo {year} {2019})},\ \bibinfo {note} {[Erratum: Phys.Rev.D 102, 019902
  (2020)]},\ \Eprint {http://arxiv.org/abs/1811.07292} {arXiv:1811.07292
  [hep-lat]} \BibitemShut {NoStop}%
\bibitem [{\citenamefont {Alexandrou}\ \emph {et~al.}(2019)\citenamefont
  {Alexandrou}, \citenamefont {Bacchio}, \citenamefont {Constantinou},
  \citenamefont {Finkenrath}, \citenamefont {Hadjiyiannakou}, \citenamefont
  {Jansen}, \citenamefont {Koutsou},\ and\ \citenamefont {Vaquero
  Aviles-Casco}}]{Alexandrou:2018sjm}%
  \BibitemOpen
  \bibfield  {author} {\bibinfo {author} {\bibfnamefont {C.}~\bibnamefont
  {Alexandrou}}, \bibinfo {author} {\bibfnamefont {S.}~\bibnamefont {Bacchio}},
  \bibinfo {author} {\bibfnamefont {M.}~\bibnamefont {Constantinou}}, \bibinfo
  {author} {\bibfnamefont {J.}~\bibnamefont {Finkenrath}}, \bibinfo {author}
  {\bibfnamefont {K.}~\bibnamefont {Hadjiyiannakou}}, \bibinfo {author}
  {\bibfnamefont {K.}~\bibnamefont {Jansen}}, \bibinfo {author} {\bibfnamefont
  {G.}~\bibnamefont {Koutsou}}, \ and\ \bibinfo {author} {\bibfnamefont
  {A.}~\bibnamefont {Vaquero Aviles-Casco}},\ }\href {\doibase
  10.1103/PhysRevD.100.014509} {\bibfield  {journal} {\bibinfo  {journal}
  {Phys. Rev. D}\ }\textbf {\bibinfo {volume} {100}},\ \bibinfo {pages}
  {014509} (\bibinfo {year} {2019})},\ \Eprint
  {http://arxiv.org/abs/1812.10311} {arXiv:1812.10311 [hep-lat]} \BibitemShut
  {NoStop}%
\bibitem [{\citenamefont {Jang}\ \emph {et~al.}(2020)\citenamefont {Jang},
  \citenamefont {Gupta}, \citenamefont {Lin}, \citenamefont {Yoon},\ and\
  \citenamefont {Bhattacharya}}]{Jang:2019jkn}%
  \BibitemOpen
  \bibfield  {author} {\bibinfo {author} {\bibfnamefont {Y.-C.}\ \bibnamefont
  {Jang}}, \bibinfo {author} {\bibfnamefont {R.}~\bibnamefont {Gupta}},
  \bibinfo {author} {\bibfnamefont {H.-W.}\ \bibnamefont {Lin}}, \bibinfo
  {author} {\bibfnamefont {B.}~\bibnamefont {Yoon}}, \ and\ \bibinfo {author}
  {\bibfnamefont {T.}~\bibnamefont {Bhattacharya}},\ }\href {\doibase
  10.1103/PhysRevD.101.014507} {\bibfield  {journal} {\bibinfo  {journal}
  {Phys. Rev. D}\ }\textbf {\bibinfo {volume} {101}},\ \bibinfo {pages}
  {014507} (\bibinfo {year} {2020})},\ \Eprint
  {http://arxiv.org/abs/1906.07217} {arXiv:1906.07217 [hep-lat]} \BibitemShut
  {NoStop}%
\bibitem [{\citenamefont {Alexandrou}\ \emph {et~al.}(2020)\citenamefont
  {Alexandrou}, \citenamefont {Hadjiyiannakou}, \citenamefont {Koutsou},
  \citenamefont {Ottnad},\ and\ \citenamefont
  {Petschlies}}]{Alexandrou:2020aja}%
  \BibitemOpen
  \bibfield  {author} {\bibinfo {author} {\bibfnamefont {C.}~\bibnamefont
  {Alexandrou}}, \bibinfo {author} {\bibfnamefont {K.}~\bibnamefont
  {Hadjiyiannakou}}, \bibinfo {author} {\bibfnamefont {G.}~\bibnamefont
  {Koutsou}}, \bibinfo {author} {\bibfnamefont {K.}~\bibnamefont {Ottnad}}, \
  and\ \bibinfo {author} {\bibfnamefont {M.}~\bibnamefont {Petschlies}},\
  }\href {\doibase 10.1103/PhysRevD.101.114504} {\bibfield  {journal} {\bibinfo
   {journal} {Phys. Rev. D}\ }\textbf {\bibinfo {volume} {101}},\ \bibinfo
  {pages} {114504} (\bibinfo {year} {2020})},\ \Eprint
  {http://arxiv.org/abs/2002.06984} {arXiv:2002.06984 [hep-lat]} \BibitemShut
  {NoStop}%
\bibitem [{\citenamefont {Wang}\ \emph {et~al.}(2021)\citenamefont {Wang},
  \citenamefont {Liang}, \citenamefont {Draper}, \citenamefont {Liu},\ and\
  \citenamefont {Yang}}]{Wang:2020nbf}%
  \BibitemOpen
  \bibfield  {author} {\bibinfo {author} {\bibfnamefont {G.}~\bibnamefont
  {Wang}}, \bibinfo {author} {\bibfnamefont {J.}~\bibnamefont {Liang}},
  \bibinfo {author} {\bibfnamefont {T.}~\bibnamefont {Draper}}, \bibinfo
  {author} {\bibfnamefont {K.-F.}\ \bibnamefont {Liu}}, \ and\ \bibinfo
  {author} {\bibfnamefont {Y.-B.}\ \bibnamefont {Yang}} (\bibinfo
  {collaboration} {chiQCD}),\ }\href {\doibase 10.1103/PhysRevD.104.074502}
  {\bibfield  {journal} {\bibinfo  {journal} {Phys. Rev. D}\ }\textbf {\bibinfo
  {volume} {104}},\ \bibinfo {pages} {074502} (\bibinfo {year} {2021})},\
  \Eprint {http://arxiv.org/abs/2006.05431} {arXiv:2006.05431 [hep-ph]}
  \BibitemShut {NoStop}%
\bibitem [{\citenamefont {Park}\ \emph {et~al.}(2022)\citenamefont {Park},
  \citenamefont {Gupta}, \citenamefont {Yoon}, \citenamefont {Mondal},
  \citenamefont {Bhattacharya}, \citenamefont {Jang}, \citenamefont {Jo\'o},\
  and\ \citenamefont {Winter}}]{Park:2021ypf}%
  \BibitemOpen
  \bibfield  {author} {\bibinfo {author} {\bibfnamefont {S.}~\bibnamefont
  {Park}}, \bibinfo {author} {\bibfnamefont {R.}~\bibnamefont {Gupta}},
  \bibinfo {author} {\bibfnamefont {B.}~\bibnamefont {Yoon}}, \bibinfo {author}
  {\bibfnamefont {S.}~\bibnamefont {Mondal}}, \bibinfo {author} {\bibfnamefont
  {T.}~\bibnamefont {Bhattacharya}}, \bibinfo {author} {\bibfnamefont {Y.-C.}\
  \bibnamefont {Jang}}, \bibinfo {author} {\bibfnamefont {B.}~\bibnamefont
  {Jo\'o}}, \ and\ \bibinfo {author} {\bibfnamefont {F.}~\bibnamefont {Winter}}
  (\bibinfo {collaboration} {Nucleon Matrix Elements (NME)}),\ }\href {\doibase
  10.1103/PhysRevD.105.054505} {\bibfield  {journal} {\bibinfo  {journal}
  {Phys. Rev. D}\ }\textbf {\bibinfo {volume} {105}},\ \bibinfo {pages}
  {054505} (\bibinfo {year} {2022})},\ \Eprint
  {http://arxiv.org/abs/2103.05599} {arXiv:2103.05599 [hep-lat]} \BibitemShut
  {NoStop}%
\bibitem [{\citenamefont {Ishikawa}\ \emph {et~al.}(2021)\citenamefont
  {Ishikawa}, \citenamefont {Kuramashi}, \citenamefont {Sasaki}, \citenamefont
  {Shintani},\ and\ \citenamefont {Yamazaki}}]{Ishikawa:2021eut}%
  \BibitemOpen
  \bibfield  {author} {\bibinfo {author} {\bibfnamefont {K.-I.}\ \bibnamefont
  {Ishikawa}}, \bibinfo {author} {\bibfnamefont {Y.}~\bibnamefont {Kuramashi}},
  \bibinfo {author} {\bibfnamefont {S.}~\bibnamefont {Sasaki}}, \bibinfo
  {author} {\bibfnamefont {E.}~\bibnamefont {Shintani}}, \ and\ \bibinfo
  {author} {\bibfnamefont {T.}~\bibnamefont {Yamazaki}} (\bibinfo
  {collaboration} {PACS}),\ }\href {\doibase 10.1103/PhysRevD.104.074514}
  {\bibfield  {journal} {\bibinfo  {journal} {Phys. Rev. D}\ }\textbf {\bibinfo
  {volume} {104}},\ \bibinfo {pages} {074514} (\bibinfo {year} {2021})},\
  \Eprint {http://arxiv.org/abs/2107.07085} {arXiv:2107.07085 [hep-lat]}
  \BibitemShut {NoStop}%
\bibitem [{\citenamefont {Bar}\ and\ \citenamefont
  {Colic}(2021)}]{Bar:2021crj}%
  \BibitemOpen
  \bibfield  {author} {\bibinfo {author} {\bibfnamefont {O.}~\bibnamefont
  {Bar}}\ and\ \bibinfo {author} {\bibfnamefont {H.}~\bibnamefont {Colic}},\
  }\href {\doibase 10.1103/PhysRevD.103.114514} {\bibfield  {journal} {\bibinfo
   {journal} {Phys. Rev. D}\ }\textbf {\bibinfo {volume} {103}},\ \bibinfo
  {pages} {114514} (\bibinfo {year} {2021})},\ \Eprint
  {http://arxiv.org/abs/2104.00329} {arXiv:2104.00329 [hep-lat]} \BibitemShut
  {NoStop}%
\bibitem [{\citenamefont {Djukanovic}\ \emph {et~al.}(2021)\citenamefont
  {Djukanovic}, \citenamefont {Harris}, \citenamefont {von Hippel},
  \citenamefont {Junnarkar}, \citenamefont {Meyer}, \citenamefont {Mohler},
  \citenamefont {Ottnad}, \citenamefont {Schulz}, \citenamefont {Wilhelm},\
  and\ \citenamefont {Wittig}}]{Djukanovic:2021cgp}%
  \BibitemOpen
  \bibfield  {author} {\bibinfo {author} {\bibfnamefont {D.}~\bibnamefont
  {Djukanovic}}, \bibinfo {author} {\bibfnamefont {T.}~\bibnamefont {Harris}},
  \bibinfo {author} {\bibfnamefont {G.}~\bibnamefont {von Hippel}}, \bibinfo
  {author} {\bibfnamefont {P.~M.}\ \bibnamefont {Junnarkar}}, \bibinfo {author}
  {\bibfnamefont {H.~B.}\ \bibnamefont {Meyer}}, \bibinfo {author}
  {\bibfnamefont {D.}~\bibnamefont {Mohler}}, \bibinfo {author} {\bibfnamefont
  {K.}~\bibnamefont {Ottnad}}, \bibinfo {author} {\bibfnamefont
  {T.}~\bibnamefont {Schulz}}, \bibinfo {author} {\bibfnamefont
  {J.}~\bibnamefont {Wilhelm}}, \ and\ \bibinfo {author} {\bibfnamefont
  {H.}~\bibnamefont {Wittig}},\ }\href {\doibase 10.1103/PhysRevD.103.094522}
  {\bibfield  {journal} {\bibinfo  {journal} {Phys. Rev. D}\ }\textbf {\bibinfo
  {volume} {103}},\ \bibinfo {pages} {094522} (\bibinfo {year} {2021})},\
  \Eprint {http://arxiv.org/abs/2102.07460} {arXiv:2102.07460 [hep-lat]}
  \BibitemShut {NoStop}%
\bibitem [{\citenamefont {Djukanovic}(2022)}]{Djukanovic:2021qxp}%
  \BibitemOpen
  \bibfield  {author} {\bibinfo {author} {\bibfnamefont {D.}~\bibnamefont
  {Djukanovic}},\ }\href {\doibase 10.22323/1.396.0009} {\bibfield  {journal}
  {\bibinfo  {journal} {PoS}\ }\textbf {\bibinfo {volume} {LATTICE2021}},\
  \bibinfo {pages} {009} (\bibinfo {year} {2022})},\ \Eprint
  {http://arxiv.org/abs/2112.00128} {arXiv:2112.00128 [hep-lat]} \BibitemShut
  {NoStop}%
\bibitem [{\citenamefont {Alexandrou}\ \emph {et~al.}(2022)\citenamefont
  {Alexandrou}, \citenamefont {Bacchio}, \citenamefont {Cloet}, \citenamefont
  {Constantinou}, \citenamefont {Delmar}, \citenamefont {Hadjiyiannakou},
  \citenamefont {Koutsou}, \citenamefont {Lauer},\ and\ \citenamefont
  {Vaquero}}]{Alexandrou:2021ztx}%
  \BibitemOpen
  \bibfield  {author} {\bibinfo {author} {\bibfnamefont {C.}~\bibnamefont
  {Alexandrou}}, \bibinfo {author} {\bibfnamefont {S.}~\bibnamefont {Bacchio}},
  \bibinfo {author} {\bibfnamefont {I.}~\bibnamefont {Cloet}}, \bibinfo
  {author} {\bibfnamefont {M.}~\bibnamefont {Constantinou}}, \bibinfo {author}
  {\bibfnamefont {J.}~\bibnamefont {Delmar}}, \bibinfo {author} {\bibfnamefont
  {K.}~\bibnamefont {Hadjiyiannakou}}, \bibinfo {author} {\bibfnamefont
  {G.}~\bibnamefont {Koutsou}}, \bibinfo {author} {\bibfnamefont
  {C.}~\bibnamefont {Lauer}}, \ and\ \bibinfo {author} {\bibfnamefont
  {A.}~\bibnamefont {Vaquero}} (\bibinfo {collaboration} {ETM}),\ }\href
  {\doibase 10.1103/PhysRevD.105.054502} {\bibfield  {journal} {\bibinfo
  {journal} {Phys. Rev. D}\ }\textbf {\bibinfo {volume} {105}},\ \bibinfo
  {pages} {054502} (\bibinfo {year} {2022})},\ \Eprint
  {http://arxiv.org/abs/2111.08135} {arXiv:2111.08135 [hep-lat]} \BibitemShut
  {NoStop}%
\bibitem [{\citenamefont {Arrington}\ \emph
  {et~al.}(2007{\natexlab{b}})\citenamefont {Arrington}, \citenamefont
  {Roberts},\ and\ \citenamefont {Zanotti}}]{Arrington:2006zm}%
  \BibitemOpen
  \bibfield  {author} {\bibinfo {author} {\bibfnamefont {J.}~\bibnamefont
  {Arrington}}, \bibinfo {author} {\bibfnamefont {C.~D.}\ \bibnamefont
  {Roberts}}, \ and\ \bibinfo {author} {\bibfnamefont {J.~M.}\ \bibnamefont
  {Zanotti}},\ }\href {\doibase 10.1088/0954-3899/34/7/S03} {\bibfield
  {journal} {\bibinfo  {journal} {J. Phys. G}\ }\textbf {\bibinfo {volume}
  {34}},\ \bibinfo {pages} {S23} (\bibinfo {year} {2007}{\natexlab{b}})},\
  \Eprint {http://arxiv.org/abs/nucl-th/0611050} {arXiv:nucl-th/0611050}
  \BibitemShut {NoStop}%
\bibitem [{\citenamefont {Perdrisat}\ \emph {et~al.}(2007)\citenamefont
  {Perdrisat}, \citenamefont {Punjabi},\ and\ \citenamefont
  {Vanderhaeghen}}]{Perdrisat:2006hj}%
  \BibitemOpen
  \bibfield  {author} {\bibinfo {author} {\bibfnamefont {C.~F.}\ \bibnamefont
  {Perdrisat}}, \bibinfo {author} {\bibfnamefont {V.}~\bibnamefont {Punjabi}},
  \ and\ \bibinfo {author} {\bibfnamefont {M.}~\bibnamefont {Vanderhaeghen}},\
  }\href {\doibase 10.1016/j.ppnp.2007.05.001} {\bibfield  {journal} {\bibinfo
  {journal} {Prog. Part. Nucl. Phys.}\ }\textbf {\bibinfo {volume} {59}},\
  \bibinfo {pages} {694} (\bibinfo {year} {2007})},\ \Eprint
  {http://arxiv.org/abs/hep-ph/0612014} {arXiv:hep-ph/0612014} \BibitemShut
  {NoStop}%
\bibitem [{\citenamefont {Pacetti}\ \emph {et~al.}(2015)\citenamefont
  {Pacetti}, \citenamefont {Baldini~Ferroli},\ and\ \citenamefont
  {Tomasi-Gustafsson}}]{Pacetti:2014jai}%
  \BibitemOpen
  \bibfield  {author} {\bibinfo {author} {\bibfnamefont {S.}~\bibnamefont
  {Pacetti}}, \bibinfo {author} {\bibfnamefont {R.}~\bibnamefont
  {Baldini~Ferroli}}, \ and\ \bibinfo {author} {\bibfnamefont {E.}~\bibnamefont
  {Tomasi-Gustafsson}},\ }\href {\doibase 10.1016/j.physrep.2014.09.005}
  {\bibfield  {journal} {\bibinfo  {journal} {Phys. Rept.}\ }\textbf {\bibinfo
  {volume} {550-551}},\ \bibinfo {pages} {1} (\bibinfo {year}
  {2015})}\BibitemShut {NoStop}%
\bibitem [{\citenamefont {Ernst}\ \emph {et~al.}(1960)\citenamefont {Ernst},
  \citenamefont {Sachs},\ and\ \citenamefont {Wali}}]{Ernst:1960zza}%
  \BibitemOpen
  \bibfield  {author} {\bibinfo {author} {\bibfnamefont {F.~J.}\ \bibnamefont
  {Ernst}}, \bibinfo {author} {\bibfnamefont {R.~G.}\ \bibnamefont {Sachs}}, \
  and\ \bibinfo {author} {\bibfnamefont {K.~C.}\ \bibnamefont {Wali}},\ }\href
  {\doibase 10.1103/PhysRev.119.1105} {\bibfield  {journal} {\bibinfo
  {journal} {Phys. Rev.}\ }\textbf {\bibinfo {volume} {119}},\ \bibinfo {pages}
  {1105} (\bibinfo {year} {1960})}\BibitemShut {NoStop}%
\bibitem [{\citenamefont {Sachs}(1962)}]{Sachs:1962zzc}%
  \BibitemOpen
  \bibfield  {author} {\bibinfo {author} {\bibfnamefont {R.~G.}\ \bibnamefont
  {Sachs}},\ }\href {\doibase 10.1103/PhysRev.126.2256} {\bibfield  {journal}
  {\bibinfo  {journal} {Phys. Rev.}\ }\textbf {\bibinfo {volume} {126}},\
  \bibinfo {pages} {2256} (\bibinfo {year} {1962})}\BibitemShut {NoStop}%
\bibitem [{\citenamefont {Yennie}\ \emph {et~al.}(1957)\citenamefont {Yennie},
  \citenamefont {L\'evy},\ and\ \citenamefont {Ravenhall}}]{Yennie:1957rmp}%
  \BibitemOpen
  \bibfield  {author} {\bibinfo {author} {\bibfnamefont {D.~R.}\ \bibnamefont
  {Yennie}}, \bibinfo {author} {\bibfnamefont {M.~M.}\ \bibnamefont {L\'evy}},
  \ and\ \bibinfo {author} {\bibfnamefont {D.~G.}\ \bibnamefont {Ravenhall}},\
  }\href {\doibase 10.1103/RevModPhys.29.144} {\bibfield  {journal} {\bibinfo
  {journal} {Rev. Mod. Phys.}\ }\textbf {\bibinfo {volume} {29}},\ \bibinfo
  {pages} {144} (\bibinfo {year} {1957})}\BibitemShut {NoStop}%
\bibitem [{\citenamefont {Breit}(1966)}]{Breit:1964ga}%
  \BibitemOpen
  \bibfield  {author} {\bibinfo {author} {\bibfnamefont {G.}~\bibnamefont
  {Breit}},\ }in\ \href {https://inspirehep.net/literature/1670085} {\emph
  {\bibinfo {booktitle} {Proceedings of the XII International Conference on
  High Energy Physics (ICHEP 1964)}}}\ (\bibinfo  {publisher} {Atomizdat},\
  \bibinfo {address} {Moscow},\ \bibinfo {year} {1966})\ pp.\ \bibinfo {pages}
  {985--987}\BibitemShut {NoStop}%
\bibitem [{\citenamefont {Kelly}(2002)}]{Kelly:2002if}%
  \BibitemOpen
  \bibfield  {author} {\bibinfo {author} {\bibfnamefont {J.~J.}\ \bibnamefont
  {Kelly}},\ }\href {\doibase 10.1103/PhysRevC.66.065203} {\bibfield  {journal}
  {\bibinfo  {journal} {Phys. Rev. C}\ }\textbf {\bibinfo {volume} {66}},\
  \bibinfo {pages} {065203} (\bibinfo {year} {2002})},\ \Eprint
  {http://arxiv.org/abs/hep-ph/0204239} {arXiv:hep-ph/0204239} \BibitemShut
  {NoStop}%
\bibitem [{\citenamefont {Burkardt}(2000)}]{Burkardt:2000za}%
  \BibitemOpen
  \bibfield  {author} {\bibinfo {author} {\bibfnamefont {M.}~\bibnamefont
  {Burkardt}},\ }\href {\doibase 10.1103/PhysRevD.62.071503} {\bibfield
  {journal} {\bibinfo  {journal} {Phys. Rev. D}\ }\textbf {\bibinfo {volume}
  {62}},\ \bibinfo {pages} {071503} (\bibinfo {year} {2000})},\ \bibinfo {note}
  {[Erratum: Phys.Rev.D 66, 119903 (2002)]},\ \Eprint
  {http://arxiv.org/abs/hep-ph/0005108} {arXiv:hep-ph/0005108} \BibitemShut
  {NoStop}%
\bibitem [{\citenamefont {Belitsky}\ \emph {et~al.}(2004)\citenamefont
  {Belitsky}, \citenamefont {Ji},\ and\ \citenamefont
  {Yuan}}]{Belitsky:2003nz}%
  \BibitemOpen
  \bibfield  {author} {\bibinfo {author} {\bibfnamefont {A.~V.}\ \bibnamefont
  {Belitsky}}, \bibinfo {author} {\bibfnamefont {X.-d.}\ \bibnamefont {Ji}}, \
  and\ \bibinfo {author} {\bibfnamefont {F.}~\bibnamefont {Yuan}},\ }\href
  {\doibase 10.1103/PhysRevD.69.074014} {\bibfield  {journal} {\bibinfo
  {journal} {Phys. Rev. D}\ }\textbf {\bibinfo {volume} {69}},\ \bibinfo
  {pages} {074014} (\bibinfo {year} {2004})},\ \Eprint
  {http://arxiv.org/abs/hep-ph/0307383} {arXiv:hep-ph/0307383} \BibitemShut
  {NoStop}%
\bibitem [{\citenamefont {Jaffe}(2021)}]{Jaffe:2020ebz}%
  \BibitemOpen
  \bibfield  {author} {\bibinfo {author} {\bibfnamefont {R.~L.}\ \bibnamefont
  {Jaffe}},\ }\href {\doibase 10.1103/PhysRevD.103.016017} {\bibfield
  {journal} {\bibinfo  {journal} {Phys. Rev. D}\ }\textbf {\bibinfo {volume}
  {103}},\ \bibinfo {pages} {016017} (\bibinfo {year} {2021})},\ \Eprint
  {http://arxiv.org/abs/2010.15887} {arXiv:2010.15887 [hep-ph]} \BibitemShut
  {NoStop}%
\bibitem [{\citenamefont {Burkardt}(2003)}]{Burkardt:2002hr}%
  \BibitemOpen
  \bibfield  {author} {\bibinfo {author} {\bibfnamefont {M.}~\bibnamefont
  {Burkardt}},\ }\href {\doibase 10.1142/S0217751X03012370} {\bibfield
  {journal} {\bibinfo  {journal} {Int. J. Mod. Phys. A}\ }\textbf {\bibinfo
  {volume} {18}},\ \bibinfo {pages} {173} (\bibinfo {year} {2003})},\ \Eprint
  {http://arxiv.org/abs/hep-ph/0207047} {arXiv:hep-ph/0207047} \BibitemShut
  {NoStop}%
\bibitem [{\citenamefont {Miller}(2007)}]{Miller:2007uy}%
  \BibitemOpen
  \bibfield  {author} {\bibinfo {author} {\bibfnamefont {G.~A.}\ \bibnamefont
  {Miller}},\ }\href {\doibase 10.1103/PhysRevLett.99.112001} {\bibfield
  {journal} {\bibinfo  {journal} {Phys. Rev. Lett.}\ }\textbf {\bibinfo
  {volume} {99}},\ \bibinfo {pages} {112001} (\bibinfo {year} {2007})},\
  \Eprint {http://arxiv.org/abs/0705.2409} {arXiv:0705.2409 [nucl-th]}
  \BibitemShut {NoStop}%
\bibitem [{\citenamefont {Carlson}\ and\ \citenamefont
  {Vanderhaeghen}(2008)}]{Carlson:2007xd}%
  \BibitemOpen
  \bibfield  {author} {\bibinfo {author} {\bibfnamefont {C.~E.}\ \bibnamefont
  {Carlson}}\ and\ \bibinfo {author} {\bibfnamefont {M.}~\bibnamefont
  {Vanderhaeghen}},\ }\href {\doibase 10.1103/PhysRevLett.100.032004}
  {\bibfield  {journal} {\bibinfo  {journal} {Phys. Rev. Lett.}\ }\textbf
  {\bibinfo {volume} {100}},\ \bibinfo {pages} {032004} (\bibinfo {year}
  {2008})},\ \Eprint {http://arxiv.org/abs/0710.0835} {arXiv:0710.0835
  [hep-ph]} \BibitemShut {NoStop}%
\bibitem [{\citenamefont {Alexandrou}\ \emph
  {et~al.}(2009{\natexlab{a}})\citenamefont {Alexandrou}, \citenamefont
  {Korzec}, \citenamefont {Koutsou}, \citenamefont {Leontiou}, \citenamefont
  {Lorc\'e}, \citenamefont {Negele}, \citenamefont {Pascalutsa}, \citenamefont
  {Tsapalis},\ and\ \citenamefont {Vanderhaeghen}}]{Alexandrou:2008bn}%
  \BibitemOpen
  \bibfield  {author} {\bibinfo {author} {\bibfnamefont {C.}~\bibnamefont
  {Alexandrou}}, \bibinfo {author} {\bibfnamefont {T.}~\bibnamefont {Korzec}},
  \bibinfo {author} {\bibfnamefont {G.}~\bibnamefont {Koutsou}}, \bibinfo
  {author} {\bibfnamefont {T.}~\bibnamefont {Leontiou}}, \bibinfo {author}
  {\bibfnamefont {C.}~\bibnamefont {Lorc\'e}}, \bibinfo {author} {\bibfnamefont
  {J.~W.}\ \bibnamefont {Negele}}, \bibinfo {author} {\bibfnamefont
  {V.}~\bibnamefont {Pascalutsa}}, \bibinfo {author} {\bibfnamefont
  {A.}~\bibnamefont {Tsapalis}}, \ and\ \bibinfo {author} {\bibfnamefont
  {M.}~\bibnamefont {Vanderhaeghen}},\ }\href {\doibase
  10.1103/PhysRevD.79.014507} {\bibfield  {journal} {\bibinfo  {journal} {Phys.
  Rev. D}\ }\textbf {\bibinfo {volume} {79}},\ \bibinfo {pages} {014507}
  (\bibinfo {year} {2009}{\natexlab{a}})},\ \Eprint
  {http://arxiv.org/abs/0810.3976} {arXiv:0810.3976 [hep-lat]} \BibitemShut
  {NoStop}%
\bibitem [{\citenamefont {Alexandrou}\ \emph
  {et~al.}(2009{\natexlab{b}})\citenamefont {Alexandrou}, \citenamefont
  {Korzec}, \citenamefont {Koutsou}, \citenamefont {Lorc\'e}, \citenamefont
  {Negele}, \citenamefont {Pascalutsa}, \citenamefont {Tsapalis},\ and\
  \citenamefont {Vanderhaeghen}}]{Alexandrou:2009hs}%
  \BibitemOpen
  \bibfield  {author} {\bibinfo {author} {\bibfnamefont {C.}~\bibnamefont
  {Alexandrou}}, \bibinfo {author} {\bibfnamefont {T.}~\bibnamefont {Korzec}},
  \bibinfo {author} {\bibfnamefont {G.}~\bibnamefont {Koutsou}}, \bibinfo
  {author} {\bibfnamefont {C.}~\bibnamefont {Lorc\'e}}, \bibinfo {author}
  {\bibfnamefont {J.~W.}\ \bibnamefont {Negele}}, \bibinfo {author}
  {\bibfnamefont {V.}~\bibnamefont {Pascalutsa}}, \bibinfo {author}
  {\bibfnamefont {A.}~\bibnamefont {Tsapalis}}, \ and\ \bibinfo {author}
  {\bibfnamefont {M.}~\bibnamefont {Vanderhaeghen}},\ }\href {\doibase
  10.1016/j.nuclphysa.2009.04.005} {\bibfield  {journal} {\bibinfo  {journal}
  {Nucl. Phys. A}\ }\textbf {\bibinfo {volume} {825}},\ \bibinfo {pages} {115}
  (\bibinfo {year} {2009}{\natexlab{b}})},\ \Eprint
  {http://arxiv.org/abs/0901.3457} {arXiv:0901.3457 [hep-ph]} \BibitemShut
  {NoStop}%
\bibitem [{\citenamefont {Gorchtein}\ \emph {et~al.}(2010)\citenamefont
  {Gorchtein}, \citenamefont {Lorc\'e}, \citenamefont {Pasquini},\ and\
  \citenamefont {Vanderhaeghen}}]{Gorchtein:2009qq}%
  \BibitemOpen
  \bibfield  {author} {\bibinfo {author} {\bibfnamefont {M.}~\bibnamefont
  {Gorchtein}}, \bibinfo {author} {\bibfnamefont {C.}~\bibnamefont {Lorc\'e}},
  \bibinfo {author} {\bibfnamefont {B.}~\bibnamefont {Pasquini}}, \ and\
  \bibinfo {author} {\bibfnamefont {M.}~\bibnamefont {Vanderhaeghen}},\ }\href
  {\doibase 10.1103/PhysRevLett.104.112001} {\bibfield  {journal} {\bibinfo
  {journal} {Phys. Rev. Lett.}\ }\textbf {\bibinfo {volume} {104}},\ \bibinfo
  {pages} {112001} (\bibinfo {year} {2010})},\ \Eprint
  {http://arxiv.org/abs/0911.2882} {arXiv:0911.2882 [hep-ph]} \BibitemShut
  {NoStop}%
\bibitem [{\citenamefont {Carlson}\ and\ \citenamefont
  {Vanderhaeghen}(2009)}]{Carlson:2009ovh}%
  \BibitemOpen
  \bibfield  {author} {\bibinfo {author} {\bibfnamefont {C.~E.}\ \bibnamefont
  {Carlson}}\ and\ \bibinfo {author} {\bibfnamefont {M.}~\bibnamefont
  {Vanderhaeghen}},\ }\href {\doibase 10.1140/epja/i2009-10800-0} {\bibfield
  {journal} {\bibinfo  {journal} {Eur. Phys. J. A}\ }\textbf {\bibinfo {volume}
  {41}},\ \bibinfo {pages} {1} (\bibinfo {year} {2009})},\ \Eprint
  {http://arxiv.org/abs/0807.4537} {arXiv:0807.4537 [hep-ph]} \BibitemShut
  {NoStop}%
\bibitem [{\citenamefont {Miller}(2010)}]{Miller:2010nz}%
  \BibitemOpen
  \bibfield  {author} {\bibinfo {author} {\bibfnamefont {G.~A.}\ \bibnamefont
  {Miller}},\ }\href {\doibase 10.1146/annurev.nucl.012809.104508} {\bibfield
  {journal} {\bibinfo  {journal} {Ann. Rev. Nucl. Part. Sci.}\ }\textbf
  {\bibinfo {volume} {60}},\ \bibinfo {pages} {1} (\bibinfo {year} {2010})},\
  \Eprint {http://arxiv.org/abs/1002.0355} {arXiv:1002.0355 [nucl-th]}
  \BibitemShut {NoStop}%
\bibitem [{\citenamefont {Miller}(2019)}]{Miller:2018ybm}%
  \BibitemOpen
  \bibfield  {author} {\bibinfo {author} {\bibfnamefont {G.~A.}\ \bibnamefont
  {Miller}},\ }\href {\doibase 10.1103/PhysRevC.99.035202} {\bibfield
  {journal} {\bibinfo  {journal} {Phys. Rev. C}\ }\textbf {\bibinfo {volume}
  {99}},\ \bibinfo {pages} {035202} (\bibinfo {year} {2019})},\ \Eprint
  {http://arxiv.org/abs/1812.02714} {arXiv:1812.02714 [nucl-th]} \BibitemShut
  {NoStop}%
\bibitem [{\citenamefont {Panteleeva}\ and\ \citenamefont
  {Polyakov}(2021)}]{Panteleeva:2021iip}%
  \BibitemOpen
  \bibfield  {author} {\bibinfo {author} {\bibfnamefont {J.~Y.}\ \bibnamefont
  {Panteleeva}}\ and\ \bibinfo {author} {\bibfnamefont {M.~V.}\ \bibnamefont
  {Polyakov}},\ }\href {\doibase 10.1103/PhysRevD.104.014008} {\bibfield
  {journal} {\bibinfo  {journal} {Phys. Rev. D}\ }\textbf {\bibinfo {volume}
  {104}},\ \bibinfo {pages} {014008} (\bibinfo {year} {2021})},\ \Eprint
  {http://arxiv.org/abs/2102.10902} {arXiv:2102.10902 [hep-ph]} \BibitemShut
  {NoStop}%
\bibitem [{\citenamefont {Freese}\ and\ \citenamefont
  {Miller}(2022{\natexlab{a}})}]{Freese:2021mzg}%
  \BibitemOpen
  \bibfield  {author} {\bibinfo {author} {\bibfnamefont {A.}~\bibnamefont
  {Freese}}\ and\ \bibinfo {author} {\bibfnamefont {G.~A.}\ \bibnamefont
  {Miller}},\ }\href {\doibase 10.1103/PhysRevD.105.014003} {\bibfield
  {journal} {\bibinfo  {journal} {Phys. Rev. D}\ }\textbf {\bibinfo {volume}
  {105}},\ \bibinfo {pages} {014003} (\bibinfo {year} {2022}{\natexlab{a}})},\
  \Eprint {http://arxiv.org/abs/2108.03301} {arXiv:2108.03301 [hep-ph]}
  \BibitemShut {NoStop}%
\bibitem [{\citenamefont {Kim}\ and\ \citenamefont {Kim}(2021)}]{Kim:2021kum}%
  \BibitemOpen
  \bibfield  {author} {\bibinfo {author} {\bibfnamefont {J.-Y.}\ \bibnamefont
  {Kim}}\ and\ \bibinfo {author} {\bibfnamefont {H.-C.}\ \bibnamefont {Kim}},\
  }\href {\doibase 10.1103/PhysRevD.104.074003} {\bibfield  {journal} {\bibinfo
   {journal} {Phys. Rev. D}\ }\textbf {\bibinfo {volume} {104}},\ \bibinfo
  {pages} {074003} (\bibinfo {year} {2021})},\ \Eprint
  {http://arxiv.org/abs/2106.10986} {arXiv:2106.10986 [hep-ph]} \BibitemShut
  {NoStop}%
\bibitem [{\citenamefont {Kim}(2022)}]{Kim:2022bia}%
  \BibitemOpen
  \bibfield  {author} {\bibinfo {author} {\bibfnamefont {J.-Y.}\ \bibnamefont
  {Kim}},\ }\href {\doibase 10.1103/PhysRevD.106.014022} {\bibfield  {journal}
  {\bibinfo  {journal} {Phys. Rev. D}\ }\textbf {\bibinfo {volume} {106}},\
  \bibinfo {pages} {014022} (\bibinfo {year} {2022})},\ \Eprint
  {http://arxiv.org/abs/2204.08248} {arXiv:2204.08248 [hep-ph]} \BibitemShut
  {NoStop}%
\bibitem [{\citenamefont {Epelbaum}\ \emph {et~al.}(2022)\citenamefont
  {Epelbaum}, \citenamefont {Gegelia}, \citenamefont {Lange}, \citenamefont
  {Mei\ss{}ner},\ and\ \citenamefont {Polyakov}}]{Epelbaum:2022fjc}%
  \BibitemOpen
  \bibfield  {author} {\bibinfo {author} {\bibfnamefont {E.}~\bibnamefont
  {Epelbaum}}, \bibinfo {author} {\bibfnamefont {J.}~\bibnamefont {Gegelia}},
  \bibinfo {author} {\bibfnamefont {N.}~\bibnamefont {Lange}}, \bibinfo
  {author} {\bibfnamefont {U.~G.}\ \bibnamefont {Mei\ss{}ner}}, \ and\ \bibinfo
  {author} {\bibfnamefont {M.~V.}\ \bibnamefont {Polyakov}},\ }\href {\doibase
  10.1103/PhysRevLett.129.012001} {\bibfield  {journal} {\bibinfo  {journal}
  {Phys. Rev. Lett.}\ }\textbf {\bibinfo {volume} {129}},\ \bibinfo {pages}
  {012001} (\bibinfo {year} {2022})},\ \Eprint
  {http://arxiv.org/abs/2201.02565} {arXiv:2201.02565 [hep-ph]} \BibitemShut
  {NoStop}%
\bibitem [{\citenamefont {Panteleeva}\ \emph {et~al.}(2022)\citenamefont
  {Panteleeva}, \citenamefont {Epelbaum}, \citenamefont {Gegelia},\ and\
  \citenamefont {Mei\ss{}ner}}]{Panteleeva:2022khw}%
  \BibitemOpen
  \bibfield  {author} {\bibinfo {author} {\bibfnamefont {J.~Y.}\ \bibnamefont
  {Panteleeva}}, \bibinfo {author} {\bibfnamefont {E.}~\bibnamefont
  {Epelbaum}}, \bibinfo {author} {\bibfnamefont {J.}~\bibnamefont {Gegelia}}, \
  and\ \bibinfo {author} {\bibfnamefont {U.~G.}\ \bibnamefont {Mei\ss{}ner}},\
  }\href {\doibase 10.1103/PhysRevD.106.056019} {\bibfield  {journal} {\bibinfo
   {journal} {Phys. Rev. D}\ }\textbf {\bibinfo {volume} {106}},\ \bibinfo
  {pages} {056019} (\bibinfo {year} {2022})},\ \Eprint
  {http://arxiv.org/abs/2205.15061} {arXiv:2205.15061 [hep-ph]} \BibitemShut
  {NoStop}%
\bibitem [{\citenamefont {Li}\ \emph {et~al.}(2022)\citenamefont {Li},
  \citenamefont {Dong}, \citenamefont {Yin}, \citenamefont {Wang},\ and\
  \citenamefont {Vary}}]{Li:2022ldb}%
  \BibitemOpen
  \bibfield  {author} {\bibinfo {author} {\bibfnamefont {Y.}~\bibnamefont
  {Li}}, \bibinfo {author} {\bibfnamefont {W.-b.}\ \bibnamefont {Dong}},
  \bibinfo {author} {\bibfnamefont {Y.-l.}\ \bibnamefont {Yin}}, \bibinfo
  {author} {\bibfnamefont {Q.}~\bibnamefont {Wang}}, \ and\ \bibinfo {author}
  {\bibfnamefont {J.~P.}\ \bibnamefont {Vary}},\ }\href@noop {} {\  (\bibinfo
  {year} {2022})},\ \Eprint {http://arxiv.org/abs/2206.12903} {arXiv:2206.12903
  [hep-ph]} \BibitemShut {NoStop}%
\bibitem [{\citenamefont {Carlson}(2022)}]{Carlson:2022eps}%
  \BibitemOpen
  \bibfield  {author} {\bibinfo {author} {\bibfnamefont {C.~E.}\ \bibnamefont
  {Carlson}},\ }\href@noop {} {\  (\bibinfo {year} {2022})},\ \Eprint
  {http://arxiv.org/abs/2208.00826} {arXiv:2208.00826 [hep-ph]} \BibitemShut
  {NoStop}%
\bibitem [{\citenamefont {Freese}\ and\ \citenamefont
  {Miller}(2022{\natexlab{b}})}]{Freese:2022fat}%
  \BibitemOpen
  \bibfield  {author} {\bibinfo {author} {\bibfnamefont {A.}~\bibnamefont
  {Freese}}\ and\ \bibinfo {author} {\bibfnamefont {G.~A.}\ \bibnamefont
  {Miller}},\ }\href@noop {} {\  (\bibinfo {year} {2022}{\natexlab{b}})},\
  \Eprint {http://arxiv.org/abs/2210.03807} {arXiv:2210.03807 [hep-ph]}
  \BibitemShut {NoStop}%
\bibitem [{\citenamefont {Wigner}(1932)}]{Wigner:1932eb}%
  \BibitemOpen
  \bibfield  {author} {\bibinfo {author} {\bibfnamefont {E.~P.}\ \bibnamefont
  {Wigner}},\ }\href {\doibase 10.1103/PhysRev.40.749} {\bibfield  {journal}
  {\bibinfo  {journal} {Phys. Rev.}\ }\textbf {\bibinfo {volume} {40}},\
  \bibinfo {pages} {749} (\bibinfo {year} {1932})}\BibitemShut {NoStop}%
\bibitem [{\citenamefont {Hillery}\ \emph {et~al.}(1984)\citenamefont
  {Hillery}, \citenamefont {O'Connell}, \citenamefont {Scully},\ and\
  \citenamefont {Wigner}}]{Hillery:1983ms}%
  \BibitemOpen
  \bibfield  {author} {\bibinfo {author} {\bibfnamefont {M.}~\bibnamefont
  {Hillery}}, \bibinfo {author} {\bibfnamefont {R.~F.}\ \bibnamefont
  {O'Connell}}, \bibinfo {author} {\bibfnamefont {M.~O.}\ \bibnamefont
  {Scully}}, \ and\ \bibinfo {author} {\bibfnamefont {E.~P.}\ \bibnamefont
  {Wigner}},\ }\href {\doibase 10.1016/0370-1573(84)90160-1} {\bibfield
  {journal} {\bibinfo  {journal} {Phys. Rept.}\ }\textbf {\bibinfo {volume}
  {106}},\ \bibinfo {pages} {121} (\bibinfo {year} {1984})}\BibitemShut
  {NoStop}%
\bibitem [{\citenamefont {Lorc\'e}\ \emph {et~al.}(2018)\citenamefont
  {Lorc\'e}, \citenamefont {Mantovani},\ and\ \citenamefont
  {Pasquini}}]{Lorce:2017wkb}%
  \BibitemOpen
  \bibfield  {author} {\bibinfo {author} {\bibfnamefont {C.}~\bibnamefont
  {Lorc\'e}}, \bibinfo {author} {\bibfnamefont {L.}~\bibnamefont {Mantovani}},
  \ and\ \bibinfo {author} {\bibfnamefont {B.}~\bibnamefont {Pasquini}},\
  }\href {\doibase 10.1016/j.physletb.2017.11.018} {\bibfield  {journal}
  {\bibinfo  {journal} {Phys. Lett. B}\ }\textbf {\bibinfo {volume} {776}},\
  \bibinfo {pages} {38} (\bibinfo {year} {2018})},\ \Eprint
  {http://arxiv.org/abs/1704.08557} {arXiv:1704.08557 [hep-ph]} \BibitemShut
  {NoStop}%
\bibitem [{\citenamefont {Lorc\'e}(2018{\natexlab{a}})}]{Lorce:2018zpf}%
  \BibitemOpen
  \bibfield  {author} {\bibinfo {author} {\bibfnamefont {C.}~\bibnamefont
  {Lorc\'e}},\ }\href {\doibase 10.1140/epjc/s10052-018-6249-3} {\bibfield
  {journal} {\bibinfo  {journal} {Eur. Phys. J. C}\ }\textbf {\bibinfo {volume}
  {78}},\ \bibinfo {pages} {785} (\bibinfo {year} {2018}{\natexlab{a}})},\
  \Eprint {http://arxiv.org/abs/1805.05284} {arXiv:1805.05284 [hep-ph]}
  \BibitemShut {NoStop}%
\bibitem [{\citenamefont {Lorc\'e}\ \emph {et~al.}(2019)\citenamefont
  {Lorc\'e}, \citenamefont {Moutarde},\ and\ \citenamefont
  {Trawi\'nski}}]{Lorce:2018egm}%
  \BibitemOpen
  \bibfield  {author} {\bibinfo {author} {\bibfnamefont {C.}~\bibnamefont
  {Lorc\'e}}, \bibinfo {author} {\bibfnamefont {H.}~\bibnamefont {Moutarde}}, \
  and\ \bibinfo {author} {\bibfnamefont {A.~P.}\ \bibnamefont {Trawi\'nski}},\
  }\href {\doibase 10.1140/epjc/s10052-019-6572-3} {\bibfield  {journal}
  {\bibinfo  {journal} {Eur. Phys. J. C}\ }\textbf {\bibinfo {volume} {79}},\
  \bibinfo {pages} {89} (\bibinfo {year} {2019})},\ \Eprint
  {http://arxiv.org/abs/1810.09837} {arXiv:1810.09837 [hep-ph]} \BibitemShut
  {NoStop}%
\bibitem [{\citenamefont {Lorc\'e}(2020)}]{Lorce:2020onh}%
  \BibitemOpen
  \bibfield  {author} {\bibinfo {author} {\bibfnamefont {C.}~\bibnamefont
  {Lorc\'e}},\ }\href {\doibase 10.1103/PhysRevLett.125.232002} {\bibfield
  {journal} {\bibinfo  {journal} {Phys. Rev. Lett.}\ }\textbf {\bibinfo
  {volume} {125}},\ \bibinfo {pages} {232002} (\bibinfo {year} {2020})},\
  \Eprint {http://arxiv.org/abs/2007.05318} {arXiv:2007.05318 [hep-ph]}
  \BibitemShut {NoStop}%
\bibitem [{\citenamefont {Lorc\'e}\ and\ \citenamefont
  {Wang}(2022)}]{Lorce:2022jyi}%
  \BibitemOpen
  \bibfield  {author} {\bibinfo {author} {\bibfnamefont {C.}~\bibnamefont
  {Lorc\'e}}\ and\ \bibinfo {author} {\bibfnamefont {P.}~\bibnamefont {Wang}},\
  }\href {\doibase 10.1103/PhysRevD.105.096032} {\bibfield  {journal} {\bibinfo
   {journal} {Phys. Rev. D}\ }\textbf {\bibinfo {volume} {105}},\ \bibinfo
  {pages} {096032} (\bibinfo {year} {2022})},\ \Eprint
  {http://arxiv.org/abs/2204.01465} {arXiv:2204.01465 [hep-ph]} \BibitemShut
  {NoStop}%
\bibitem [{\citenamefont {Lorc\'e}\ \emph {et~al.}(2022)\citenamefont
  {Lorc\'e}, \citenamefont {Schweitzer},\ and\ \citenamefont
  {Tezgin}}]{Lorce:2022cle}%
  \BibitemOpen
  \bibfield  {author} {\bibinfo {author} {\bibfnamefont {C.}~\bibnamefont
  {Lorc\'e}}, \bibinfo {author} {\bibfnamefont {P.}~\bibnamefont {Schweitzer}},
  \ and\ \bibinfo {author} {\bibfnamefont {K.}~\bibnamefont {Tezgin}},\ }\href
  {\doibase 10.1103/PhysRevD.106.014012} {\bibfield  {journal} {\bibinfo
  {journal} {Phys. Rev. D}\ }\textbf {\bibinfo {volume} {106}},\ \bibinfo
  {pages} {014012} (\bibinfo {year} {2022})},\ \Eprint
  {http://arxiv.org/abs/2202.01192} {arXiv:2202.01192 [hep-ph]} \BibitemShut
  {NoStop}%
\bibitem [{\citenamefont {Lorc\'e}(2021)}]{Lorce:2021gxs}%
  \BibitemOpen
  \bibfield  {author} {\bibinfo {author} {\bibfnamefont {C.}~\bibnamefont
  {Lorc\'e}},\ }\href {\doibase 10.1140/epjc/s10052-021-09207-4} {\bibfield
  {journal} {\bibinfo  {journal} {Eur. Phys. J. C}\ }\textbf {\bibinfo {volume}
  {81}},\ \bibinfo {pages} {413} (\bibinfo {year} {2021})},\ \Eprint
  {http://arxiv.org/abs/2103.10100} {arXiv:2103.10100 [hep-ph]} \BibitemShut
  {NoStop}%
\bibitem [{\citenamefont {Kogut}\ and\ \citenamefont
  {Soper}(1970)}]{Kogut:1969xa}%
  \BibitemOpen
  \bibfield  {author} {\bibinfo {author} {\bibfnamefont {J.~B.}\ \bibnamefont
  {Kogut}}\ and\ \bibinfo {author} {\bibfnamefont {D.~E.}\ \bibnamefont
  {Soper}},\ }\href {\doibase 10.1103/PhysRevD.1.2901} {\bibfield  {journal}
  {\bibinfo  {journal} {Phys. Rev. D}\ }\textbf {\bibinfo {volume} {1}},\
  \bibinfo {pages} {2901} (\bibinfo {year} {1970})}\BibitemShut {NoStop}%
\bibitem [{\citenamefont {Friar}\ and\ \citenamefont
  {Negele}(1975)}]{Friar:1975bk}%
  \BibitemOpen
  \bibfield  {author} {\bibinfo {author} {\bibfnamefont {J.~L.}\ \bibnamefont
  {Friar}}\ and\ \bibinfo {author} {\bibfnamefont {J.~W.}\ \bibnamefont
  {Negele}},\ }\enquote {\bibinfo {title} {Theoretical and experimental
  determination of nuclear charge distributions},}\ in\ \href {\doibase
  10.1007/978-1-4757-4398-2_3} {\emph {\bibinfo {booktitle} {Advances in
  Nuclear Physics: Volume 8}}},\ \bibinfo {editor} {edited by\ \bibinfo
  {editor} {\bibfnamefont {M.}~\bibnamefont {Baranger}}\ and\ \bibinfo {editor}
  {\bibfnamefont {E.}~\bibnamefont {Vogt}}}\ (\bibinfo  {publisher} {Springer
  US},\ \bibinfo {address} {Boston, MA},\ \bibinfo {year} {1975})\ pp.\
  \bibinfo {pages} {219--376}\BibitemShut {NoStop}%
\bibitem [{\citenamefont {Durand}\ \emph {et~al.}(1962)\citenamefont {Durand},
  \citenamefont {DeCelles},\ and\ \citenamefont {Marr}}]{Durand:1962zza}%
  \BibitemOpen
  \bibfield  {author} {\bibinfo {author} {\bibfnamefont {L.}~\bibnamefont
  {Durand}}, \bibinfo {author} {\bibfnamefont {P.~C.}\ \bibnamefont
  {DeCelles}}, \ and\ \bibinfo {author} {\bibfnamefont {R.~B.}\ \bibnamefont
  {Marr}},\ }\href {\doibase 10.1103/PhysRev.126.1882} {\bibfield  {journal}
  {\bibinfo  {journal} {Phys. Rev.}\ }\textbf {\bibinfo {volume} {126}},\
  \bibinfo {pages} {1882} (\bibinfo {year} {1962})}\BibitemShut {NoStop}%
\bibitem [{\citenamefont {Scadron}(1968)}]{Scadron:1968zz}%
  \BibitemOpen
  \bibfield  {author} {\bibinfo {author} {\bibfnamefont {M.~D.}\ \bibnamefont
  {Scadron}},\ }\href {\doibase 10.1103/PhysRev.165.1640} {\bibfield  {journal}
  {\bibinfo  {journal} {Phys. Rev.}\ }\textbf {\bibinfo {volume} {165}},\
  \bibinfo {pages} {1640} (\bibinfo {year} {1968})}\BibitemShut {NoStop}%
\bibitem [{\citenamefont {Lorc\'e}(2009)}]{Lorce:2009bs}%
  \BibitemOpen
  \bibfield  {author} {\bibinfo {author} {\bibfnamefont {C.}~\bibnamefont
  {Lorc\'e}},\ }\href {\doibase 10.1103/PhysRevD.79.113011} {\bibfield
  {journal} {\bibinfo  {journal} {Phys. Rev. D}\ }\textbf {\bibinfo {volume}
  {79}},\ \bibinfo {pages} {113011} (\bibinfo {year} {2009})},\ \Eprint
  {http://arxiv.org/abs/0901.4200} {arXiv:0901.4200 [hep-ph]} \BibitemShut
  {NoStop}%
\bibitem [{\citenamefont {Cotogno}\ \emph {et~al.}(2020)\citenamefont
  {Cotogno}, \citenamefont {Lorc\'e}, \citenamefont {Lowdon},\ and\
  \citenamefont {Morales}}]{Cotogno:2019vjb}%
  \BibitemOpen
  \bibfield  {author} {\bibinfo {author} {\bibfnamefont {S.}~\bibnamefont
  {Cotogno}}, \bibinfo {author} {\bibfnamefont {C.}~\bibnamefont {Lorc\'e}},
  \bibinfo {author} {\bibfnamefont {P.}~\bibnamefont {Lowdon}}, \ and\ \bibinfo
  {author} {\bibfnamefont {M.}~\bibnamefont {Morales}},\ }\href {\doibase
  10.1103/PhysRevD.101.056016} {\bibfield  {journal} {\bibinfo  {journal}
  {Phys. Rev. D}\ }\textbf {\bibinfo {volume} {101}},\ \bibinfo {pages}
  {056016} (\bibinfo {year} {2020})},\ \Eprint
  {http://arxiv.org/abs/1912.08749} {arXiv:1912.08749 [hep-ph]} \BibitemShut
  {NoStop}%
\bibitem [{\citenamefont {Schwartz}(1955)}]{Scwhartz:1955}%
  \BibitemOpen
  \bibfield  {author} {\bibinfo {author} {\bibfnamefont {C.}~\bibnamefont
  {Schwartz}},\ }\href {\doibase 10.1103/PhysRev.97.380} {\bibfield  {journal}
  {\bibinfo  {journal} {Phys. Rev.}\ }\textbf {\bibinfo {volume} {97}},\
  \bibinfo {pages} {380} (\bibinfo {year} {1955})}\BibitemShut {NoStop}%
\bibitem [{\citenamefont {Kleefeld}(2000)}]{Kleefeld:2000nv}%
  \BibitemOpen
  \bibfield  {author} {\bibinfo {author} {\bibfnamefont {F.}~\bibnamefont
  {Kleefeld}},\ }\href@noop {} {\  (\bibinfo {year} {2000})},\ \Eprint
  {http://arxiv.org/abs/nucl-th/0012076} {arXiv:nucl-th/0012076} \BibitemShut
  {NoStop}%
\bibitem [{\citenamefont {Jacob}\ and\ \citenamefont
  {Wick}(1959)}]{Jacob:1959at}%
  \BibitemOpen
  \bibfield  {author} {\bibinfo {author} {\bibfnamefont {M.}~\bibnamefont
  {Jacob}}\ and\ \bibinfo {author} {\bibfnamefont {G.~C.}\ \bibnamefont
  {Wick}},\ }\href {\doibase 10.1016/0003-4916(59)90051-X} {\bibfield
  {journal} {\bibinfo  {journal} {Annals Phys.}\ }\textbf {\bibinfo {volume}
  {7}},\ \bibinfo {pages} {404} (\bibinfo {year} {1959})}\BibitemShut {NoStop}%
\bibitem [{\citenamefont {Lorc\'e}\ and\ \citenamefont
  {Pasquini}(2011)}]{Lorce:2011zta}%
  \BibitemOpen
  \bibfield  {author} {\bibinfo {author} {\bibfnamefont {C.}~\bibnamefont
  {Lorc\'e}}\ and\ \bibinfo {author} {\bibfnamefont {B.}~\bibnamefont
  {Pasquini}},\ }\href {\doibase 10.1103/PhysRevD.84.034039} {\bibfield
  {journal} {\bibinfo  {journal} {Phys. Rev. D}\ }\textbf {\bibinfo {volume}
  {84}},\ \bibinfo {pages} {034039} (\bibinfo {year} {2011})},\ \Eprint
  {http://arxiv.org/abs/1104.5651} {arXiv:1104.5651 [hep-ph]} \BibitemShut
  {NoStop}%
\bibitem [{\citenamefont {Lorc\'e}\ and\ \citenamefont
  {Pasquini}(2012)}]{Lorce:2011kn}%
  \BibitemOpen
  \bibfield  {author} {\bibinfo {author} {\bibfnamefont {C.}~\bibnamefont
  {Lorc\'e}}\ and\ \bibinfo {author} {\bibfnamefont {B.}~\bibnamefont
  {Pasquini}},\ }\href {\doibase 10.1016/j.physletb.2012.03.025} {\bibfield
  {journal} {\bibinfo  {journal} {Phys. Lett. B}\ }\textbf {\bibinfo {volume}
  {710}},\ \bibinfo {pages} {486} (\bibinfo {year} {2012})},\ \Eprint
  {http://arxiv.org/abs/1111.6069} {arXiv:1111.6069 [hep-ph]} \BibitemShut
  {NoStop}%
\bibitem [{\citenamefont {Hofstadter}(1956)}]{Hofstadter:1956qs}%
  \BibitemOpen
  \bibfield  {author} {\bibinfo {author} {\bibfnamefont {R.}~\bibnamefont
  {Hofstadter}},\ }\href {\doibase 10.1103/RevModPhys.28.214} {\bibfield
  {journal} {\bibinfo  {journal} {Rev. Mod. Phys.}\ }\textbf {\bibinfo {volume}
  {28}},\ \bibinfo {pages} {214} (\bibinfo {year} {1956})}\BibitemShut
  {NoStop}%
\bibitem [{\citenamefont {Susskind}(1968)}]{Susskind:1967rg}%
  \BibitemOpen
  \bibfield  {author} {\bibinfo {author} {\bibfnamefont {L.}~\bibnamefont
  {Susskind}},\ }\href {\doibase 10.1103/PhysRev.165.1535} {\bibfield
  {journal} {\bibinfo  {journal} {Phys. Rev.}\ }\textbf {\bibinfo {volume}
  {165}},\ \bibinfo {pages} {1535} (\bibinfo {year} {1968})}\BibitemShut
  {NoStop}%
\bibitem [{\citenamefont {Miller}(2009{\natexlab{a}})}]{Miller:2009sg}%
  \BibitemOpen
  \bibfield  {author} {\bibinfo {author} {\bibfnamefont {G.~A.}\ \bibnamefont
  {Miller}},\ }\href {\doibase 10.1103/PhysRevC.80.045210} {\bibfield
  {journal} {\bibinfo  {journal} {Phys. Rev. C}\ }\textbf {\bibinfo {volume}
  {80}},\ \bibinfo {pages} {045210} (\bibinfo {year} {2009}{\natexlab{a}})},\
  \Eprint {http://arxiv.org/abs/0908.1535} {arXiv:0908.1535 [nucl-th]}
  \BibitemShut {NoStop}%
\bibitem [{\citenamefont {Diehl}(2002)}]{Diehl:2002he}%
  \BibitemOpen
  \bibfield  {author} {\bibinfo {author} {\bibfnamefont {M.}~\bibnamefont
  {Diehl}},\ }\href {\doibase 10.1007/s10052-002-1016-9} {\bibfield  {journal}
  {\bibinfo  {journal} {Eur. Phys. J. C}\ }\textbf {\bibinfo {volume} {25}},\
  \bibinfo {pages} {223} (\bibinfo {year} {2002})},\ \bibinfo {note} {[Erratum:
  Eur.Phys.J.C 31, 277--278 (2003)]},\ \Eprint
  {http://arxiv.org/abs/hep-ph/0205208} {arXiv:hep-ph/0205208} \BibitemShut
  {NoStop}%
\bibitem [{\citenamefont {Sheng}\ \emph {et~al.}(2022)\citenamefont {Sheng},
  \citenamefont {Li}, \citenamefont {Pu},\ and\ \citenamefont
  {Wang}}]{Sheng:2022iyn}%
  \BibitemOpen
  \bibfield  {author} {\bibinfo {author} {\bibfnamefont {X.-L.}\ \bibnamefont
  {Sheng}}, \bibinfo {author} {\bibfnamefont {Y.}~\bibnamefont {Li}}, \bibinfo
  {author} {\bibfnamefont {S.}~\bibnamefont {Pu}}, \ and\ \bibinfo {author}
  {\bibfnamefont {Q.}~\bibnamefont {Wang}},\ }\href {\doibase
  10.3390/sym14081641} {\bibfield  {journal} {\bibinfo  {journal} {Symmetry}\
  }\textbf {\bibinfo {volume} {14}},\ \bibinfo {pages} {1641} (\bibinfo {year}
  {2022})},\ \Eprint {http://arxiv.org/abs/2202.03122} {arXiv:2202.03122
  [physics.class-ph]} \BibitemShut {NoStop}%
\bibitem [{\citenamefont {Bradford}\ \emph {et~al.}(2006)\citenamefont
  {Bradford}, \citenamefont {Bodek}, \citenamefont {Budd},\ and\ \citenamefont
  {Arrington}}]{Bradford:2006yz}%
  \BibitemOpen
  \bibfield  {author} {\bibinfo {author} {\bibfnamefont {R.}~\bibnamefont
  {Bradford}}, \bibinfo {author} {\bibfnamefont {A.}~\bibnamefont {Bodek}},
  \bibinfo {author} {\bibfnamefont {H.~S.}\ \bibnamefont {Budd}}, \ and\
  \bibinfo {author} {\bibfnamefont {J.}~\bibnamefont {Arrington}},\ }\href
  {\doibase 10.1016/j.nuclphysbps.2006.08.028} {\bibfield  {journal} {\bibinfo
  {journal} {Nucl. Phys. B Proc. Suppl.}\ }\textbf {\bibinfo {volume} {159}},\
  \bibinfo {pages} {127} (\bibinfo {year} {2006})},\ \Eprint
  {http://arxiv.org/abs/hep-ex/0602017} {arXiv:hep-ex/0602017} \BibitemShut
  {NoStop}%
\bibitem [{\citenamefont {Lorc\'e}(2018{\natexlab{b}})}]{Lorce:2017isp}%
  \BibitemOpen
  \bibfield  {author} {\bibinfo {author} {\bibfnamefont {C.}~\bibnamefont
  {Lorc\'e}},\ }\href {\doibase 10.1103/PhysRevD.97.016005} {\bibfield
  {journal} {\bibinfo  {journal} {Phys. Rev. D}\ }\textbf {\bibinfo {volume}
  {97}},\ \bibinfo {pages} {016005} (\bibinfo {year} {2018}{\natexlab{b}})},\
  \Eprint {http://arxiv.org/abs/1705.08370} {arXiv:1705.08370 [hep-ph]}
  \BibitemShut {NoStop}%
\bibitem [{\citenamefont {Rinehimer}\ and\ \citenamefont
  {Miller}(2009)}]{Rinehimer:2009yv}%
  \BibitemOpen
  \bibfield  {author} {\bibinfo {author} {\bibfnamefont {J.~A.}\ \bibnamefont
  {Rinehimer}}\ and\ \bibinfo {author} {\bibfnamefont {G.~A.}\ \bibnamefont
  {Miller}},\ }\href {\doibase 10.1103/PhysRevC.80.015201} {\bibfield
  {journal} {\bibinfo  {journal} {Phys. Rev. C}\ }\textbf {\bibinfo {volume}
  {80}},\ \bibinfo {pages} {015201} (\bibinfo {year} {2009})},\ \Eprint
  {http://arxiv.org/abs/0902.4286} {arXiv:0902.4286 [nucl-th]} \BibitemShut
  {NoStop}%
\bibitem [{\citenamefont {Chung}\ \emph {et~al.}(1988)\citenamefont {Chung},
  \citenamefont {Polyzou}, \citenamefont {Coester},\ and\ \citenamefont
  {Keister}}]{Chung:1988my}%
  \BibitemOpen
  \bibfield  {author} {\bibinfo {author} {\bibfnamefont {P.~L.}\ \bibnamefont
  {Chung}}, \bibinfo {author} {\bibfnamefont {W.~N.}\ \bibnamefont {Polyzou}},
  \bibinfo {author} {\bibfnamefont {F.}~\bibnamefont {Coester}}, \ and\
  \bibinfo {author} {\bibfnamefont {B.~D.}\ \bibnamefont {Keister}},\ }\href
  {\doibase 10.1103/PhysRevC.37.2000} {\bibfield  {journal} {\bibinfo
  {journal} {Phys. Rev. C}\ }\textbf {\bibinfo {volume} {37}},\ \bibinfo
  {pages} {2000} (\bibinfo {year} {1988})}\BibitemShut {NoStop}%
\bibitem [{\citenamefont {Melosh}(1974)}]{Melosh:1974cu}%
  \BibitemOpen
  \bibfield  {author} {\bibinfo {author} {\bibfnamefont {H.~J.}\ \bibnamefont
  {Melosh}},\ }\href {\doibase 10.1103/PhysRevD.9.1095} {\bibfield  {journal}
  {\bibinfo  {journal} {Phys. Rev. D}\ }\textbf {\bibinfo {volume} {9}},\
  \bibinfo {pages} {1095} (\bibinfo {year} {1974})}\BibitemShut {NoStop}%
\bibitem [{\citenamefont {Soper}(1977)}]{Soper:1976jc}%
  \BibitemOpen
  \bibfield  {author} {\bibinfo {author} {\bibfnamefont {D.~E.}\ \bibnamefont
  {Soper}},\ }\href {\doibase 10.1103/PhysRevD.15.1141} {\bibfield  {journal}
  {\bibinfo  {journal} {Phys. Rev. D}\ }\textbf {\bibinfo {volume} {15}},\
  \bibinfo {pages} {1141} (\bibinfo {year} {1977})}\BibitemShut {NoStop}%
\bibitem [{\citenamefont {Beringer}\ \emph {et~al.}(2012)\citenamefont
  {Beringer} \emph {et~al.}}]{ParticleDataGroup:2012pjm}%
  \BibitemOpen
  \bibfield  {author} {\bibinfo {author} {\bibfnamefont {J.}~\bibnamefont
  {Beringer}} \emph {et~al.} (\bibinfo {collaboration} {Particle Data Group}),\
  }\href {\doibase 10.1103/PhysRevD.86.010001} {\bibfield  {journal} {\bibinfo
  {journal} {Phys. Rev. D}\ }\textbf {\bibinfo {volume} {86}},\ \bibinfo
  {pages} {010001} (\bibinfo {year} {2012})}\BibitemShut {NoStop}%
\bibitem [{\citenamefont {Carmignotto}\ \emph {et~al.}(2014)\citenamefont
  {Carmignotto}, \citenamefont {Horn},\ and\ \citenamefont
  {Miller}}]{Carmignotto:2014rqa}%
  \BibitemOpen
  \bibfield  {author} {\bibinfo {author} {\bibfnamefont {M.}~\bibnamefont
  {Carmignotto}}, \bibinfo {author} {\bibfnamefont {T.}~\bibnamefont {Horn}}, \
  and\ \bibinfo {author} {\bibfnamefont {G.~A.}\ \bibnamefont {Miller}},\
  }\href {\doibase 10.1103/PhysRevC.90.025211} {\bibfield  {journal} {\bibinfo
  {journal} {Phys. Rev. C}\ }\textbf {\bibinfo {volume} {90}},\ \bibinfo
  {pages} {025211} (\bibinfo {year} {2014})},\ \Eprint
  {http://arxiv.org/abs/1404.1539} {arXiv:1404.1539 [nucl-ex]} \BibitemShut
  {NoStop}%
\bibitem [{\citenamefont {Miller}(2009{\natexlab{b}})}]{Miller:2009qu}%
  \BibitemOpen
  \bibfield  {author} {\bibinfo {author} {\bibfnamefont {G.~A.}\ \bibnamefont
  {Miller}},\ }\href {\doibase 10.1103/PhysRevC.79.055204} {\bibfield
  {journal} {\bibinfo  {journal} {Phys. Rev. C}\ }\textbf {\bibinfo {volume}
  {79}},\ \bibinfo {pages} {055204} (\bibinfo {year} {2009}{\natexlab{b}})},\
  \Eprint {http://arxiv.org/abs/0901.1117} {arXiv:0901.1117 [nucl-th]}
  \BibitemShut {NoStop}%
\bibitem [{\citenamefont {Geshkenbein}(2000)}]{Geshkenbein:1998gu}%
  \BibitemOpen
  \bibfield  {author} {\bibinfo {author} {\bibfnamefont {B.~V.}\ \bibnamefont
  {Geshkenbein}},\ }\href {\doibase 10.1103/PhysRevD.61.033009} {\bibfield
  {journal} {\bibinfo  {journal} {Phys. Rev. D}\ }\textbf {\bibinfo {volume}
  {61}},\ \bibinfo {pages} {033009} (\bibinfo {year} {2000})},\ \Eprint
  {http://arxiv.org/abs/hep-ph/9806418} {arXiv:hep-ph/9806418} \BibitemShut
  {NoStop}%
\bibitem [{\citenamefont {Cheng}\ \emph {et~al.}(2020)\citenamefont {Cheng},
  \citenamefont {Khodjamirian},\ and\ \citenamefont {Rusov}}]{Cheng:2020vwr}%
  \BibitemOpen
  \bibfield  {author} {\bibinfo {author} {\bibfnamefont {S.}~\bibnamefont
  {Cheng}}, \bibinfo {author} {\bibfnamefont {A.}~\bibnamefont {Khodjamirian}},
  \ and\ \bibinfo {author} {\bibfnamefont {A.~V.}\ \bibnamefont {Rusov}},\
  }\href {\doibase 10.1103/PhysRevD.102.074022} {\bibfield  {journal} {\bibinfo
   {journal} {Phys. Rev. D}\ }\textbf {\bibinfo {volume} {102}},\ \bibinfo
  {pages} {074022} (\bibinfo {year} {2020})},\ \Eprint
  {http://arxiv.org/abs/2007.05550} {arXiv:2007.05550 [hep-ph]} \BibitemShut
  {NoStop}%
\bibitem [{\citenamefont {Chai}\ \emph {et~al.}(2022)\citenamefont {Chai},
  \citenamefont {Cheng},\ and\ \citenamefont {Hua}}]{Chai:2022ipu}%
  \BibitemOpen
  \bibfield  {author} {\bibinfo {author} {\bibfnamefont {J.}~\bibnamefont
  {Chai}}, \bibinfo {author} {\bibfnamefont {S.}~\bibnamefont {Cheng}}, \ and\
  \bibinfo {author} {\bibfnamefont {J.}~\bibnamefont {Hua}},\ }\href@noop {} {\
   (\bibinfo {year} {2022})},\ \Eprint {http://arxiv.org/abs/2209.13312}
  {arXiv:2209.13312 [hep-ph]} \BibitemShut {NoStop}%
\bibitem [{\citenamefont {Jackson}(1977)}]{Jackson:1977mta}%
  \BibitemOpen
  \bibfield  {author} {\bibinfo {author} {\bibfnamefont {D.~R.}\ \bibnamefont
  {Jackson}},\ }\emph {\bibinfo {title} {{Light-cone behavior of hadronic
  wavefunctions in QCD and experimental consequences}}},\ \href@noop {} {Ph.D.
  thesis},\ \bibinfo  {school} {Caltech} (\bibinfo {year} {1977})\BibitemShut
  {NoStop}%
\bibitem [{\citenamefont {Farrar}\ and\ \citenamefont
  {Jackson}(1979)}]{Farrar:1979aw}%
  \BibitemOpen
  \bibfield  {author} {\bibinfo {author} {\bibfnamefont {G.~R.}\ \bibnamefont
  {Farrar}}\ and\ \bibinfo {author} {\bibfnamefont {D.~R.}\ \bibnamefont
  {Jackson}},\ }\href {\doibase 10.1103/PhysRevLett.43.246} {\bibfield
  {journal} {\bibinfo  {journal} {Phys. Rev. Lett.}\ }\textbf {\bibinfo
  {volume} {43}},\ \bibinfo {pages} {246} (\bibinfo {year} {1979})}\BibitemShut
  {NoStop}%
\bibitem [{\citenamefont {Efremov}\ and\ \citenamefont
  {Radyushkin}(1978)}]{Efremov:1978fi}%
  \BibitemOpen
  \bibfield  {author} {\bibinfo {author} {\bibfnamefont {A.~V.}\ \bibnamefont
  {Efremov}}\ and\ \bibinfo {author} {\bibfnamefont {A.~V.}\ \bibnamefont
  {Radyushkin}},\ }\href {https://inspirehep.net/literature/130756} {\bibfield
  {journal} {\bibinfo  {journal} {JINR-E2-11535}\ } (\bibinfo {year}
  {1978})}\BibitemShut {NoStop}%
\bibitem [{\citenamefont {Efremov}\ and\ \citenamefont
  {Radyushkin}(1980{\natexlab{a}})}]{Efremov:1978rn}%
  \BibitemOpen
  \bibfield  {author} {\bibinfo {author} {\bibfnamefont {A.~V.}\ \bibnamefont
  {Efremov}}\ and\ \bibinfo {author} {\bibfnamefont {A.~V.}\ \bibnamefont
  {Radyushkin}},\ }\href {\doibase 10.1007/BF01032111} {\bibfield  {journal}
  {\bibinfo  {journal} {Theor. Math. Phys.}\ }\textbf {\bibinfo {volume}
  {42}},\ \bibinfo {pages} {97} (\bibinfo {year}
  {1980}{\natexlab{a}})}\BibitemShut {NoStop}%
\bibitem [{\citenamefont {Efremov}\ and\ \citenamefont
  {Radyushkin}(1979)}]{Efremov:1979sn}%
  \BibitemOpen
  \bibfield  {author} {\bibinfo {author} {\bibfnamefont {A.~V.}\ \bibnamefont
  {Efremov}}\ and\ \bibinfo {author} {\bibfnamefont {A.~V.}\ \bibnamefont
  {Radyushkin}},\ }\href {https://inspirehep.net/literature/141992} {\bibfield
  {journal} {\bibinfo  {journal} {JINR-E2-12384}\ } (\bibinfo {year}
  {1979})}\BibitemShut {NoStop}%
\bibitem [{\citenamefont {Efremov}\ and\ \citenamefont
  {Radyushkin}(1980{\natexlab{b}})}]{Efremov:1979qk}%
  \BibitemOpen
  \bibfield  {author} {\bibinfo {author} {\bibfnamefont {A.~V.}\ \bibnamefont
  {Efremov}}\ and\ \bibinfo {author} {\bibfnamefont {A.~V.}\ \bibnamefont
  {Radyushkin}},\ }\href {\doibase 10.1016/0370-2693(80)90869-2} {\bibfield
  {journal} {\bibinfo  {journal} {Phys. Lett. B}\ }\textbf {\bibinfo {volume}
  {94}},\ \bibinfo {pages} {245} (\bibinfo {year}
  {1980}{\natexlab{b}})}\BibitemShut {NoStop}%
\bibitem [{\citenamefont {Lepage}\ and\ \citenamefont
  {Brodsky}(1979{\natexlab{a}})}]{Lepage:1979zb}%
  \BibitemOpen
  \bibfield  {author} {\bibinfo {author} {\bibfnamefont {G.~P.}\ \bibnamefont
  {Lepage}}\ and\ \bibinfo {author} {\bibfnamefont {S.~J.}\ \bibnamefont
  {Brodsky}},\ }\href {\doibase 10.1016/0370-2693(79)90554-9} {\bibfield
  {journal} {\bibinfo  {journal} {Phys. Lett. B}\ }\textbf {\bibinfo {volume}
  {87}},\ \bibinfo {pages} {359} (\bibinfo {year}
  {1979}{\natexlab{a}})}\BibitemShut {NoStop}%
\bibitem [{\citenamefont {Lepage}\ and\ \citenamefont
  {Brodsky}(1979{\natexlab{b}})}]{Lepage:1979za}%
  \BibitemOpen
  \bibfield  {author} {\bibinfo {author} {\bibfnamefont {G.~P.}\ \bibnamefont
  {Lepage}}\ and\ \bibinfo {author} {\bibfnamefont {S.~J.}\ \bibnamefont
  {Brodsky}},\ }\href {\doibase 10.1103/PhysRevLett.43.545} {\bibfield
  {journal} {\bibinfo  {journal} {Phys. Rev. Lett.}\ }\textbf {\bibinfo
  {volume} {43}},\ \bibinfo {pages} {545} (\bibinfo {year}
  {1979}{\natexlab{b}})},\ \bibinfo {note} {[Erratum: Phys.Rev.Lett. 43,
  1625--1626 (1979)]}\BibitemShut {NoStop}%
\bibitem [{\citenamefont {Lepage}\ and\ \citenamefont
  {Brodsky}(1980)}]{Lepage:1980fj}%
  \BibitemOpen
  \bibfield  {author} {\bibinfo {author} {\bibfnamefont {G.~P.}\ \bibnamefont
  {Lepage}}\ and\ \bibinfo {author} {\bibfnamefont {S.~J.}\ \bibnamefont
  {Brodsky}},\ }\href {\doibase 10.1103/PhysRevD.22.2157} {\bibfield  {journal}
  {\bibinfo  {journal} {Phys. Rev. D}\ }\textbf {\bibinfo {volume} {22}},\
  \bibinfo {pages} {2157} (\bibinfo {year} {1980})}\BibitemShut {NoStop}%
\bibitem [{\citenamefont {Parisi}(1979)}]{Parisi:1979jp}%
  \BibitemOpen
  \bibfield  {author} {\bibinfo {author} {\bibfnamefont {G.}~\bibnamefont
  {Parisi}},\ }\href {\doibase 10.1016/0370-2693(79)90291-0} {\bibfield
  {journal} {\bibinfo  {journal} {Phys. Lett. B}\ }\textbf {\bibinfo {volume}
  {84}},\ \bibinfo {pages} {225} (\bibinfo {year} {1979})}\BibitemShut
  {NoStop}%
\bibitem [{\citenamefont {Workman}(2022)}]{Workman:2022ynf}%
  \BibitemOpen
  \bibfield  {author} {\bibinfo {author} {\bibfnamefont {R.~L.}\ \bibnamefont
  {Workman}} (\bibinfo {collaboration} {Particle Data Group}),\ }\href
  {\doibase 10.1093/ptep/ptac097} {\bibfield  {journal} {\bibinfo  {journal}
  {PTEP}\ }\textbf {\bibinfo {volume} {2022}},\ \bibinfo {pages} {083C01}
  (\bibinfo {year} {2022})}\BibitemShut {NoStop}%
\bibitem [{\citenamefont {Brodsky}\ and\ \citenamefont
  {Farrar}(1973)}]{Brodsky:1973kr}%
  \BibitemOpen
  \bibfield  {author} {\bibinfo {author} {\bibfnamefont {S.~J.}\ \bibnamefont
  {Brodsky}}\ and\ \bibinfo {author} {\bibfnamefont {G.~R.}\ \bibnamefont
  {Farrar}},\ }\href {\doibase 10.1103/PhysRevLett.31.1153} {\bibfield
  {journal} {\bibinfo  {journal} {Phys. Rev. Lett.}\ }\textbf {\bibinfo
  {volume} {31}},\ \bibinfo {pages} {1153} (\bibinfo {year}
  {1973})}\BibitemShut {NoStop}%
\bibitem [{\citenamefont {Matveev}\ \emph {et~al.}(1973)\citenamefont
  {Matveev}, \citenamefont {Muradian},\ and\ \citenamefont
  {Tavkhelidze}}]{Matveev:1973ra}%
  \BibitemOpen
  \bibfield  {author} {\bibinfo {author} {\bibfnamefont {V.~A.}\ \bibnamefont
  {Matveev}}, \bibinfo {author} {\bibfnamefont {R.~M.}\ \bibnamefont
  {Muradian}}, \ and\ \bibinfo {author} {\bibfnamefont {A.~N.}\ \bibnamefont
  {Tavkhelidze}},\ }\href {\doibase 10.1007/BF02728133} {\bibfield  {journal}
  {\bibinfo  {journal} {Lett. Nuovo Cim.}\ }\textbf {\bibinfo {volume} {7}},\
  \bibinfo {pages} {719} (\bibinfo {year} {1973})}\BibitemShut {NoStop}%
\bibitem [{\citenamefont {Brodsky}\ \emph {et~al.}(1979)\citenamefont
  {Brodsky}, \citenamefont {Carlson},\ and\ \citenamefont
  {Lipkin}}]{Brodsky:1979nc}%
  \BibitemOpen
  \bibfield  {author} {\bibinfo {author} {\bibfnamefont {S.~J.}\ \bibnamefont
  {Brodsky}}, \bibinfo {author} {\bibfnamefont {C.~E.}\ \bibnamefont
  {Carlson}}, \ and\ \bibinfo {author} {\bibfnamefont {H.~J.}\ \bibnamefont
  {Lipkin}},\ }\href {\doibase 10.1103/PhysRevD.20.2278} {\bibfield  {journal}
  {\bibinfo  {journal} {Phys. Rev. D}\ }\textbf {\bibinfo {volume} {20}},\
  \bibinfo {pages} {2278} (\bibinfo {year} {1979})}\BibitemShut {NoStop}%
\bibitem [{\citenamefont {Dominguez}(2001)}]{Dominguez:2001zu}%
  \BibitemOpen
  \bibfield  {author} {\bibinfo {author} {\bibfnamefont {C.~A.}\ \bibnamefont
  {Dominguez}},\ }\href {\doibase 10.1016/S0370-2693(01)00576-7} {\bibfield
  {journal} {\bibinfo  {journal} {Phys. Lett. B}\ }\textbf {\bibinfo {volume}
  {512}},\ \bibinfo {pages} {331} (\bibinfo {year} {2001})},\ \Eprint
  {http://arxiv.org/abs/hep-ph/0102190} {arXiv:hep-ph/0102190} \BibitemShut
  {NoStop}%
\bibitem [{\citenamefont {Bruch}\ \emph {et~al.}(2005)\citenamefont {Bruch},
  \citenamefont {Khodjamirian},\ and\ \citenamefont {Kuhn}}]{Bruch:2004py}%
  \BibitemOpen
  \bibfield  {author} {\bibinfo {author} {\bibfnamefont {C.}~\bibnamefont
  {Bruch}}, \bibinfo {author} {\bibfnamefont {A.}~\bibnamefont {Khodjamirian}},
  \ and\ \bibinfo {author} {\bibfnamefont {J.~H.}\ \bibnamefont {Kuhn}},\
  }\href {\doibase 10.1140/epjc/s2004-02064-3} {\bibfield  {journal} {\bibinfo
  {journal} {Eur. Phys. J. C}\ }\textbf {\bibinfo {volume} {39}},\ \bibinfo
  {pages} {41} (\bibinfo {year} {2005})},\ \Eprint
  {http://arxiv.org/abs/hep-ph/0409080} {arXiv:hep-ph/0409080} \BibitemShut
  {NoStop}%
\bibitem [{\citenamefont {Newton}\ and\ \citenamefont
  {Wigner}(1949)}]{Newton:1949cq}%
  \BibitemOpen
  \bibfield  {author} {\bibinfo {author} {\bibfnamefont {T.~D.}\ \bibnamefont
  {Newton}}\ and\ \bibinfo {author} {\bibfnamefont {E.~P.}\ \bibnamefont
  {Wigner}},\ }\href {\doibase 10.1103/RevModPhys.21.400} {\bibfield  {journal}
  {\bibinfo  {journal} {Rev. Mod. Phys.}\ }\textbf {\bibinfo {volume} {21}},\
  \bibinfo {pages} {400} (\bibinfo {year} {1949})}\BibitemShut {NoStop}%
\bibitem [{\citenamefont {Pav\v{s}i\v{c}}(2018)}]{Pavsic:2017orp}%
  \BibitemOpen
  \bibfield  {author} {\bibinfo {author} {\bibfnamefont {M.}~\bibnamefont
  {Pav\v{s}i\v{c}}},\ }\href {\doibase 10.1007/s00006-018-0904-5} {\bibfield
  {journal} {\bibinfo  {journal} {Adv. Appl. Clifford Algebras}\ }\textbf
  {\bibinfo {volume} {28}},\ \bibinfo {pages} {89} (\bibinfo {year} {2018})},\
  \Eprint {http://arxiv.org/abs/1705.02774} {arXiv:1705.02774 [hep-th]}
  \BibitemShut {NoStop}%
\end{thebibliography}%
%%%%%%%%%%%%%%%%%%%%%%%%%%%%%%%%%%%%%%%%%%%%%%%%%%%%%%%%%%%%%%%%%%%%%%%%%%%%%%%%%%
\end{document}